\documentstyle[fancyheadings,epsfig,floatfig,12pt,epsf]{book}
\renewcommand{\chaptermark}[1]%
          {\markboth{#1}{}}
\renewcommand{\sectionmark}[1]
      {\markright{\thesection\ #1}}
\newcommand{\nc}{\newcommand}
\def\frac#1#2{{\textstyle {#1 \over #2}}}

\nc{\beq}{\begin{equation}}
\nc{\eeq}{\end{equation}}
\nc{\beqa}{\begin{eqnarray}}
\nc{\eeqa}{\end{eqnarray}}
\newcommand{\mysection}[1]{\setcounter{equation}{0}\section{#1}}

\def\ZZ{\hbox{\it Z\hskip -4.pt Z}}
\def\CC{\hbox{\it l\hskip -5.5pt C\/}}
\def\RR{\hbox{\it I\hskip -2.pt R }}
\def\HH{\hbox{\it I\hskip -2.pt H }}

\def\C{{\cal C}}
\def\GG{{\cal G}}
\def\LL{{\cal L}}
\def\m{{\mathrm {mus}}(\phi)}
\def\p{\partial_\theta}
\def\d{\delta_\theta}
\def\g{\mathrm{gr}}
\def\H{{\cal H}}
\def\MC{\hbox{\it I\hskip -2.pt M \hskip -7 pt I \hskip - 3.pt \CC}_n}
\def\tl{\theta_L}
\def\tr{\theta_R}
\def\tvi{\vrule height 12pt depth 6pt width 0pt}
\def\tv{\tvi\vrule}
\def\cc#1{\kern .7em\hfill #1 \hfill\kern .7em}
\begin{document}
\initfloatingfigs
%\addlanguage{french}
%\selectlanguage{french}
\begin{titlepage}
\thispagestyle{empty}
 \epsfxsize = 10cm
$$
%\hskip10truecm
\hskip - 10cm
\epsffile{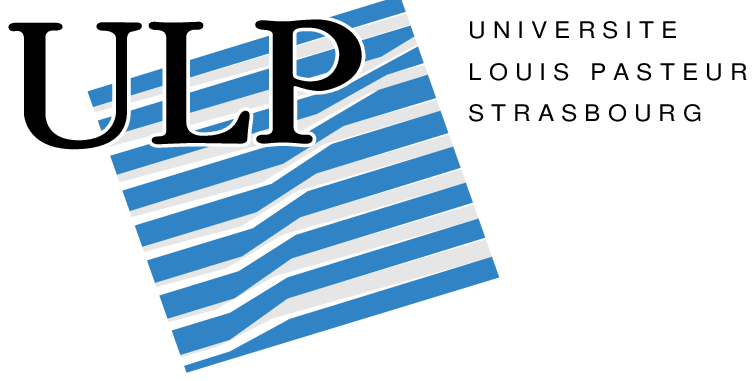}
$$

\begin{center}
\Large{\bf HABILITATION A DIRIGER DES RECHERCHES}
\vskip 2.cm
\Large{\bf Universit\'e Louis Pasteur}\\
\large{\bf Laboratoire de Physique Th\'eorique}
\vskip 2.cm
\large{\bf Pr\'esent\'ee par {\it Michel Rausch de Traubenberg}}
\vskip 2. cm
{\LARGE  \bf Alg\`ebres de Clifford }\\
%\vskip .5cm
{\LARGE \bf  Supersym\'etrie ~~et ~~Sym\'etries 
$\hbox{\it Z\hskip -7.7pt Z}_n$}\\
%\vskip .5 cm
{\LARGE \bf Applications en Th\'eorie des Champs }
\end{center}
\vskip 1.5truecm
\centerline{{\it Soutenue le 23 Octobre 1997 devant la commission 
d'examen}}
\vskip .5truecm
\tabskip 1em plus 2em minus .5em
\halign to\hsize{\hfil#\hfil&#\hfil&#\hfil\cr
& N.~Fleury& \cr
&A.~Comtet& \cr
&R.~Kerner&Rapporteur externe \cr
&J.~Richert& \cr
&M.~Rosso&Rapporteur interne \cr
&M.~G.~Schmidt&Rapporteur externe \cr
&J.~-B.~Zuber&Pr\'esident \cr
&&\cr}

\eject
\thispagestyle{empty}

%\epsfxsize = 10cm
%$$
%\epsfysize = 5cm
%hskip10truecm
%\hskip - 10cm
%\epsffile{logo.eps}
%$$
\textsc{}
\par
\begin{floatingfigure}{3cm}
\epsfysize = 3.cm
\hskip - 15cm
%\epsfysize = 5cm
%\mbox{\epsffile{b.eps}}
\mbox{\epsfig{file=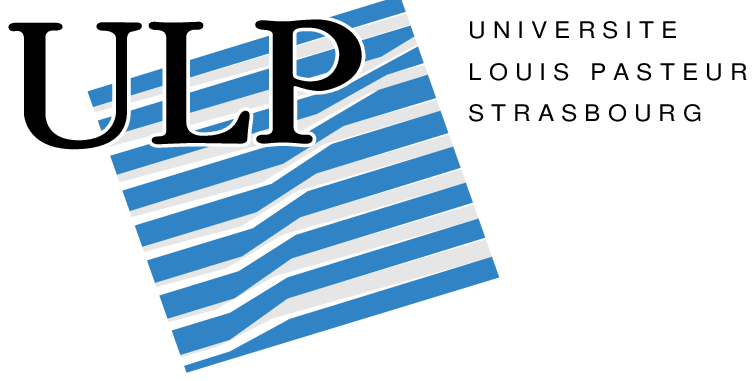}}
  \end{floatingfigure}
\par
\quad \mbox{\hskip 12cm  LPT 97-02}
\vspace{2cm}
\begin{center}

%\hfill  hep-th/  \\
\hfill
\vskip 1 cm
{\LARGE  \bf Alg\`ebres de Clifford }
\vskip 1cm
{\LARGE \bf  Supersym\'etrie ~~et ~~Sym\'etries 
$\hbox{\it Z\hskip -7.7pt Z}_n$}
\vskip 1 cm
{\LARGE \bf Applications en Th\'eorie des Champs }
\vskip 2 cm
{ \large
  {\bf M.~Rausch de Traubenberg}\footnote{rausch@lpt1.u-strasbg.fr}
   \vskip 0.7 cm
   {\it Laboratoire de Physique Th\'eorique}
\vskip .3 truecm
      { \it  3 rue de l'Universit\'e,
        67084 Strasbourg cedex, FRANCE}\\ }
  \vskip 0.3 cm
\end{center}
\vfill
\eject
\thispagestyle{empty}

\hfill
\vspace{ 10.cm}

\begin{flushright}
{\it A Monique et Yourik,}\\
{\it A V\'eronique,}\\
{\it Au petit \^etre que j'attends avec impatience ...}\\
{\it Aur\'elien, qui est venu  entre temps.~~~~~~~~~~~~~~}
\end{flushright}
\vfill
\eject
\thispagestyle{empty}
\begin{center}
{ \LARGE \bf REMERCIEMENTS}
\end{center}

\vskip 1truecm

Cette habilitation \`a diriger des recherches est le fruit d'une dizaine
d'ann\'ees de recherche. Je voudrais tout d'abord exprimer mes
remerciements \`a tous mes
collaborateurs, Norbert Fleury, Asher Perez, Pascal Simon, Marcus Slupinski 
et Robert Yamaleev ainsi que Ulrich Ellwanger et Carlos Savoy.
A cet \'egard, une pens\'ee toute particuli\`ere va \`a Norbert Fleury
que je consid\`ere comme un de mes plus ``anciens collaborateurs'',
je ne pourrai que regretter ses fonctions de plus en plus prenantes \`a
l'IUFM. Je tiens \'egalement \`a remercier tout sp\'ecialement
Carlos Savoy qui, lors de ma th\`ese de doctorat, a su me transmettre
une certaine fa\c con d'aborder la physique.\\

Je   souhaite 
remercier tous les membres de mon jury, A.~Comtet, R.~Kerner,
J.~Richert, M.~Rosso, M.~G.~Schmidt et J.-B.~Zuber, pour l'int\'er\^et
qu'ils ont manifest\'e pour ce travail et pour les discussions que nous
avons pu avoir, discussions qui m'ont permis de progresser et qui m'ont ouvert
de nouvelles perspectives.\\

Durant toutes ces ann\'ees, j'ai profit\'e de nombreuses discussions et 
conseils. Remercier nominativement tous le monde prendrait trop de
place. Cependant, j'aimerais, parmi toutes ces personnes anonymes,
mentionner P.~Grang\'e, C.~Schubert et L.~Wills~Toro.\\

Je voudrais \'egalement exprimer mes remerciements \`a  tous mes coll\`egues,
et en particulier \`a J.~-L.~Jacquot et N.~Rivier.\\

En dernier lieu, je tiens \`a remercier parents et amis, et parmi eux
tout sp\'ecialement Monique et Yourik. Une pens\'ee toute particuli\`erement
affectueuse va \`a V\'eronique.  
   
\vfill
\eject
\begin{center}
\vskip 3truecm
%{\bf R\'esum\'e}
\end{center}
\vskip 2truecm
%\footnotesize{
%\begin{abstract}
\thispagestyle{empty}
Apr\`es une br\`eve introduction concernant les alg\`ebres de 
Clifford associ\'ees \`a un
polyn\^ome quelconque, nous indiquons une m\'ethode syst\'ematique permettant
d'en obtenir une repr\'esen\-tation matricielle. Les matrices ainsi construites
peuvent \^etre consid\'er\'ees comme une extension des matrices de Dirac
car elles permettent la lin\'earisation d'un polyn\^ome arbitraire.
Cette m\'ethode  met en \'evidence deux structures 
fondamentales: les alg\`ebres de Grassmann et de Clifford 
g\'en\'eralis\'ees. Ces derni\`eres permettent de d\'efinir une 
g\'en\'eralisation naturelle des nombres complexes et des quaternions
que nous \'etudions.
Il appara\^{\i}t, en outre, que si on munit les alg\`ebres de Grassmann  
g\'en\'eralis\'ees  d'une structure diff\'erentielle, on obtient  une 
alg\`ebre de Heisenberg $q-$d\'eform\'ee (ou alg\`ebre 
des $q-$oscillateurs); celle-ci permet de construire une 
repr\'esentation de l'alg\`ebre de Clifford du polyn\^ome $x^{F-1}y$. 
Ensuite, en mettant en avant les propri\'et\'es des espaces de dimension un,
deux et trois, et en utilisant l'alg\`ebre de Heisenberg $q-$d\'eform\'ee,
nous construisons explicitement une extension des th\'eories 
supersym\'etriques, appel\'ee supersym\'etrie fractionnaire d'ordre 
$F$ (FSUSY). La FSUSY se trouve donc, de ce fait, naturellement associ\'ee
au polyn\^ome $x^{F-1}y$.
Nous appliquons alors la FSUSY en dimension $1$, dans le  formalisme
de la ligne d'univers, et en suivant la m\'ethode de quantification avec
contraintes,
nous d\'erivons une extension de l'\'equation de Dirac. En dimension $2$,
il appara\^{\i}t que la FSUSY est une th\'eorie conforme connectant des 
champs primaires de poids conformes $(0,1/F,\cdots,(F-1)/F)$, et outre
le tenseur \'energie-impulsion, on construit un courant conserv\'e de poids
conforme $1+1/F$. L'alg\`ebre (OPE)
est alors compl\`etement d\'etermin\'ee, et on montre
que la charge centrale est rationnelle pour $F=2$ (la supersym\'etrie usuelle),
$3$ et $4$.
Puis, nous nous attachons \`a construire une extension
non-triviale de l'alg\`ebre de Poincar\'e en dimension $1+2$. Apr\`es avoir
\'etudi\'e les repr\'esentations de cette alg\`ebre, on montre que la FSUSY
en $3d$ est une sym\'etrie qui agit sur des champs de spin fractionnaire
ou anyons. Pour terminer, 
et ceci quelque soit la dimension, il appara\^{\i}t qu'une 
classification naturelle des th\'eories FSUSY \'emerge en d\'ecomposant $F$,
l'ordre de la FSUSY, en ses produits de facteurs premiers; ainsi,
une th\'eorie FSUSY d'ordre $F_1F_2$ sera invariante d'ordre $F_1$
($F_2$).  On peut donc d\'efinir des sous-syst\`emes pr\'esentant
des sous-sym\'etries.
%\end{abstract}
%}
%\vskip .5 in

%\begin{abstract}
%\end{abstract}
\eject
\end{titlepage}
\setcounter{page}{1}
\tableofcontents

\chapter*{\center{ Introduction}}
\addcontentsline{toc}{chapter}{\numberline{} \hskip -.78cmIntroduction}
%\stepcounter{chapter}

Notre interpr\'etation des propri\'et\'es physiques des particules 
\'el\'ementaires
passe, par exem\-ple, par une description  alg\'ebrique. Le concept d'alg\`ebre
devient alors intimement corr\'el\'e au comportement des particules.
De ce fait,  la connaissance des sym\'etries fondamentales permet  de 
mettre en avant un certain nombre de caract\'eristiques et de faire 
des pr\'edictions que l'on pourra   confronter \`a 
l'exp\'erience. Ainsi,  ce principe directeur a permis de construire
des th\'eories plus ou moins bien confirm\'ees 
exp\'erimentalement. Etant donn\'e que les m\^emes grands principes 
sous-tendent ces diverses approches, on peut noter une certaine coh\'erence 
dans la fa\c con dont, \`a partir du mod\`ele standard, on construit la 
th\'eorie des cordes.

Toutes ces descriptions ont cependant une caract\'eristique commune car elles
sont bas\'ees sur des structures alg\'ebriques similaires. Les unes sont
associ\'ees aux alg\`ebres de Lie et les autres aux super-alg\`ebres de Lie.
Les premi\`eres  sont donc li\'ees \`a des g\'en\'erateurs bosoniques et 
les secondes, \`a des g\'en\'erateurs fermioniques. 

Avant que la th\'eorie des cordes ne devienne une th\'eorie susceptible
de d\'ecrire les particules \'el\'ementaires, une certaine libert\'e 
semblait de mise: 
on recherchait, dans le cadre des th\'eories de 
Grand-Unification, un groupe de jauge plus ou moins
grand mais reproduisant au moins celui du mod\`ele standard,
ou bien, suivant l'approche des th\'eories de Kaluza-Klein, on augmentait la
dimension de l'espace-temps ou encore on imposait une th\'eorie 
supersym\'etrique, etc..  Beaucoup de bonnes raisons allaient dans le sens de 
telle ou telle description mais aucun argument n'imposait une approche plut\^ot
qu'une autre.

Avec la th\'eorie des cordes, de nouvelles perspectives
sont apparues dans le formalisme fondamental, et les propri\'et\'es
des particules pouvaient \^etre comprises \`a partir de l'\'etude des 
sym\'etries sur la surface d'univers de la corde. Autrement dit, on a,
avec cette alternative, une corr\'elation entre la physique en dimension $2$ et
la physique dans l'espace-temps, ce qui revient \`a dire que la physique dans 
l'espace-temps est contrainte par les propri\'et\'es d'invariance \`a deux
dimensions. Parall\`element aux th\'eories de cordes, une approche similaire
s'est d\'evelopp\'ee et une description relativiste des particules, voire une
th\'eorie des champs, peut \^etre induite par
une description des particules  sur la ligne d'univers.
On voit donc que, suivant cette logique, les petites dimensions tiennent un 
r\^ole particulier. En outre, dans les espaces de petites dimensions, il est
possible de d\'efinir des \'etats qui ne pr\'esentent ni les 
ca\-ract\'eristiques
des bosons, ni celles des fermions. Cette propri\'et\'e permet
donc de d\'efinir  de nouvelles  structures alg\'ebriques ne
reposant ni sur des bosons ni sur des fermions. On peut alors se poser la
question de l'incidence de telles alg\`ebres sur les propri\'et\'es physiques
des syst\`emes qu'ils d\'ecrivent. Tout comme les super-alg\`ebres admettent
comme structures fondamentales les alg\`ebres de Clifford ou de Grassmann,
il va falloir introduire des structures analogues engendrant ces nouveaux types
d'alg\`ebres.

D'un point de vue math\'ematique, et ind\'ependamment de toutes alternatives
et perspectives physiques, s'est d\'evelopp\'ee une g\'en\'eralisation 
des alg\`ebres de Clifford et de Grassmann. Ces extensions, associ\'ees
\`a un polyn\^ome de degr\'e plus grand que deux, ont \'et\'e
appel\'ees alg\`ebres de Clifford d'un polyn\^ome et alg\`ebres
$n-$ext\'erieures. Elles  permettent de mettre en avant deux structures
de base, les alg\`ebres de Clifford et  de Grassmann g\'en\'eralis\'ees.
Ces structures permettront de construire des th\'eories que l'on peut
comprendre comme  des extensions des th\'eories supersym\'etriques.
De plus, ces g\'en\'eralisations pr\'esentent une propri\'et\'e int\'eressante:
une alg\`ebre de Clifford ou de Grassmann g\'en\'eralis\'ee,  
associ\'ee \`a un polyn\^ome de degr\'e $n_1n_2$ contient une alg\`ebre
de Clifford ou de Grassmann d'ordre $n_1$ (ou $n_2$) comme sous-alg\`ebre.
Cette classification se r\'epercutera au niveau des th\'eories g\'en\'eralisant
la supersym\'etrie. 

La premi\`ere \'etape, que nous d\'evelopperons dans le premier chapitre,
consistera \`a d\'efinir cette structure alg\'ebrique fondamentale 
associ\'ee \`a un polyn\^ome non-quadratique. On \'etudie\-ra 
alors en d\'etail
les propri\'et\'es math\'ematiques de ces structures, propri\'et\'es qui
nous permet\-tront de construire des extensions naturelles des th\'eories
quadratiques (nombres complexes, quaternions, etc.).

Dans une seconde \'etape, nous mettrons en \'evidence le fait que ces 
nouvelles
alg\`ebres induisent naturellement une structure diff\'erentielle sur
des variables non-commutatives, ou des oscillateurs
pr\'esentant des $q-$d\'eformations, et donc ouvrent  des
perspectives en g\'eom\'etrie non-commutative ou dans les groupes quantiques,
o\`u  de tels objets sont \'egalement d\'efinis. Nous n'approfondirons pas
cette possibilit\'e.

Il appara\^{\i}tra alors, dans le troisi\`eme chapitre, que ces structures 
diff\'erentielles, connect\'ees aux propri\'et\'es exceptionnelles des espaces
de dimensions $1,2,3$, vont nous permettre de d\'efinir une th\'eorie allant
au-del\`a de la supersym\'etrie, th\'eorie appel\'ee supersym\'etrie 
frac\-tionnaire. Nous allons  exploiter cette extension pour construire une
th\'eorie des champs en dimension $1$, dans le langage du formalisme
de la ligne d'univers; en dimension deux, en mettant en avant l'invariance
conforme et en dimension $1+2$, en relation avec les anyons, ou \'etats de spin
fractionnaire.

A la diff\'erence des formes quadratiques et des 
alg\`ebres $\ZZ_2-$gradu\'ees, o\`u la structure math\'ematique est unique,
d\`es que l'on consid\`ere des polyn\^omes de degr\'e sup\'erieur \`a $2$,
tous les polyn\^omes ne sont pas \'equivalents ($f$ et $g$ sont \'equivalents
si on peut trouver une transformation $U$ telle que $f(x)=g(Ux)$), et donc
de grandes classes de th\'eories sont alors envisageables. De ce fait, 
d'un point de vue formel, il est bien \'evidemment primordial d'\'etudier
ce type de th\'eories ainsi que  de mettre en \'evidence les 
concepts de base qui permettent de  construire des extensions de
la supersym\'etrie. 
En effet, potentiellement, \`a chaque famille de polyn\^omes, et pour chaque
repr\'esentation de son alg\`ebre de Clifford associ\'ee, une extension des
th\'eories supersym\'etriques est envisageable. La FSUSY, quant \`a elle, 
est \'etroitement reli\'ee au polyn\^ome $x^{F-1}y$ (ou \`a l'alg\`ebre de
Heisenberg $q-$d\'eform\'ee), et \`a sa version classique, l'alg\`ebre de 
Grassmann g\'en\'eralis\'ee.
Il est
\'egalement important de pouvoir \'etudier les  cons\'equences 
physiques 
que ces g\'en\'eralisations impliquent. 

Enfin, une approche
alg\'ebrique g\'en\'erale peut, peut-\^etre, mettre un peu d'ordre
devant le foisonnement des id\'ees ayant permis d'aller au-del\`a la 
supersym\'etrie.

\chapter[Extensions des Alg\`ebres de Clifford et de Grassmann]{\center{ 
Extensions des Alg\`ebres de Clifford et de Grassmann}}

\pagestyle{fancyplain}
La th\'eorie des nombres a toujours \'et\'e un sujet riche et fascinant,
particuli\`erement lorsque celle-ci est sous-tendue par le concept
d'alg\`ebre. L'importance de la  notion d'alg\`ebre, tant en physique qu'en
math\'ematique, n'est en effet plus \`a d\'emontrer. Ainsi, par exemple,
on peut noter le r\^ole central des alg\`ebres dans la description des
sym\'etries des lois de la physique. Pouss\'ee \` a son paroxysme, cette
id\'ee a conduit du mod\`ele standard de la physique des parti\-cules
\`a la th\'eorie des cordes, en passant par les th\'eories de Grand Unification
ou les th\'eories dites supersym\'etriques. Toutes ces descriptions ont
cependant une caract\'eristique commune: en effet, les relations de 
fermeture d\'efinissant la structure math\'ematique de ces groupes
de sym\'etrie sont toujours d\'efinies \`a partir de relations 
{\it bilin\'eaires} 
(commutateurs ou anticommutateurs). La raison profonde de cette propri\'et\'e
remarquable provient du fait que les op\'erateurs de la sym\'etrie
agissent sur un espace muni d'une m\'etrique {\it i.e.} d'une forme
quadratique. C'est pourquoi les formes quadratiques ont \'et\'e amen\'ees
\`a jouer un r\^ole essentiel.

Historiquement, lors de la construction des diff\'erents types d'alg\`ebres,
le th\'eor\`eme d'Hur\-witz (1898) a limit\'e de  fa{\c c}on drastique les 
extensions possibles des nombres r\'eels. En effet, si on munit une alg\`ebre
${\cal A}$ d'une norme,     on ne peut construire que quatre alg\`ebres
sans diviseurs de z\'ero et dont le produit des normes est la norme
du produit pour tout \'el\'ement de ${ \cal A}$ : les nombres r\'eels, les 
complexes, les quaternions et les octonions. La prise en compte de structures
math\'ematiques plus g\'en\'erales {\it n\'ecessite} que l'on abandonne
au moins l'une des conditions de validit\'e du th\'eor\`eme d'Hurwitz.

Parmi les extensions qui ont jou\'e un r\^ole crucial, on peut relever les
alg\`ebres dites de Clifford ou de Grassmann. Ces deux extensions sont
d\'efinies {\it primo} \`a partir de relations de fermeture quadratiques,
et {\it secundo} sont associ\'ees \`a un polyn\^ome quadratique, comme nous
allons le revoir ci-apr\`es. De ce fait, bien que les conditions de validit\'e
du th\'eor\`eme d'Hurwitz aient \'et\'e rel\^ach\'ees, ces g\'en\'eralisations
pr\'esentent des similarit\'es avec les quaternions, elles en constituent donc
une extension (on note ${\cal C}^p_2$ l'alg\`ebre engendr\'ee par $p$ 
g\'en\'erateurs)

$$ \RR  \longrightarrow \CC  \longrightarrow \HH \longrightarrow \cdots
\longrightarrow {\cal C}^p_2,$$
%$$\NN  ~~\ZZ~~\QQ~~\RR~~~\CC~~~\HH$$
\noindent
dans le sens o\`u les r\'eels (resp. complexes  et quaternions) s'identifient
\`a ${\cal C}^0_2$ (resp. ${\cal C}^1_2, {\cal C}^2_2)$ c'est-\`a-dire sont
engendr\'es par $0$ (resp. $1,2$) g\'en\'erateur(s). Notons $\Gamma_i,i=1,
\cdots,p$ les $p$ g\'en\'erateurs de ${\cal C}^p_2$. Ils sont soumis \`a
la contrainte suivante\footnote{{\it Sricto sensu}, il faudrait tenir compte
de la signature de la m\'etrique, mais cela ne change en rien la discussion.}

$$\left\{ \Gamma_i, \Gamma_j\right\}=2\delta_{ij}.$$

\noindent
Ces structures alg\'ebriques sont associ\'ees au
polyn\^ome quadratique $~~~~S_2(x)=\sum \limits_{i=1}^p x_i^2 =$
$\left(\sum\limits_{i=1}^p \Gamma_i x_i\right)^2$ et sont \`a la base de 
l'op\'erateur de Dirac ainsi que de la repr\'esentation spinorielle de
$SO(p)$.

Les alg\`ebres de Grassmann, quant \`a elles, se construisent \`a partir
des alg\`ebres de Clifford ${\cal C}^{2p}_2 \longrightarrow {\cal G}^p_2$.

\beqa
{\cal G}^p_2&:& \theta_i = {\Gamma_i + i \Gamma_{i+p} \nonumber \over \sqrt{2}} \\
{\cal G^\prime}^p_2&:& \theta_i^\prime ={ \Gamma_i - i \Gamma_{i+p}
\over \sqrt{2}}.
 \nonumber
\eeqa

\noindent
Pour un espace Euclidien, $\theta_i$ et $\theta_i^\prime$ sont conjugu\'ees
l'une de l'autre.
Il est facile de v\'erifier que l'on a
\beqa
\left\{\theta_i,\theta_j\right\}&=&0    \nonumber \\
\left\{\theta^\prime_i,\theta^\prime_j\right\}&=&0  \nonumber \\
\left\{\theta_i,\theta^\prime_j\right\}&=& \delta_{ij}. \nonumber
\eeqa
\noindent
Les deux premi\`eres relations constituent la d\'efinition de l'alg\`ebre de
Grassmann et nous montrent qu'une telle alg\`ebre est  associ\'ee au polyn\^ome
 nul, vu en tant que polyn\^ome quadratique car 
$ \left(\sum\limits_{i=1}^p \theta_i x_i\right)^2=0$. La derni\`ere nous 
montre explicitement que les deux s\'eries de variables constituent
en fait des oscillateurs fermioniques et donc permettent de construire
la repr\'esentation spinorielle \`a partir d'un vide de 
Clifford annihil\'e par les $a$ ($a_i \equiv \theta_i,
a^\dag_i \equiv \theta^\prime_i$).

Comme nous l'avons rappel\'e, les alg\`ebres de Clifford et de Grassmann sont
sous-tendues par des formes quadratiques; en outre, 
ces structures alg\'ebriques
admettent  une structure $\ZZ_2-$gradu\'ee, et 
apparaissent naturellement dans les th\'eories supersym\'etriques (o\`u
les bosons sont pairs et les fermions impairs). Ceci nous sugg\`ere la prise
en compte d'alg\`ebres associ\'ees \`a des polyn\^omes homog\`enes de degr\'e
$n$ et admettant une structure $\ZZ_n-$gradu\'ee. Cependant, il
existe une diff\'erence essentielle entre les polyn\^omes quadratiques et ceux
qui ne le sont pas. Les premiers, quand $n=2$, sont toujours \'equivalents au 
polyn\^ome $x_1^2 + x_2^2 + \cdots$ (modulo le nombre de signes plus et
moins), alors, pour les formes non-quadratiques ($n > 2$), 
il n'est pas toujours possible
de r\'e\'ecrire $f$ sous forme diagonale. 
On aura donc affaire \`a
des familles d'alg\`ebres. De telles
extensions ont  \'et\'e
consid\'er\'ees par les math\'ematiciens et constituent les alg\`ebres
de Clifford d'un polyn\^ome \cite{ro1,re1,c,re2,h,vdb,ht,re3,he} 
ou les alg\`ebres $n$-ext\'erieures \cite{ro2,re4}. 
Dans une premi\`ere section, nous allons rappeler
la d\'efinition de ces alg\`ebres. A la diff\'erence des 
alg\`ebres
de Clifford usuelles, les alg\`ebres de Clifford d'un polyn\^ome
{\it n'admettent pas de repr\'esentation fid\`ele de dimension finie}.
Dans une seconde section, nous allons analyser cette propri\'et\'e  et donner
une m\'ethode qui permette dans tous les cas, \- 
d' obtenir une repr\'esentation
matricielle {\it donc non fid\`ele} \cite{fr1}. Les alg\`ebres fondamentales
qui \'emergeront lors de ce processus sont les alg\`ebres dites de Clifford
g\'en\'eralis\'ees \cite{mn,y,m,pg,tl,k,gcap}; elles constitueront une
g\'en\'eralisation directe du concept d'alg\`ebre de Clifford. 
Ces familles d'alg\`ebres, associ\'ees aux polyn\^omes 
$x_1^n+\cdots $, vont nous
permettre d'obtenir une extension de la notion de nombre complexe ainsi
que de celle de quaternion. Nous avons appel\'e les premiers multicomplexes
alors que les seconds ont \'et\'e nomm\'es nonions (pour des polyn\^omes
cubiques) par Sylvester  et $n^2-$ions dans le cas g\'en\'eral par Cartan. 

\section{Alg\`ebres de Clifford d'un polyn\^ome}
\subsection{D\'efinition}
L'id\'ee naturelle qui vient \`a l'esprit pour \'etendre la notion  
d'alg\`ebre de Clifford, associ\'ee \`a un  polyn\^ome, consiste tout 
simplement \`a 
introduire une s\'erie de g\'en\'erateurs permettant d'\'ecrire notre
polyn\^ome comme une puissance parfaite. Consid\'erons  un polyn\^ome $f$
de degr\'e $n$ \`a $p$ variables. Comme tout polyn\^ome non-homog\`ene
peut \^etre rendu homog\`ene  en consid\'erant les coordonn\'ees 
projectives $x_o^nf\left({x_1\over x_0},{x_2 \over x_0},\cdots,
{x_p \over x_0}\right)$, on peut
se limiter aux polyn\^omes homog\`enes. Cette structure 
alg\'ebrique a \'et\'e introduite
pour la premi\`ere fois pour les polyn\^omes  
homog\`enes par Roby \cite{ro1} et les polyn\^omes
quelconques par Revoy \cite{re1} (en fait, les premiers, et de fa\c con
ind\'ependante, \`a avoir introduit
de telles alg\`ebres sont, Heerema \cite{he}, mais uniquement pour des 
polyn\^omes cubiques \`a trois variables, ainsi que Morinaga et N\~ono, qui
ont \'etudi\'e les polyn\^omes diagonaux \cite{mn}).
 Le polyn\^ome $f$ peut s'\'ecrire soit \`a l'aide
d'une s\'erie de mon\^omes
\beq
\label{eq:fmono}
f(x)=\sum \limits_{i_1 + \cdots +i_p=n}
  \left[ \alpha_{(i_1,\cdots,i_p)} x_1^{i_1} \cdots x_p^{i_p}\right],
\eeq
\noindent
ou bien, profitant de l'isomorphisme entre les polyn\^omes et les tenseurs
sym\'etriques d'ordre $n$, comme suit
\beq
\label{eq:fg}
f(x) =\sum \limits_{\{i\}} x_{i_1} x_{i_2} \cdots x_{i_n} g_{i_1 \cdots i_n}.
\eeq
\noindent
Par abus de langage, nous appellerons le tenseur $g_{i_1 \cdots i_n}$, 
la $n-$m\'etrique associ\'ee \`a $f$.
Introduisons alors $p$ g\'en\'erateurs $g_1,\cdots,g_p$ qui vont nous permettre
de r\'e\'ecrire {\it par d\'efinition } $f$ sous la forme d'une 
puissance $n$-i\`eme
\beq
\label{eq:line}
f(x) =\sum \limits_{\{i\}} x_{i_1} x_{i_2} \cdots x_{i_n} g_{i_1 \cdots i_n}
=\left(\sum \limits_{i=1}^p x_i g_i \right)^n.
\eeq
\noindent
Si on d\'eveloppe explicitement la puissance apparaissant dans (\ref{eq:line})
(et que l'on identifie termes \`a termes avec l'expression  (\ref{eq:fg})),
alors on obtient la relation de d\'efinition des $g$

\beq
\label{eq:g}
\left\{g_{i_1},g_{i_2},\cdots,g_{i_n}\right\} ={1 \over n!} \sum 
\limits_{{\mathrm {perm.}} \sigma} g_{i_{\sigma(1)}}g_{i_{\sigma(2)}}\cdots
g_{i_{\sigma(n)}} = g_{i_1 \cdots i_n}.
\eeq

L'alg\`ebre de Clifford du polyn\^ome $f$, not\'ee ${\cal C}_f(n,p)$, est
donc engendr\'ee par les $p$ g\'en\'era\-teurs $g_i$.
Les math\'ematiciens ont une construction  plus formelle de $C_f(n,p)$.
Nous allons bri\`evement la mentionner,  ce qui nous permettra de mettre en
avant l'unification qui sous-tend ces diff\'erentes structures alg\'ebriques.
A cet \'egard, introduisons l'ensemble des  polyn\^o\-mes sur $\CC$ \`a
$p$ ind\'etermin\'ees $\CC[x_1,\cdots,x_p]$. Utilisant l'isomorphisme
entre les tenseurs sym\'etriques $T(\CC^p)$ et  $\CC[x_1,\cdots,x_p]$, on d\'efinit
l'id\'eal bilat\`ere de $T(\CC^p)$

\beq
I=\left\{{1 \over n!} \sum 
\limits_{{\mathrm {perm.}} \sigma} g_{i_{\sigma(1)}}g_{i_{\sigma(2)}}\cdots
g_{i_{\sigma(n)}}- g_{i_1 \cdots i_n}=0\right\}, \nonumber
\eeq
engendr\'e
par les $g_i$. La contrainte de lin\'earisation peut alors se d\'efinir \`a
l'aide des espaces quotients \cite{ro1} 

$${\cal C}_f(n,p) \equiv T(\CC^p)/I.$$

Le cas quadratique reproduit une d\'efinition plus formelle des
alg\`ebres de Clifford usuelles \cite{clif}.

Le passage aux alg\`ebres  $n-$ext\'erieures, $\Lambda_n^p$, se fait  
comme pour les polyn\^omes  quadratiques, 
c'est-\`a-dire  en consid\'erant $f$ comme
\'egal \`a z\'ero. Ceci revient donc \`a poser $g_{i_1 \cdots i_n}=0$
 $\Lambda_n^p \equiv \C_{(f = 0)}(n,p)$ \cite{ro2,re4}.

On voit  appara\^{\i}tre une diff\'erence essentielle 
avec les polyn\^omes quadratiques;
en effet, les contraintes conduisant aux $g$ sont $n-$lin\'eaires et de ce
fait, on voit imm\'ediatement que, d\`es que $n > 2$, le nombre
de mon\^omes augmente avec leur degr\'e :
$g_1g_2g_1$ est {\it ind\'ependant } de $g_1^2 g_2$ car (\ref{eq:g}) ne
nous permet pas de passer $g_1$ \`a gauche de $g_2$. Or, lorsque
 $n=2$, la dimension de $\C_f(2,p) \equiv \C_2^p$ est de
$2^{E(p/2)}$ ($E$ \'etant la partie enti\`ere) et 
les $g_i$ sont repr\'esent\'es
par les matrices $\Gamma$ de Dirac, pour un polyn\^ome arbitraire, la dimension
de $\C_f(n,p)$ est {\it infinie}. Childs a  \'etabli que $\C_f(n,p)$ 
admet des repr\'esentations de dimension finie si l'une des deux 
conditions suivantes est v\'erifi\'ee $n=2$ ou $p \le 2$ \cite{c}.

\subsection{Repr\'esentations}

Cependant, nous avons \'etabli que tout polyn\^ome peut \^etre lin\'earis\'e
par des matrices appropri\'ees \cite{fr1}.  
Ce qui revient \`a dire qu'il est toujours
possible de mettre la main sur une repr\'esen\-tation {\it non-fid\`ele}
de $\C_f(n,p)$. De prime abord, la consid\'eration de telles repr\'esentations
peut para\^{\i}tre surprenante (on verra cependant leur utilit\'e 
au chapitre 3).
En outre, le fait de consid\'erer de telles repr\'esentations  
en physique n'est
pas neuf. A titre d'illustration, nous ne citerons que deux exemples. Toute 
alg\`ebre de Lie compacte peut \^etre repr\'esent\'ee par des matrices de 
taille
appropri\'ee et il est bien connu qu'une telle repr\'esentation est fid\`ele.
Toutefois, dans l'alg\`ebre enveloppante, une telle repr\'esentation est
bien entendu {\it non-fid\`ele}. Ainsi, par exemple, si on consid\`ere  
$SU(2)$, il n'y a 
aucune raison pour que $t_i$, les g\'en\'erateurs de $SU(2)$, satisfassent
 $t_i^2=1$, alors que pour la repr\'esentation
spinorielle $t_i \longrightarrow \sigma_i$, on a bien $\sigma_i^2=1$. Comme
autre exemple, nous consid\'ererons le groupe des rotations (ou plus exactement
son groupe de recouvrement universel). Pour un tel groupe, une rotation
de $2 \pi$ conduit, pour les rep\'resentations de spin demi-entier, \`a un
changement de signe, alors que pour celles de spin entier,
on {\it identifie} la rotation de $2 \pi$ avec l'identit\'e. 

Ceci nous am\`ene donc \`a la d\'efinition d'une repr\'esentation non-fid\`ele.
Une repr\'esentation d'une alg\`ebre ${\cal A}$ est non-fid\`ele si on peut
construire un homomorphisme surjectif  (dont le noyau est non-nul). Appliqu\'e
aux alg\`ebres $\C_f$, cet homomorphisme surjectif $f_m$ nous permettra de
remplacer (\ref{eq:g}) par des contraintes quadratiques. Notons
$\LL_m(f,n,p)$ une repr\'esentation de dimension $m$ de $\C_f(n,p)$:
$ \LL_m(f,n,p)= \C_f(n,p)/{\mathrm {Ker}}f_m$.

Avant de nous attaquer \`a la lin\'earisation d'un polyn\^ome arbitraire,
int\'eressons-nous aux deux polyn\^omes particuliers:

-le polyn\^ome somme: $S(x)=\sum \limits_{i=1}^p x_i^n$;

-le polyn\^ome produit: $P(x) = \prod \limits_{i=1}^n x_i$.

\noindent
$S$ et $P$  s'av\`ereront \^etre les deux polyn\^omes fondamentaux apparaissant
dans le processus de lin\'earisation.
 
On peut noter que Childs, lors de son
\'etude des alg\`ebres de Clifford d'un polyn\^ome,  a \'etabli un r\'esultat
proche du n\^otre, pour la  somme et le produit de polyn\^omes,
\`a  savoir que si on
conna\^{\i}t une repr\'esenta\-tion finie de $\C_f(n,p)$ et 
$\C_g(n^\prime,p^\prime)$ alors il en est de m\^eme pour
une repr\'esentation de dimension finie de $\C_{f+g}(n^\prime=n,p+p^\prime)$ 
et de
$\C_{fg}(nn^\prime,p+p^\prime)$. Les matrices
apparaissant dans sa d\'emonstration sont les m\^emes que celles que nous
introduirons dans les deux  sections suivantes.  Il \'etait donc tr\`es proche
de la conclusion quant \`a la lin\'earisation d'un polyn\^ome arbitraire.
Remarquons que sa conclusion sur les repr\'esentations de dimension finie
d\`es que $p \le 2$ d\'ecoule automatiquement de sa propri\'et\'e sur les
polyn\^omes produits.\\

\noindent
\addcontentsline{toc}{subsection}{\numberline{} 1.1.2.1. Polyn\^ome produit}
{\large {\bf 1.1.2.1. Polyn\^ome produit}}\\
%\subsubsection{Polyn\^ome produit}

Nous voulons trouver des matrices, de taille minimale, permettant de 
lin\'eariser le poly\-n\^ome produit. A cet \'egard, notons
 $h_a$ les g\'en\'erateurs associ\'es \`a la d\'efinition de 
$\C_P(n,n)$.
Consi\-d\'erons alors les $n^2$ \'el\'ements
$H_{ij}, i,j=1 \cdots n$  assujettis \`a la loi de multiplication 
$H_{ij} H_{kl}=\delta_{jk} H_{il}$ et d\'efinissons l'application
$h_i \longrightarrow H_{i i+1}$.  Un calcul direct nous convainc que les
$H_{i i+i}$ sa\-tisfont la contrainte (\ref{eq:g}) associ\'ee \`a $P$ et donc
constituent une repr\'esentation non fid\`ele de $\C_P(n,n)$. Le noyau d'une
telle repr\'esentation est donn\'e par $h_i h_{j} \neq 0 \longrightarrow 
H_{i i+i} H_{j j+1}=0 {\mathrm { ~pour~}} j \neq i+1$. Cette application a, en
outre, le m\'erite de souligner comment les relations de contraintes
(\ref{eq:g}), d'ordre $n$, ont \'et\'e remplac\'ees par des contraintes 
quadratiques,
avec comme corollaire l'existence d'une repr\'esentation de dimension finie
pour les $H_{ii+1}$. En fait, la relation de multiplication donn\'ee ci-dessus
constitue tout simplement la loi de multiplication des \'el\'ements de
la base canonique des matrices $n \times n$: $\left(H_{ab}\right)_{ij}=
\delta_{ai} \delta_{bj}$ et donc la matrice repr\'esentative de $h_a$ est 
donn\'ee par $ \left(H_{aa+1}\right)_{ij}= \delta_{ai}\delta_{a+1,j}$.
Il est aussi trivial de se rendre compte que la repr\'esentation exhib\'ee 
est la repr\'esentation de {\it dimension minimale} de $\C_P(n,n)$.\\

\noindent
{\large {\bf 1.1.2.2. Polyn\^ome somme-Alg\`ebre de Clifford 
g\'en\'eralis\'ee}}\\
\addcontentsline{toc}{subsection}{\numberline{}
1.1.2.2. Polyn\^ome somme-Alg\`ebre de Clifford g\'en\'eralis\'ee }

Nous allons proc\'eder de fa\c con similaire avec le polyn\^ome somme,
c'est-\`a-dire mettre en \'evidence une s\'erie de matrices conduisant
\`a une lin\'earisation finie de $S$.
En ayant la d\'efinition des matrices $\Gamma$ en t\^ete, consid\'erons les $p$
\'el\'ements (que nous notons \'egalement $\Gamma$, par analogie) 
satisfaisant

\beqa
\label{eq:GCA}
&&\Gamma_i \Gamma_j = q \Gamma_j \Gamma_i,~i<j, \\
&&\Gamma_i^n=1, \nonumber
\eeqa

\noindent
o\`u $q$ est une racine primitive $n$-i\`eme de l'unit\'e que nous choisissons
\'egale \`a $\exp{({2 i \pi \over n})}$.
Montrons que les \'el\'ements ainsi construits constituent 
une  repr\'esentation
de $\C_S(n,p)$, c'est-\`a-dire que les $\Gamma$ satisfont la contrainte 
(\ref{eq:g}) avec $g_{i_1 \cdots i_n}=\delta_{i_1 \cdots i_n}$, le symbole
de Kronecker g\'en\'era\-lis\'e.
Si nous calculons $\left\{\Gamma_{i_1},\cdots,\Gamma_{i_n}\right\}$ avec

\begin{enumerate}
\item
l'un des $i$ \'egal \`a 1, puis deux des $i$ \'egaux \`a 1 etc.,
et que l'on ram\`ene tous les $\Gamma_1$ en premi\`ere position, on
en d\'eduit (en utilisant les relations coefficients-racines) que
$\left\{\Gamma_1,\cdots, \Gamma_1,\Gamma_{i_{a+1}},\cdots, \Gamma_{i_n}\right\}=0,
a=1, \cdots n-1$;
\item
En r\'eit\'erant le processus sans $\Gamma_1$ et avec $\Gamma_2$ et
ainsi de suite, on obtient bien la propri\'et\'e d\'esir\'ee.
\end{enumerate}

On en d\'eduit que l'ensemble des $\Gamma$ constitue une
repr\'esentation de $\C_S(n,p)$. Une telle repr\'esentation met la lumi\`ere
sur les contraintes quadratiques (\ref{eq:GCA}) induisant
une repr\'esen\-tation matricielle de l'alg\`ebre de Clifford du polyn\^ome.
Cette propri\'et\'e a d'ailleurs \'et\'e \'etablie par  Revoy \cite{re3}.
Une telle alg\`ebre, not\'ee $\C_n^p$, a \'et\'e nomm\'ee alg\`ebre
 de Clifford g\'en\'eralis\'ee
\cite{mn,y,m,pg,k,gcap}. Nous avons ensuite mis en \'evidence la 
repr\'esentation {\it de dimension minimale} de $\C_S(n,p)$ \cite{fr1};
une telle d\'emonstration \'etant
longue et fastidieuse, nous ne la reproduirons pas ici.  Nous allons nous
contenter d'en rappeler les \'etapes.

{\it Primo:} nous avons \'etabli que la taille minimale des matrices est
$n \times n$;

{\it Secundo:} uniquement si le nombre de variables est 
inf\'erieur ou \'egal \`a $3$, on peut utiliser des matrices $n \times n$.

Les matrices en question sont

\beq
\sigma_1=\pmatrix{0&1&0&\cdots&0 \cr
                  0&0&1&\cdots&0 \cr
                  \vdots&&\ddots&\ddots&\vdots  \cr
                  0&0&\cdots&0&1 \cr 
                  1&0&0&\cdots&0},~ 
\sigma_3=\pmatrix{1&0&\cdots&0 \cr
                    0&q&\cdots&0 \cr
                    \vdots&&\ddots& \cr
                     0&0&\cdots&q^{n-1}},~
\sigma_2=(\sqrt{q}) \sigma_3 \sigma_1,
\eeq
\noindent
$\sqrt{q}$ \'etant pr\'esent uniquement si $n$ est pair.

A partir des matrices $\sigma$, il devient facile de lin\'eariser $S$, pour
un nombre arbitraire de variables. En effet, si nous supposons avoir
obtenu une lin\'earisation de $S(x)=M^n(x) $, on en d\'eduit la lin\'earisation
de $S(x,y_1,y_2)$:

$$S(x,y_1,y_2)= M^n(x) + y_1^n + y_2^n =
\left[\sigma_3\otimes M(x) + \sigma_1  y_1 + \sigma_2  y_2\right]^n.$$

\noindent
Ce processus it\'eratif conduit alors aux matrices $\Gamma$ (en notant
$y_{i+2}=x_i$)

\beqa
\label{eq:gamma}
\begin{array}{ll}
\Gamma_1=\sigma_1 \otimes I^{\otimes^{(k-1)}}~~&
\Gamma_2=\sigma_2 \otimes I^{\otimes^{(k-1)}} \cr
~~~~~~~~~ \vdots & ~~~~~~~~~ \vdots \cr 
\Gamma_{2l-1}=\sigma_3^{\otimes^{(l-1)}}\otimes \sigma_1 
\otimes I^{\otimes^{(k-l-1)}}~~&
\Gamma_{2l}=\sigma_3^{\otimes^{(l-1)}}\otimes\sigma_2 
\otimes I^{\otimes^{(k-l-1)}} \cr
~~~~~~~~~ \vdots & ~~~~~~~~~ \vdots \cr  
\Gamma_{2k-1}=\sigma_3^{\otimes^{(k-1)}}\otimes\sigma_1~~&
\Gamma_{2k}=\sigma_3^{\otimes^{(k-1)}}\otimes \sigma_2 \cr
\hskip 4.5cm \Gamma_{2k+1}= \sigma_3^{\otimes^{k}}. 
\end{array} 
%\hskip - 7cm   \Gamma_{2k+1}= \sigma_3^{\otimes^{k}}, 
\eeqa
\noindent
$I$ \'etant la matrice unit\'e $n \times n$.

A cette \'etape, un certain nombre de remarques s'imposent. Il est d'abord 
frappant de noter que les relations qui d\'efinissent les matrices engendrant
l'alg\`ebre de Clifford g\'en\'eralis\'ee correspondent aux expressions 
rencontr\'ees dans les alg\`ebres de Clifford usuelles, comme
l'ont montr\'e par exemple Brauer et Weyl \cite{wb}. De ce fait, la dimension
minimale de la repr\'esentation est $n^k,k=E(p/2)$. La classification de telles
alg\`ebres a \'et\'e faite par Morris \cite{m}, 
Popovici et Gh\'eorghe \cite{pg},
et pr\'esente un certain nombre de similarit\'es compar\'ee aux alg\`ebres de 
Clifford; cette classification a d'ailleurs \'et\'e \'etendue au cas 
o\`u le corps de base n'est pas celui des nombres complexes \cite{tl}. 
Les matrices $\Gamma$ avaient en outre d\'ej\`a \'et\'e mises en \'evidence par
certains auteurs \cite{mn,m} et dans les r\'ef\'erences \cite{re2,y,mn,k},
 il y 
est mentionn\'e qu'une telle alg\`ebre conduit \`a une lin\'earisation
de $x_1^n + \cdots$. 
A ce niveau, il
subsiste une petite diff\'erence d\`es que $n>2$, du fait des relations 
(\ref{eq:GCA}), $\C_n^{p>1}$ ne peut \^etre comprise que comme une $\CC$ 
alg\`ebre. Enfin, si $p$ est pair, on voit appara\^{\i}tre  
une matrice analogue \`a celle
de chiralit\'e. C'est cette analogie qui nous a fait choisir la notation
de $\Gamma$ pour les matrices (\ref{eq:gamma}).

Notons enfin que les matrices $H$, introduites dans le paragraphe concernant
la lin\'earisa\-tion du polyn\^ome produit, peuvent s'exprimer \`a partir
de $\sigma_1$ et de $\sigma_3$

\beq
\label{eq:hij}
H_{ij}= { 1 \over n} \sum \limits_{k=0}^{n-1} 
q^{-(i-1)k}\sigma_3^k\sigma_1^{i-j},
\eeq

\noindent
et de ce fait (pour les repr\'esentations exhib\'ees), on a 
$\LL_n(S,n,3) \sim \LL_n(P,n,n) \sim {\cal M}_n(\CC)$.

Comme cela \'et\'e montr\'e dans \cite{m,pg}, on a plus exactement les
isomorphismes suivants:

quand $p=2\nu$,~~$\C_n^p \sim {\cal M}_{n^\nu}(\CC)$;

quand $p=2\nu$ +1,~~ $\C_n^p \sim \oplus^p C_n^{p-1} \sim \oplus^p
{\cal M}_{n^\nu}(\CC)$.

\noindent
Ces alg\`ebres admettent naturellement une $\ZZ_n$ graduation, h\'erit\'ee de 
celle de ses g\'en\'erateurs (${\mathrm {gr}}(\Gamma)=1$), et on a 
la propri\'et\'e
remarquable $\left(\C_n^p\right)_0 \sim \C_n^{p-1}$ \cite{m,pg}, propri\'et\'e
\`a la base de l'isomorphisme pr\'ec\'edent ($\left(\C_n^p\right)_0$ 
repr\'esente les \'el\'ements de graduation z\'ero de $\C_n^p$). 
En outre, on peut noter  que le centre de $\C_n^{p=2\nu}$ est engendr\'e 
par les  $\Gamma_1^{-a} \Gamma_2^a \cdots \Gamma_p^a,~~ a=0,\cdots,n-1$ 
\cite{pg}. Finalement, il est facile de se rendre compte que
$\C_n^p \subset \C_{nn^\prime}^p$, cette propri\'et\'e sera exploit\'ee par la
suite.\\

\noindent
{\large {\bf 1.1.2.3. Polyn\^ome quelconque}}\\
\addcontentsline{toc}{subsection}{\numberline{} 
1.1.2.3. Polyn\^ome quelconque }

La lin\'earisation d'un polyn\^ome arbitraire est maintenant triviale et se
fait \`a partir des deux alg\`ebres pr\'ec\'edemment construites. 
Partant de l'expression de $f$ sous forme d'une somme de mon\^omes, chacun
de ceux-ci appara\^{\i}t alors comme un cas particulier du polyn\^ome $P$
\vfill \eject

\beqa
&&(x_1)^{a_1} (x_2)^{a_2} \cdots (x_p)^{a_p}= \nonumber \\
&&\left[x_1 \left(\sum\limits_{i=1}^{a_1}   H_{ii+1}\right)  +
     x_2  \left( \sum\limits_{i=1+a_1}^{a_1+a_2} H_{ii+1} \right)  + \cdots+
     x_p \left( \sum\limits_{i=1+a_1+\cdots+a_{p-1}}^n H_{ii+1} \right) 
        \right]^n. \nonumber
\eeqa        
Dans un premier, temps la lin\'earisation est donc faite \`a l'aide des
matrices $H$. En notant $m_l=M^n_l(x)$ le $l$-i\`eme mon\^ome
apparaissant dans (\ref{eq:fmono}),
 on peut r\'e\'ecrire $f(x)=\sum_l M^n_l(x)$.
On applique ensuite la m\'ethode de lin\'earisation aux $M_l$ 
(apr\`es les avoir
rendues commutantes en substituant $\tilde M_l = I^{\otimes^{l-1}}\otimes
M_l\otimes I^{\otimes^{k-l-1}}$ \`a $M_l$) en utilisant 
les matrices (\ref{eq:gamma})

$$f(x)= \sum_{l=1}^k\left[ \tilde M_l^n(x) \right] =
        \left[\sum_{l=1}^k \Gamma_l \otimes \tilde M_l(x)\right]^n
         =\left[ G_1 x_1 + \cdots + G_p x_p  \right]^n.$$ 
         
\noindent
Ceci termine le processus de  lin\'earisation. Lors de celui-ci,
\`a la diff\'erence des polyn\^omes $S$ et $P$, nous
ne sommes absolument pas s\^urs d'avoir obtenu la repr\'esentation de dimension
minimale de $\C_f$. En effet, suivant le chemin choisi pour calculer les
matrices $G$, on construit des matrices de taille diff\'erente. 
La raison profonde pour laquelle les matrices $G$ ne sont pas, 
{\it  a priori},  minimales est la suivante: si $\C_{f_1}(n_1,p_1)$ et 
$\C_{f_2}(n_2,p_2)$ sont
repr\'esent\'ees par des alg\`ebres de dimensions $m_1$ et $m_2$ (que 
l'on suppose minimales), alors le processus d\'evelopp\'e ci-dessus conduit
pour ($f+g$ et $fg$) \`a $\LL_{m_1m_2(n_1+n_2)}(f_1f_2,n_1n_2,p_1+p_2)$
et $\LL_{n_1m_1m_2}(f_1+f_2,n_1=n_2,p_1+p_2)$, qui ne sont pas obligatoirement
minimales, comme on peut le v\'erifier sur des exemples concrets.
Quelques
indications quant \`a l'optimisation du processus de lin\'earisation
sont propos\'ees dans \cite{fr1}. Ceci  sous-tend un probl\`eme
beaucoup plus fondamental, celui de la classification des diff\'erentes
repr\'esentations de $\C_f$, probl\`eme toujours non r\'esolu \`a l'heure
actuelle, sauf pour les polyn\^omes cubiques \cite{re3}. Un certain
nombre de r\'esultats, en vue d'une classification g\'en\'erale,
ont cependant \'et\'e obtenus.  Dans cette optique, Haille et 
Tesser ont \'etabli
que, la dimension de la repr\'esentation est un multiple
du degr\'e du polyn\^ome \cite{ht}. La plupart des \'etudes ont \'et\'e 
concentr\'ees sur les  polyn\^omes cubiques \cite{h,he,vdb}. 

Au vu de la m\'ethode propos\'ee, on peut noter le r\^ole central jou\'e
par les alg\`ebres de Clifford g\'en\'eralis\'ees $\C_n^p$ car, en
d\'efinitive, elles apparaissent comme les \'el\'ements fondamentaux li\'es
\`a la lin\'earisation. Bien que pour chaque classe de polyn\^ome, on ait
une alg\`ebre diff\'erente, on voit transpara\^{\i}tre une certaine unit\'e.
D'un point de vue historique, on peut remarquer que les deux approches,
alg\`ebres de Clifford d'un polyn\^ome \cite{ro1,re1,c,re2,h,vdb,ht,
re3,he,ro2,re4} et alg\`ebres de Clifford g\'en\'eralis\'ees 
\cite{mn,y,m,pg}, ont \'et\'e faites de mani\`ere relativement ind\'ependante
l'une de l'autre. D'ailleurs, Kwasniewski a, en quelque sorte, red\'ecouvert 
les premi\`eres alg\`ebres \`a partir des secondes \cite{k}. 
Avant de conclure cette section, mentionnons que, dans  la r\'ef\'erence 
\cite{mc},
une m\'ethode syst\'ematique permettant d'exprimer $f(x)$ comme un produit
de $n$ matrices a \'et\'e propos\'ee. Ces deux m\'ethodes (lin\'earisation
et factorisation) sont donc
 compl\'ementaires, comme le sont la lin\'erisation de Dirac et la 
factorisation quaternionique de la m\'etrique de Minkowski.\\ 

{\large {\bf 1.1.2.4. Alg\`ebre de Grassmann g\'en\'eralis\'ee}}\\
\addcontentsline{toc}{subsection}{\numberline{}  
1.1.2.4. Alg\`ebre de Grassmann g\'en\'eralis\'ee}

Le pendant des alg\`ebres de Clifford g\'en\'eralis\'ees s'av\`ere  \^etre
les alg\`ebres de Grassmann g\'en\'eralis\'ees. Cette structure, not\'ee
$\GG_n^p$, \'emerge  naturellement comme repr\'esentation de $\Lambda_n^p$.
A partir des $\Gamma$, on peut construire  $n$ s\'eries
de matrices $\theta_i^{(l)}, i=1, \cdots, p; l=0, \cdots, n-1$
v\'erifiant 

\beqa
\label{eq:qtheta}
&&\theta_i^{(l)} \theta_j^{(k)} = q\theta_j^{(k)}\theta_i^{(l)},~~ i<j,~~
\forall k,l=0,\cdots n-1 \\
&&\left(\theta_i^{(l)}\right)^n=0, \nonumber
\eeqa
\noindent
et de ce fait, en utilisant des arguments similaires aux pr\'ec\'edents,
il est facile de s'assurer que l'on a bien

$$\left\{\theta_{i_1}^{(l)}, \cdots, \theta_{i_n}^{(l)}\right\}=0.$$
\noindent
Les matrices $\theta$ s' expriment simplement \`a partir des matrices
$\Gamma$

\beq
\label{eq:theta}
\theta_i^{(l)} = \Gamma_{2i-1} +\sqrt{q} q^l \Gamma_{2i},
\eeq

\noindent
ces expressions renforcent encore l'analogie avec les formes quadratiques.
Cependant, on verra appara\^{\i}tre une diff\'erence au 
chapitre suivant.

En remarquant que la  matrice $\sigma_1+\sqrt{q}\sigma_2$ admet la valeur
propre z\'ero d\'eg\'en\'er\'ee $n$ fois, 
et que $n$ est la premi\`ere puissance qui annule $\theta$,
d'apr\`es le lemme de r\'eduction
de Jordan, il est possible d'\'ecrire 
\beq
\label{eq:thetamat}
\theta \equiv  \sigma_1+\sqrt{q}\sigma_2  \sim 
          \pmatrix{0&0&0&\cdots&0 \cr
                   1&0&0&\cdots&0 \cr
                   0&1&0&\cdots&0 \cr
                  \vdots&&\ddots&\ddots&\vdots  \cr
                  0&0&\cdots&1&0 },
\eeq

\noindent
et donc de d\'efinir
\beqa
\label{eq:theta2}
&&\theta_1=\theta \otimes I^{\otimes^{p-1} }\nonumber \\
&&~~~~~~~\vdots \nonumber \\
&&\theta_k=\sigma_3^{\otimes^{k-1}} \otimes \theta \otimes I^{\otimes^{p-k-1} }\\
&&~~~~~~~\vdots \nonumber \\
&&\theta_p=\sigma_3^{\otimes^{p-1}} \otimes \theta. \nonumber
\eeqa

%Il serait tentant d'identifier les diff\'erentes familles de $\theta$, ainsi
%construites, \`a une construction analogue $n-$dimensionnel 
%d'un oscillateur fermionique.
Kwasniewski \cite{k} a \'etudi\'e la structure alg\'ebrique engendr\'ee
par ces familles d'alg\`ebres de Grassmann g\'en\'eralis\'ees:

$$
\left\{\theta_{i_1}^{(l_1)}, \cdots, \theta_{i_n}^{(l_n)}\right\}=
(1-q^{l_1 + \cdots + \l_n}) \delta_{i_1 \cdots i_n}.
$$
\noindent
En outre, nous avons la loi de transformation inverse permettant, \`a partir
des variables $\theta$, de construire les variables $\Gamma$

\beqa
\Gamma_{2i-1}&=& {1 \over n} \sum \limits_{l=0}^{n-1} \theta_i^{(l)} \nonumber \\
\Gamma_{2i} &=&{1 \over n \sqrt{q}} 
\sum \limits_{l=0}^{n-1} q^{-l} \theta_i^{(l)}.
\nonumber
\eeqa

\noindent
Tous ceci milite en faveur d'une analogie tr\`es forte avec les formes
quadratiques et donne envie de d\'efinir une construction
analogue \`a  la repr\'esentation
spinorielle, pour les polyn\^omes arbitraires. Comme nous allons le
voir au cours du chapitre ult\'erieur, de~ telles ~ extensions 
 existent bien,
mais on se trouve dans l'impossibilit\'e de d\'efinir {\it simultan\'ement}
des op\'erateurs de cr\'eation et d'annihilation par des relations
analogues au cas spinoriel. C'est cette limitation qui donne aux alg\`ebres 
quadratiques un statut privil\'egi\'e. 

Il existe une autre repr\'esentation de $\Lambda_n^p$, l'alg\`ebre de
Para-Grassmann. Les g\'en\'erateurs de cette alg\`ebre sont astreints \`a
v\'erifier la relation cubique \cite{pgr} 
$\left[\left[\theta_i,\theta_j\right],\theta_k\right]=0$. Apr\`es 
quantification, on obtient  les parafermions introduits par
Green \cite{g} et conduisant aux parastatistiques \cite{ko}. Dans la 
litt\'erature, il appara\^{\i}t une certaine confusion entre ces deux 
extensions  alors qu'elles n'ont {\it aucun rapport}. Les premi\`eres variables
sont dans une repr\'esentation du groupe des tresses (cf. les relations 
(\ref{eq:qtheta})) alors que les secondes sont 
dans une repr\'esentation r\'eductible
du groupe des permutations (un parafermion pouvant s'exprimer comme
une somme de champs spinoriels \cite{ko}).

\subsection{ Alg\`ebre de Clifford du polyn\^ome ${ \mathbf {x^{n-1}y}}$} 

Lorsque nous \'etudierons les extensions de la supersym\'etrie ainsi que leurs
repr\'esentations, nous serons amen\'es \`a consid\'erer les repr\'esentations
{\it finies} de l'alg\`ebre de Clifford du polyn\^ome $x^{n-1}y$.
Nous rechercherons donc deux matrices $X$ et $Y$ satisfaisant

\beqa
\label{eq:XY}
&&\left\{X,X,\cdots,X,Y\right\}=1 \\
&&\left\{{\mathrm {autres} }\right\}=0, \nonumber
\eeqa
\noindent
Les relations
(\ref{eq:XY}) imposent que $X^n=0$ car, parmi tous les crochets, le seul 
qui soit non-nul est celui o\`u il y a $(n-1)$ fois la matrice $X$ et
une fois $Y$; c'est ce que traduit $\{$autres$\}$. 
Si on suppose qu'il existe $a <n$ tel que
$X^a=0$, et que l'on multiplie la premi\`ere \'equation de (\ref{eq:XY})
\`a gauche par $X$ et \`a droite par $X^{a-2}$, on arrive \`a  une 
contradiction. De ce fait, le rang de $X$ est $n-1$.  On sait que la
dimension de la repr\'esentation de $\C_{x^{n-1}y}$ est un multiple de $n$.
La dimension minimale
de $X$ est donc $n$, et  par des arguments similaires \`a celui d\'evelopp\'e
pour les 
variables de Grassmann g\'en\'eralis\'ees, on peut \'ecrire $X$ sous une
forme identique \`a celle choisie pour $\theta$ au paragraphe pr\'ec\'edent.
Pour mettre la main sur $Y$, il suffit de r\'esoudre (\ref{eq:XY}).
Une solution imm\'ediate est 

\beq
\label{eq:XYm}
X=\pmatrix{0&0&0&\cdots&0 \cr
                   1&0&0&\cdots&0 \cr
                   0&1&0&\cdots&0 \cr
                  \vdots&&\ddots&\ddots&\vdots  \cr
                  0&0&\cdots&1&0 }, ~~~
Y= \pmatrix{0&0&0&\ldots&0&1& \cr
              0&0&0&\ldots&0&0& \cr
              0&0&0&\ldots&0&0& \cr
                    &\cr\
              \vdots&\vdots&&&\ddots&\vdots& \cr
              0&0&0&\ldots&0&0&}. 
\eeq

Ph. Revoy \cite{re4} a mis en \'evidence, quand $n=3$, une famille
de repr\'esentations de dimension $3$ 

  $$     X_3=\pmatrix{0&0&0 \cr
                      1&0&0 \cr
                      0&1&0},~~~
         Y_3= \pmatrix{\mu&-\lambda \mu &1&\cr
                       0&0&\lambda\mu & \cr
                      -\lambda^{-2}\mu&\lambda^{-1}\mu  &-\mu &},
                       {\mathrm {~avec~}} 
                        2\lambda^2\mu=1.
                        $$ 
                    
Un tel r\'esultat est tr\`es probablement g\'en\'eralisable pour $n$
quelconque. Mais, fort heu\-reusement, d'autres arguments permettront, dans
le chapitre trois, de ne consid\'erer que le cas le plus simple. Celui-ci
est obtenu, lorsque $n=3$,  en posant $t=\lambda \mu$ 
(comme $2\lambda^2\mu=1$, on a
$\mu=2 t^2, \lambda={1 \over 2t}$) et en prenant la limite $t \to 0$.
Si nous avions consid\'er\'e une repr\'esentation de dimension $kn$, alors
nous aurions obtenu une repr\'esentation r\'eductible ($I_k \otimes X$ et 
$I_k \otimes Y$, avec $I_k$ la matrice identit\'e $k \times k$). 
En anticipant sur le chapitre 3, nous aurons \`a consid\'erer une seule 
repr\'esentation pour $\C_{x^{n-1}y}$.\\

\mysection{ Extension des nombres complexes}

Dans la section consacr\'ee aux extensions des alg\`ebres de Clifford,
nous avons  vu le r\^ole central jou\'e par les alg\`ebres de Clifford dites
g\'en\'eralis\'ees. Nous avons \'egalement insist\'e sur les analogies 
existant entre $n=2$  et  $n$ arbitraire. 
On rappelle que, suivant le nombre de g\'en\'erateurs, on
a, de fa\c con analogue aux  alg\`ebres de Clifford,
$$\RR \longrightarrow \C_n^1  \longrightarrow \C_n^2   \longrightarrow 
\cdots \longrightarrow \C_n^p.$$
\noindent
Parmi les alg\`ebres de Clifford, les nombres complexes jouissent 
d'un statut sp\'ecial
par rapport \`a leurs extensions. Il est donc naturel de se demander si
cette particularit\'e subsiste lorsque $n$ est quelconque. D\`es que le
nombre de g\'en\'erateurs est sup\'erieur \`a $2$, au vu des relations
de commutation (\ref{eq:GCA}), $\C_n^p$ se d\'efinit comme une $\CC-$alg\`ebre.
Par contre, quand $p=1$, on peut, sans contradiction, d\'efinir
$\C_n^1$ en tant que $\RR-$alg\`ebre. Mais, suivant que l'on
prenne $e^n=1$ ou $e^n=-1$ ($e$ \'etant le g\'en\'erateur de $\C_n^1$),
on aura deux structures diff\'erentes. Pour \^etre le plus proche possible
des nombres complexes, nous avons consid\'er\'e le second cas, et par analogie,
nous avons appel\'e multicomplexes de tels nombres \cite{fry2,fry3}.  
Notons $\MC=\left\{x= \sum \limits_{i=0}^{n-1} x_i e^i,
 e^n=-1, x_i \in \RR\right\}$ dont 
nous allons rappeler un certain nombre de propri\'et\'es. On mettra en 
\'evidence le fait que la plupart des r\'esultats 
concernant les nombres complexes
seront transposables aux multicomplexes, des propri\'et\'es alg\'ebriques
\cite{fry2} (extensions des fonctions trigonom\'etriques usuelles ...) 
aux th\'eor\`emes de l'analyse complexe \cite{fry3} (th\'eor\`emes  de 
d\'erivation, th\'eor\`emes d'int\'egration ...). Il sera m\^eme possible 
de d\'efinir des transformations conformes appropri\'ees, qui seront  
engendr\'ees par $n$ copies de l'alg\`ebre de Virasoro \cite{fry3}. Par
la suite, nous allons dresser un panorama des propri\'et\'es principales de
$\MC$; pour plus de pr\'ecisions, on pourra consulter les r\'ef\'erences
\cite{fry2,fry3}. Ind\'ependamment, et pour
$n=4$, Gervais \cite{ger} a d\'efini une alg\`ebre analogue, aboutissant
\`a des conclusions identiques aux n\^otres. Il a ensuite appliqu\'e cette
structure \`a la th\'eorie des $3-$branes, retrouvant l'amplitude de
Shapiro-Virasoro. Sa relation n'\'etait cependant pas enti\`erement 
l\'egitime car il n'avait pas obtenu les transformations
homographiques adapt\'ees.

\subsection{Propri\'et\'es alg\'ebriques des multicomplexes}

En utilisant les r\'esultats g\'en\'eraux concernant les alg\`ebres de 
Clifford g\'en\'eralis\'ees \'etablis par Morris \cite{m}, on obtient 
suivant la parit\'e de $n$, les isomorphismes suivants si:

$n$ pair : $\MC \sim \oplus^{n/2} \CC$

$n$ impair : $\MC \sim \oplus^{n} \RR$.

\noindent
Cependant, si d\`es le d\'epart, on utilise cet isomorphisme, on perd un
certain nombre de propri\'et\'es inh\'erentes \`a la structure  $n-$lin\'eaire
de $\MC$.  Pour \'etablir les propri\'et\'es li\'ees \`a la structure 
$\ZZ_n$, nous utiliserons, suivant le cas, les deux repr\'esentations 
matricielles \'equivalentes de $\MC$

\beq
\label{eq:mc}
e \longrightarrow E=\pmatrix{q^{{1 \over 2}}&0&0&\cdots&0 \cr
                  0&q^{{3 \over 2}}&0&\cdots&0 \cr
                  \vdots&&\ddots&\ddots&\vdots  \cr
                  0&0&\cdots&q^{-{3 \over 2}}&0 \cr 
                  0&0&0&\cdots&q^{-{1 \over 2}}},~ 
 E^\prime=\pmatrix{0&1&0&\cdots&0 \cr
                  0&0&1&\cdots&0 \cr
                  \vdots&&\ddots&\ddots&\vdots  \cr
                  0&0&\cdots&0&1 \cr 
                  -1&0&0&\cdots&0}. 
\eeq
\noindent
Si $e \longrightarrow E$, $x$ sera repr\'esent\'e par une matrice
$n \times n$ diagonale not\'ee $X$,
telle que $X_{aa}=X^*_{n+1-a~~ n+1-a}$. Alors
qu'avec le second choix 
$$e \longrightarrow E^\prime, 
x \longrightarrow X^\prime=\pmatrix{x_0&x_1& \cdots& x_{n-1} \cr
                                     -x_{n-1} &x_0&\cdots&x_{n-2} \cr
                                       \vdots &&&\vdots \cr
                                      \vdots&&\ddots&x_1 \cr
                                     - x_1&\cdots& -x_{n-1}& x_0}.$$
La plupart des d\'emonstrations pourront \^etre faites \`a partir de cette
constatation excessivement simple.\\

{\large {\bf 1.2.1.1 Norme.}}\\
\addcontentsline{toc}{subsection}{\numberline{}  
1.2.1.1  Norme}
 
Dans un premier temps, nous munissons $\MC$ d'une norme 
(ou plus pr\'ecis\'ement, d'une pseudo-norme)

\beq
\label{eq:norme1} 
||x||^n=\det X^\prime = \det X,
\eeq
 
\noindent
pour les nombres complexes, nous retrouvons bien $||x||^2=x_0^2+x_1^2$ alors
que, lorsque $n=3$, on obtient  $||x||^3=x_0^3-x_1^3+x_2^3+3x_0x_1x_2$.
De par la d\'efinition de la norme, on observe que, si $x$ est un diviseur
de z\'ero, sa norme est nulle. Si $n$ est pair (resp. impair),
la norme de $x$ est positive ou nulle (resp. quelconque). Cette d\'efinition 
peut, tout comme pour les nombres complexes, se r\'e\'ecrire en 
consid\'erant une  conjugaison {\it ad hoc}. Introduisons le conjugu\'e 
$\ell-$i\`eme de $\MC \equiv\MC^0$ par 

\beqa
 \MC^0 &&\longrightarrow \MC^\ell \nonumber \\
 x\equiv \left[x\right]_0=\sum\limits_{i=0}^{n-1} x_i e^i&& \longmapsto  
\left[x\right]_\ell=\sum\limits_{i=0}^{n-1} x_i q^{i\ell}e^i. \nonumber
\eeqa
\noindent
On voit que quand $n =2$, on a identification entre $\CC$ et $\CC^*$, 
propri\'et\'e qui se perd d\`es que $n \ne 2$. Les espaces ainsi  construits
constituent,  par ailleurs, des d\'efinitions isomorphes des multicomplexes.
En utilisant la repr\'esentation matricielle $x \longrightarrow X$, on
obtient sans peine les matrices repr\'esentatives de $\left[x\right]_\ell$,
et il est facile de d\'emontrer que l'on a

\beq
\label{eq:norme2} 
||x||^n= \prod \limits_{\ell=0}^{n-1}\left[x\right]_\ell.
\eeq
\noindent
On voit donc que tout comme pour les nombres complexes, il est possible
de d\'efinir la norme de $x$ \`a partir du produit de ses conjugu\'es,
mais uniquement si $n=2$, $x,x^* \in \CC$. Ceci n'est
en fait pas  compl\`etement exact car parmi les $\MC$, d\`es que
$n$ est une puissance de deux on peut substituer
\`a la d\'efinition du conjugu\'e $\ell-$i\`eme pr\'ec\'edente,
une d\'efinition alternative  $\left[x\right]_\ell=\sum\limits_{i=0}^{n-1} 
x_i e^{i(2\ell+1)} \in \MC$ \cite{fry2}.\\

{\large {\bf 1.2.1.2 Repr\'esentations polaire et cart\'esienne.}}\\
\addcontentsline{toc}{subsection}{\numberline{}  
1.2.1.2 Repr\'esentations polaire et cart\'esienne }

Peut-on pousser l'analogie avec les nombres complexes un peu plus loin,
identifier la norme  avec le ``module'' de $x$,  
et  d\'efinir une repr\'esentation exponentielle?
En utilisant les matrices 
$X$ et le fait  que l'on puisse d\'efinir le logarithme
de toute matrice non-singuli\`ere, on peut r\'e\'ecrire (et calculer), lorsque
$x$ est inversible, c'est-\`a-dire quand $||x|| \ne 0$ 

\beq
\label{eq:expo}
x=\sum \limits_{i=0}^{n-1}x_i e^i = \rho 
\exp{\left(\sum \limits_{i=1}^{n-1} \phi_i e^i\right)}. 
\eeq
\noindent
Du fait que la fonction logarithme est multi-valu\'ee, on
a plusieurs repr\'esenta\-tions possibles pour la forme  exponentielle
de $x$ \cite{fry2}. Etant donn\'e que  
$$ {\mathrm {si~~}} x \longrightarrow
\rho  \exp{\left(\sum \limits_{i=1}^{n-1} \phi_i e^i\right)}
 {\mathrm {~~ alors~~} }
\left[x\right]_\ell \to
\rho  \exp{\left(\sum \limits_{i=1}^{n-1} \phi_iq^{i\ell} e^i\right)},$$ 
\noindent
on peut  conclure imm\'ediatement 

\beq
\label{eq:norme3}
||x||^n=\rho^n.
\eeq
\noindent
De plus, en utilisant la  repr\'esentation $e \longrightarrow E$, 
on observe que parmi les $n-1$ variables $\phi$, on peut en extraire
$E({n \over 2})$ qui sont compactes. La substitution
$\phi_i \longrightarrow \phi_i + \epsilon_i^a, a=1,\cdots,E({n \over 2})$
laisse donc invariante la 
forme exponentielle. Les $\epsilon_i^a$ peuvent \^etre
calcul\'ees en expli\-citant 
$\epsilon^a = \sum \limits_{i=1}^{n-1}\epsilon_i^a
e^i$. Les $\epsilon^a$ 
s'expriment directement \`a partir des \'elements $e,\cdots, e^{n-1}$  et
leur matrices repr\'esentatives s'\'ecrivent  $X^{\prime a}_{kl}=2 i
 \pi \delta_{kl}
\left(\delta_{ka}-\delta_{n-a+1,l}\right)$.\\

\noindent
En  outre, \`a partir de (\ref{eq:expo}), ainsi que de $||e||^n = (-1)^n$,
il appara\^{\i}t que l'on peut mettre  $(-1)^ne$ sous forme exponentielle
avec comme corollaire 
\beqa
\label{eq:inclu}
  &n& {\mathrm{ ~~ pair}}~~ 
{\hbox{\it I\hskip -2.pt M \hskip -7 pt I \hskip - 3.pt \CC}_p}  
\subset \MC,
~~p < n \\
  &n& {\mathrm { ~~ impair}}~~
{\hbox{\it I\hskip -2.pt M \hskip -7 pt I \hskip - 3.pt \CC}_{2p+1}} 
\subset
 \MC,~~ 2p+1 < n. \nonumber
\eeqa
\noindent
L'existence de la forme exponentielle pour $e$ redonne les r\'esultats sur
les  directions  compactes.\\

\vfill \eject
{\large {\bf 1.2.1.3 Fonctions trigonom\'etriques}}\\
\addcontentsline{toc}{subsection}{\numberline{}  
1.2.1.3 Fonctions trigonom\'etriques }

Ayant \'elabor\'e, pour les nombres multicomplexes, une repr\'esentation 
``cart\'esienne'' et ``polaire'', l'analogie avec les nombres complexes se 
poursuit. En effet, il est possible de d\'efinir des fonctions g\'en\'eralisant
les fonctions trigonom\'etriques usuelles. A cet \'egard, il suffit tout
simplement de d\'evelopper explicitement la forme exponentielle (\ref{eq:expo})

\beq
\label{eq:mus}
x=\rho \exp{\left(\sum \limits_{i=1}^{n-1} e^i \phi_i\right)}=
\rho \sum \limits_{i=0}^{n-1}\m_i e^i,
\eeq

De telles fonctions, constituant une g\'en\'eralisation directe des fonctions
trigonom\'etriques, v\'erifient un certain nombre de propri\'et\'es qui
se d\'eduisent sans peine \cite{fry1}. Nous les \'enoncerons sans 
d\'emonstration
\begin{enumerate}
\item
$\m_i= {1 \over n} \sum \limits_{\ell=0}^{n-1} 
\left[{x \over e^i}\right]_\ell$ lorsque $||x||=1$;
\item
les fonctions mus admettent les $E(n/2)$  directions de p\'eriodicit\'e 
$\epsilon^a$;
\item
Les fonctions mus engendrent le groupe ab\'elien qui pr\'eserve la norme;
\item
${\mathrm {mus}}(\phi+\psi)_i=\sum \limits_{j=0}^{n-1}{\mathrm {sign}}(i-j) 
{\mathrm {mus}}(\phi)_i{\mathrm {mus}}(\psi)_{i-j}$, avec sign la fonction 
signe et $i-j$ d\'efini modulo $n$;
\item
${\partial \m \over \partial \phi_j}=
{\mathrm {sign}}(i-j){\mathrm {mus}}(\phi)_{i-j}$;
\item
$$\det \pmatrix{{\mathrm {mus}}(\phi)_0&{\mathrm {mus}}(\phi)_1& \cdots&
{\mathrm {mus}}(\phi)_{n-1} \cr
              -{\mathrm {mus}}(\phi)_{n-1} &{\mathrm {mus}}(\phi)_0&\cdots&
{\mathrm {mus}}(\phi)_{n-2} \cr
               \vdots &&&\vdots \cr
               \vdots&&\ddots&{\mathrm {mus}}(\phi)_1 \cr
          -{\mathrm {mus}}(\phi)_1&\cdots& -
            {\mathrm {mus}}(\phi)_{n-1}&{\mathrm {mus}}(\phi)_0}=1,$$
Bien entendu, cette relation conduit \`a $\cos^2( \phi) +\sin^2 (\phi)=1$,
alors que si $n=3$, on a
${\mathrm {mus}}^3(\phi_1,\phi_2)_0 
  -{\mathrm {mus}}^3(\phi_1,\phi_2)_1 +
{\mathrm {mus}}^3(\phi_1,\phi_2)_2  + 3 {\mathrm {mus}}(\phi_1,\phi_2)_0  
 {\mathrm {mus}}(\phi_1,\phi_2)_1 
{\mathrm {mus}}(\phi_1,\phi_2)_2=1.$
\item
Si $n$ n'est pas un nombre premier ($n=n_1n_2$), les fonctions mus d'ordre $n$
contiennent les fonctions mus d'ordre $n_1$: il suffit de prendre $\phi_i=0$
quand $i$ n'est pas un multiple de $n_2$. 
\end{enumerate}

Les fonctions ainsi obtenues permettent de passer des coordonn\'ees
``cart\'esiennes'' aux ``polaires''; il est \'egalement
possible de construire  des fonctions g\'en\'eralisant la fonction 
arc\-tangente
et exprimant les variables ``polaires'' en fonction des cart\'esiennes 
\cite{fry1}. De mani\`ere analogue, une extension
des fonctions hyperboliques peut \^etre envisag\'ee \cite{fry1}.
                                                
Notons enfin que les fonctions ainsi construites se d\'efinissent 
\`a partir des 
fonctions circulaires d'ordre $n$
$f_i^n(x)=\sum \limits_{k=0}^\infty(-1)^k \frac{{x^{nk+i}}}{(nk+i)!}$
consid\'er\'ees dans \cite{eder}. En fait, de telles fonctions, ainsi que leurs
homologues hyperboliques d'ordre $n$, avaient d\'ej\`a \'et\'e
introduites  dans
la litt\'erature \cite{mus}. Des relations analogues \`a la contrainte
$\det({\mathrm {mus}})=1$ ainsi qu'aux formules de somme ou aux
\'equations d'Euler ont \'et\'e \'etablies. A cet \'egard, ces auteurs
avaient introduit des extensions des nombres complexes (Bruwier, $e^n=1$).
Cependant, \'etant donn\'e qu'ils consid\'eraient des fonctions \`a une
variable, ils ne pouvaient pas mettre en avant les 
propri\'et\'es de p\'eriodicit\'e que nous avons \'etablies. Tout comme
les fonctions mus, ces g\'en\'eralisations des fonctions trigonom\'etriques
\`a une variable sont utiles dans la th\'eorie des \'equations 
diff\'erentielles. Elles peuvent, en outre, \^etre \'etendues \`a n'importe
quel syst\`eme lin\'eaire \cite{bru}.

\subsection{Analyse multicomplexe}

Ayant construit une structure alg\'ebrique poss\'edant de telles 
propri\'et\'es,
on est naturellement conduit \`a s'interroger sur les propri\'et\'es
sp\'ecifiques
\`a une analyse multicomplexe. Le fait de se poser une telle question n'est
pas neuf, et un certain nombre de r\'esultats ont \'et\'e obtenus pour
d'alg\`ebres de Clifford \cite{aclif} et m\^eme pour les octonions 
\cite{aoct}. Notons finalement que dans la r\'ef\'erence \cite{hor},
 une analyse
a \'et\'e construite sur une alg\`ebre plus g\'en\'erale que celle que nous
avons consid\'er\'ee (au lieu de l'alg\`ebre engendr\'ee par le
polyn\^ome $x^n+1, \MC= \RR[x]/(x^n+1)$, l'alg\`ebre associ\'ee \`a un 
polyn\^ome arbitraire de degr\'e $n$ et \`a une variable
a \'et\'e envisag\'ee).

On consid\`ere donc $F(x)=\sum \limits f_i(x)e^i$ une fonction sur $\MC$. La 
premi\`ere diff\'erence par rapport aux nombres complexes, concerne la notion
de continuit\'e et de limite; en effet, la norme ne d\'efinissant 
pas une distance, les
notions de continuit\'e et de limite ne pourront donc \^etre d\'efinies que
composante par composante, {\it i.e} sur les fonctions $f_i$ individuellement,
et non sur la fonction $F$.  Le second point, plus d\'elicat celui-ci, concerne
l'existence de diviseurs de z\'ero et donc de nouveaux types de singularit\'es.
Cependant, un prolongement respectant la structure alg\'ebrique a \'et\'e
propos\'e \cite{fry2}. Celui-ci est bas\'e sur l'inclusion (singuli\`ere)

\beq
{\hbox{\it I\hskip -2.pt M \hskip -7 pt I \hskip - 3.pt \CC}_{n-2}}
\subset \MC, \\
\eeq
\noindent 
dont les matrices repr\'esentatives  du g\'en\'erateur  et de l'identit\'e
sont donn\'ees par
$$        E_{n-2}=
        \pmatrix{0&0&0&\cdots&0 \cr
                  0&\exp{({i\pi\over n-2})}&0&\cdots&0 \cr
                  \vdots&&\ddots&\ddots&\vdots  \cr
                  0&0&\cdots&\exp{(-{i\pi\over n-2})}&0 \cr 
                  0&0&0&\cdots&0},
             I_{n-2}=
         \pmatrix{0&0&0&\cdots&0 \cr
                  0&1&0&\cdots&0 \cr
                  \vdots&&\ddots&\ddots&\vdots  \cr
                  0&0&\cdots&1&0 \cr 
                  0&0&0&\cdots&0}.$$\\

Avant d'envisager  les th\'eor\`emes de d\'erivation et 
d'int\'egration adapt\'es,
introduisons un certain nombre de notations. Nous avons d\'ej\`a d\'efini
$\left[x\right]_\ell= \sum \limits_{i=0}^{n-1} q^{i\ell} x_i e^i$, posons
alors $\left[\partial \right]_\ell={1 \over n} \sum \limits_{i=0}^{n-1} 
q^{-i\ell} \partial_i e^{-i}$. Il est facile de se convaincre que l'on a
$\left[\partial \right]_\ell\left[x\right]_k = \delta_{k \ell}$.\\

Nous pouvons donc \'enoncer, sans d\'emonstration, les th\'eor\`emes de 
d\'erivation et d'int\'e\-gration.\\

{\bf Th\'eor\`eme de Cauchy-Riemann} \cite{fry2}

\noindent
$F$ est d\'erivable en $x$ si l'une des conditions suivantes et
 \'equivalentes  est v\'erifi\'ee (il faut que la valeur de 
${d F(x) \over dx}$ ne d\'epende pas de la fa\c con dont $x$ tend vers
z\'ero;
il faut \'egalement faire attention aux diviseurs de z\'ero) 
\begin{enumerate}
\item
$\partial_0 f_k(x_0,\cdots,x_{n-1})= {\mathrm {sign}}(l-m)
\partial_l f_m(x_0,\cdots,x_{n-1})$. Ces relations constituent une
g\'en\'e\-ralisation directe des \'equations de Cauchy-Riemann.
\item
$\left[\partial \right]_\ell \left[f(x_0,\cdots,x_{n-1})\right]_k = 
\delta_{k \ell}$. La fonction  $\left[f(x_0,\cdots,x_{n-1})\right]_0$ est donc 
holomorphe et ne d\'epend  que
de $\left[x\right]_0$.
\item
$F(x) dx= \sum \limits_{i=0}^{n-1} \omega_i e^i$ d\'efinit $n$~ une-formes
ferm\'ees.
\end{enumerate}
 
 {\bf Th\'eor\`eme de Cauchy} \cite{fry2}

\noindent
On consid\`ere  un hypervolume  ferm\'e et 
orient\'e $\partial \Omega$, ainsi que  $d\sigma= \sum \limits_{i=0}^{n-1}
(-1)^i d \hat x^i e^i$, o\`u $d \hat{x}^i=dx_0\wedge dx_1\wedge \cdots\wedge
dx_{i-1}\wedge dx_{i+1}\wedge \cdots \wedge dx_{n-1}$, 
sa $(n-1)-$forme surface.

$$\int \limits_{\partial \Omega} F(x) d\sigma = 0, $$

\noindent
si l'une des conditions \'equivalentes est satisfaite
\begin{enumerate}
\item
$DF(x) \equiv \sum \limits_{i=0}^{n-1} e^i \partial_i F(x)=0$;
\item
Les $n$~ une-formes $\omega$ sont co-ferm\'ees.
\end{enumerate}

En conclusion, une fonction $F$ sera dite analytique si elle v\'erifie le 
th\'eor\`eme de Cauchy Riemann et co-analytique si elle satisfait aux 
conditions du th\'eor\`eme de Cauchy. Bien entendu, ces deux 
notions co\"{\i}ncident sur $\CC$, mais ce n'est plus vrai d\`es que 
$n >2$. A partir de la d\'efinition de $D$, un calcul direct montre que
si  $n$ est pair, la condition d'analyticit\'e implique celle
de co-analyticit\'e alors que quand $n$ est impair, ces deux th\'eor\`emes sont
incompatibles. Autrement dit, l'analyse dans  $\MC$  pr\'esente
de l'int\'er\^et si $n$ est un nombre pair. 

On se limite donc aux nombres pairs, en
remarquant que l'on peut \'ecrire $x= \sum \limits_{i=0}^{{n \over 2}-1}z_i 
e^i$,
avec $z_i=x_1 + e^{n/2} x_{i + {n \over 2}}$ un nombre complexe usuel 
$i \equiv e^{n/2}$, si $F$ est analytique en $x$, alors $D F(x)=0$ et donc
$F(x)$ est analytique en $z_i$ au sens des nombres complexes. Profitant
de cette propri\'et\'e, on d\'efinit $ \Gamma= \partial \Omega_0 \times
\cdots \times \partial \Omega_{{n \over 2}-1} \subset \Omega$ avec
$\Omega_i=\left\{x_i^2 + x_{{n \over 2} +i}^2 < R^2\right\}$ tel 
que $F$ restreint
\`a $\partial \Omega_i$ soit holomorphe au sens des $z_i$ et on en d\'eduit le

{\bf Th\'eor\`eme des r\'esidus}

$$F(u) = {1 \over (2e^{n/2} \pi)^{n/2}} \int \limits_\Gamma
{dz_0 \over (z_0-u_0)} \cdots {dz_{{n \over 2}-1} \over
 (z_{{n \over 2}-1} -u_{{n \over 2}-1})} F(x).$$

On peut remarquer l'analogie frappante qu'il y a entre le th\'eor\`eme des 
r\'esidus ainsi obtenu et celui associ\'e \`a $\CC^{n/2}$
\cite{cn}. Compar\'e aux th\'eor\`emes analogues d\'eriv\'es pour
les alg\`ebres de Clifford, nous avons un certain nombre de diff\'erences.
Par exemple, la connaissance de $F$ sur la vari\'et\'e 
$B= \times_{i=0}^{{n \over 2}-1} \left\{x_i^2 + x_{i + {n \over 2}}^2<R^2
\right\}$ est obtenue \`a partir d'une int\'egrale sur une vari\'et\'e de 
dimension $n/2$, le squelette de $B$ au lieu de l'hypersurface
$\partial B$.

Enfin, \`a partir du th\'eor\`eme des r\'esidus, il est possible 
d'\'ecrire

\beq
\label{eq:lau1}
F(x)= \sum \limits_{\{i\}\in \ZZ^{n/2}} F_{i_0i_1\cdots i_{{n \over 2}-1}}
z_0^{i_0} \cdots z_{{n \over 2}-1}^{i_{{n \over 2}-1}}
e^{i_0+\cdots+i_{{n \over 2}-1}}.
\eeq

\noindent
Si en plus la fonction est analytique en $x$, on peut r\'e\'ecrire
\beq
\label{eq:lau}
F(x)= \sum \limits_{N \in \ZZ} F_N x^N.
\eeq

\noindent
Bien entendu, il est primordial d'avoir la propri\'et\'e d'analyticit\'e pour
pouvoir calculer les coefficients $F_N$, car si la fonction est uniquement
co-analytique, les $F_N$ n'existent pas. On peut s'en  rendre contre sur
l'exemple ($n=4$) $F(z_0 + e z_1)= z_0z_1$ qui est co-analytique sans 
\^etre analytique.
 
\subsection{Invariance conforme}
L'invariance conforme, pr\'eservant la m\'etrique \`a un facteur d'\'echelle
pr\`es, prend une dimension toute particuli\`ere dans le plan complexe. En
effet, \`a la diff\'erence des espaces de dimension plus grande, en $2D$,
le groupe conforme est de dimension infinie.  De ce fait, l'invariance
conforme est devenue un outil indispensable lors de 
l'\'etude des ph\'enom\`enes
critiques \cite{id} ou de la th\'eorie des cordes \cite{gsw}. Ce caract\`ere
singulier provient du fait que dans un espace bi-dimensionnel, le degr\'e
de la m\'etrique (forme quadratique) et celui de la forme ``volume''
co\"{\i}ncident. 

Etant donn\'e que les multicomplexes ont une norme d\'efinie par un polyn\^ome
homog\`ene de degr\'e $n$ \'equivalent au polyn\^ome $p(x)=x_1x_2\cdots x_n$
et que, en outre, un tel polyn\^ome d\'efinit parfaitement un $n-$volume,
on est en droit de se demander si des propri\'et\'es similaires \`a celles
propres aux espaces bi-dimensionnels existent. Il est  donc
int\'eressant d'\'etudier le groupe des transformations pr\'eservant
la norme (\ref{eq:norme2}) \`a un facteur d'\'echelle pr\`es, ainsi que
ses implications pour l'\'etude des
ph\'enom\`enes critiques pour des dimensions plus grandes que deux, ou
m\^eme pour l'\'etude d'objets au-del\`a des cordes, les $d-$branes.
A cet \'egard, Gervais \cite{ger} a tent\'e d'appliquer aux $3-$branes
ses r\'esultats sur les multicomplexes d'ordre $4$ et dans la r\'ef\'erence
\cite{tap}, il a \'et\'e \'etabli que pour un espace de dimension
$n$ muni d'une $n-$m\'etrique, on obtenait un groupe de sym\'etrie
infini.

En vue d'\'etudier  l'invariance conforme adapt\'ee 
aux  multicomplexes, introduisons 
la base
de l'espace tangent $e_i = \left[\partial\right]_i$. Dans cette base, la
$n-$m\'etrique se r\'e\'ecrit

\beq
\label{eq:nmetrique}
\left(e_{i_1},e_{i_2},\cdots,e_{i_n}\right)=\gamma_{i_1i_2\cdots i_n},
\eeq
\noindent
o\`u $ \gamma_{i_1i_2\cdots i_n}= {1 \over n!}$ si tous les indices sont 
diff\'erents, et z\'ero sinon.

Un calcul direct \cite{fry2} montre que les transformations qui laissent
la $n-$m\'etrique $\gamma$ invariante \`a un facteur d'\'echelle pr\`es sont
les transformations analytiques

\beqa
\label{eq:conf}
&&\left[x\right]_\ell  \longrightarrow F_\ell(\left[x\right]_\ell),{\mathrm
{~avec~}}
\left[\partial\right]_k F_\ell = \delta_{k \ell} \\
&&\gamma(x)_{i_1 \cdots i_n}  \longrightarrow 
\gamma^\prime(x^\prime)_{i_1 \cdots i_n}
=\Omega(x) \gamma(x)_{i_1 \cdots i_n}. \nonumber
\eeqa

Si on consid\`ere les transformations infinit\'esimales 
$\left[x^\prime\right]_\ell = \left[x\right]_\ell + \epsilon f_\ell(x)$,
ainsi que les th\'eor\`emes \'enonc\'es pr\'ec\'edemment (r\'esidus plus
analyticit\'e) qui nous permettent de d\'eve\-lop\-per 
$f_\ell(\left[x\right]_\ell)$ en s\'erie
de Laurent, on obtient $n$ copies de l'alg\`ebre de 
Virasoro sans charge centrale ($\left[\ell_n\right]_k = -
\left(\left[x\right]_k\right)^{1-n}\left[ \partial\right]_k $) engendrant
les transformations ``conformes'' adapt\'ees
aux multicomplexes. Donc, dans un espace dont la dimension
est \'egale au rang du tenseur m\'etrique, les transformations ``conformes''
deviennent un groupe de dimension infinie.
On peut donc, par analogie, d\'efinir un
champ  primaire $\Phi$ de poids conformes $(d_1,\cdots,d_n)$ comme 
\'etant un champ se transformant de la fa\c con suivante

\beq
\Phi^\prime(\left[x^\prime\right]_0, \cdots,\left[x^\prime\right]_{n-1}) =
\prod \limits_{\ell=0}^{n-1}
 \left({d F_\ell \over d \left[x\right]_\ell}\right)^{-d_\ell}
\Phi(\left[x\right]_0, \cdots,\left[x\right]_{n-1}).
\eeq

Pour clore cette partie,  nous mentionnerons que, tout comme dans le plan 
complexe, il est possible de d\'efinir, dans $\MC$ compl\'et\'e d'un point
\`a l'infini, un groupe globalement inversible, analogue aux transformations
homographiques \cite{fry2}, au prix du prolongement de $x^{-1}$ d\'efini
ci-dessus

\beq
x \longrightarrow (ax+b) (cx+d)^{-1}, a,b,c,d \in \MC, ab-cd=1,
\eeq

\noindent
que nous pouvons noter $SL(2,\MC)$.

\vskip.2truecm
 
Face aux grandes similitudes existant entre $\CC$ et $\MC$, il para\^{\i}t
l\'egitime de se demander si les multicomplexes peuvent avoir une utilit\'e
dans la description de certains ph\'enom\`emes physiques, 
voire math\'ematiques, lorsque ceux-ci sont
sous-tendus par des formes non-quadra\-tiques ou bien des sym\'etries $\ZZ_n$ 
par exemple.

\mysection{$n^2-$ions}

Dans notre \'etude des alg\`ebres de Clifford g\'en\'eralis\'ees,
nous avons \'et\'e amen\'es \`a consid\'erer une structure engendr\'ee par
un g\'en\'erateur et permettant de construire une extension des
nombres complexes. Dans le cas quadratique, les quaternions en constituent
la g\'en\'eralisation directe.
Suivant la m\^eme logique, nous nous sommes pench\'es sur les structures
engendr\'ees par 
deux  g\'en\'erateurs $e_1$ et $e_2$ satisfaisant $e_1e_2=q e_2 e_1, 
e_1^n=e_2^n=1$, et
nous  avons  obtenu   les $n^2-$ions introduits d'abord par 
Sylvester puis 
par Cartan. Ce dernier a d\'emontr\'e que les $n^2-$ions contenaient 
les quaternions,
r\'esultats que nous pouvons retrouver \cite{fry4} en utilisant 
la d\'ecomposition 
(\ref{eq:hij}). C'est \`a partir de cette inclusion que nous avons pu
interpr\'eter le r\'esultat de Finkelstein \cite{fin} c'est-\`a-dire
tisser un lien entre les $n^2-$ions et ce qu'il avait appel\'e les
vari\'et\'es d'hyperspin. Ces derni\`eres avaient \'et\'e introduites de 
fa\c con \`a concilier les vari\'et\'es de spin (adapt\'ees pour les espaces
de dimension $2^d$) et les th\'eories de Kaluza-Klein (introduites
pour incorporer la gravitation au c\^ot\'e des interactions fondamentales
\cite{kk}). Ces vari\'et\'es ont alors \'et\'e munies
d'une $n-$m\'etrique invariante
sous $SL(n,\CC)$,  et redonnent, dans une certaine limite, la  m\'etrique de
Minkowski. Cette limite peut \^etre interpr\'et\'ee alg\'ebriquement
par l'inclusion des quaternions dans les $n^2-$ions.

Rappelons bri\`evement nos r\'esultats \cite{fry4}.
La relation (\ref{eq:hij}) permet  de construire les  
g\'en\'era\-teurs des $m^2-$ions \`a partir de ceux des $n^2-$ions ($m<n$)

\beqa
\label{eq:mions}
&&f_1= \sum \limits_{i=0}^{m-1} h_{k+1,k} {\mathrm { ~avec~}} 
h_{m,m-1} \equiv h_{0,m-1}\nonumber \\ 
&&f_2= \sum \limits_{i=0}^{m-1} r^k h_{kk},
{\mathrm { ~avec~}} r=\exp{(2i\pi/m)} \\
&&f_1 f_2 = r f_2 f_1~~~~  
f_1^m=f_2^m= \sum  \limits_{i=0}^{m-1} h_{kk}\equiv
P^{(m|n)}, \nonumber
\eeqa

\noindent
avec $P^{(m|n)}$, l'op\'erateur de projection de $\C_n^2$ sur $\C_{(m|n)}^2$
que l'on identifie comme  \'etant l'identi\-t\'e de $\C_{(m|n)}^2$. Ainsi, il
est facile de montrer que (on note $C_{(m|n)}^2$ l'alg\`ebre des $m^2-$ions
vue en tant que sous-alg\`ebre des $n^2-$ions --voir \cite{fry4} et 
la r\'ef\'erence de Cartan donn\'ee dans  \cite{fry4} pour plus de d\'etails--)
\beq
\label{eq:incl}
C_{(m|n)}^2 \otimes  C_{(n-m|n)}^2 \subset\C_n^2 .
\eeq

Si, pour les  $n^2-$ions, on utilise la 
repr\'esentation matricielle $e_1=\sigma_1, e_2=\sigma_2$,

\beqa
\label{eq:mation}
&&x=\sum \limits_{a,b=0}^{n-1}x_{ab}e_1^ae_2^b \longrightarrow 
X=\sum \limits_{a,b=0}^{n-1}x_{ab}\sigma_1^a \sigma_3^b \\
&&{\mathrm{avec~~}}
X_{ij}=\sum_{b=0}^{n-1} q^{jb} x_{j-i,b}. \nonumber
\eeqa

On peut,   comme avant, d\'efinir une norme $||x||^{2n}
=\det\left( X X^*\right)$ ($X$ n'est \`a priori pas hermitique), il
appara\^{\i}t alors
imm\'ediatement que le groupe de transformations pr\'eservant la m\'etrique
est $SL_L(n,\CC)\otimes SL_R(n,\CC)$  \cite{fry4}

\beqa
\label{eq:sln}
&&X^\prime= LXR,~~ (\det X^\prime =\det X) \\
&&L,R=\exp{\left(\sum_{(a,b)\ne(0,0)} \phi^{L,R}_{ab}e_1^a e_2^b\right)}
~~\det L=\det R=1. \nonumber
\eeqa

\noindent
On peut donc construire les g\'en\'erateurs de $SU(n)$ \`a partir
des \'el\'ements 
de $\C_n^2$, la relation entre $SU(n)$ et les $n^2-$ions a d\'ej\`a \'et\'e
\'etudi\'ee \cite{ram}.
On voit donc que, tout comme les quaternions permettent
d'\'etudier la composition des rotations dans $\RR^3$, les quaternions
g\'en\'eralis\'es conduisent \`a des r\'esultats similaires pour $SL(n,\CC)$.
En outre, dans la limite (\ref{eq:incl}) le groupe d'invariance
se r\'eduit \`a $SL_L(m,\CC)\otimes SL_R(m,\CC)\otimes
SL_L(n-m,\CC)\otimes SL_R(n-m,\CC)$. Ceci nous permet de retrouver, dans une
certaine limite \cite{fry4}, le groupe de Lorentz $SL(2,\CC)$
ainsi que les r\'esultats de \cite{fin}. Nous n'aborderons
pas l'\'etude de cette limite.

Le fait d'avoir insist\'e sur les quaternions g\'en\'eralis\'es nous 
permet de mettre en avant un certain nombre de diff\'erences par rapport aux
quaternions

\begin{enumerate}
\item 
Les matrices de Pauli lin\'earisent le polyn\^ome $x^2+y^2+z^2$ invariant
sous $SO(3)$, or $SO(3)$ est inclus dans  $ SO(1,3)$, le groupe 
d'invariance des quaternions. Par contre, les matrices de Pauli 
g\'en\'eralis\'ees permettent la lin\'earisation de $x^n+y^n+z^n$, polyn\^ome
admettant un groupe discret de sym\'etrie.
\item 
Si on calcule,  pour $n=2$,
$$exp{\left(i \vec \phi \vec \sigma\right)}= \cos(\vec \phi . \vec \phi)
+i\sin(\vec \phi . \vec \phi) {  \vec \phi \vec \sigma \over 
\vec \phi . \vec \phi},$$ on obtient une relation ferm\'ee. Une telle 
relation n'est bien entendu plus vraie quand $n >2$.
\end{enumerate}

\mysection{ Applications}
Les applications des extensions des alg\`ebres de Clifford sont
trop nombreuses pour que nous puissions en faire une pr\'esentation 
exhaustive. Nous nous limiterons donc aux cas qui ont un rapport direct
avec notre \'etude. 
On pourra
se reporter \`a l'introduction de Kwasniewski \cite{k}, ou \`a la r\'ef\'erence
\cite{fr1} ainsi qu'\`a \cite{ram} pour plus de pr\'ecisions

\begin{itemize}
\item
H. Weyl \cite{w} a le premier  remarqu\'e que l'alg\`ebre $\C_n^2$ conduisait
naturellement \`a la m\'ecanique quantique finie. Cette id\'ee  a \'et\'e
renforc\'ee par Schwinger  \cite{sw} qui a montr\'e que $\sigma_1$ et 
$\sigma_3$
formaient un ensemble  complet d'op\'erateurs unitaires. A partir de ces points
de d\'epart, la m\'ecanique quantique a pu \^etre obtenue comme une
limite o\`u $n \to \infty$ de $\C_n^2$, autrement dit, les op\'erateurs 
d'impulsion et de position, engendrant l'alg\`ebre de Heisenberg, 
ont \'et\'e d\'eriv\'es dans cette limite \cite{fqm}. 
R\'esultats qui seront mis en lumi\`ere au cours du chapitre prochain.
\item
Ces alg\`ebres, ainsi que les fonctions trigonom\'etriques
d'ordre $n$, apparaissent
 naturellement dans des mod\`eles g\'en\'eralisant
le mod\`ele d'Is\-ing. Ainsi dans le mod\`ele de Potts, la fonction de 
partition s'exprime  \`a partir des matrices 
$\Gamma$ \cite{k2}. Le calcul n'a  cependant pas pu \^etre men\'e jusqu'au
bout.
\item
Nous avons montr\'e qu'il \'etait possible, \`a partir du proc\'ed\'e de 
lin\'earisation \cite{fry1}, de remplacer un syst\`eme de $m$ \'equations
non-lin\'eaires \`a $m$ inconnues, par un syst\`eme d'\'equations aux valeurs
propres. Cette m\'ethode constitue une g\'en\'eralisation de la m\'e\-thode  
de Cramer pour un syst\`eme quelconque. Nous avons, en outre,  \'etudi\'e la 
compatibili\-t\'e d'un syst\`eme contenant plus d'\'equations que d'inconnues,
sans avoir \`a le r\'esoudre explicitement.
\item
Gr\^ace \`a   la structure $\ZZ_n-$gradu\'ee  des extensions
d'alg\`ebres de Clifford, et parce que ses g\'en\'erateurs sont
dans une repr\'esentation du groupe des tresses (voir la relation 
(\ref{eq:GCA})),
il appara\^{\i}t que ces extensions joueront un r\^ole
consid\'erable lors de l'\'etude  des d\'eformations de certaines alg\`ebres,
voir des groupes quantiques, 
ou de l' extension des th\'eories supersym\'etriques.  Nous consacrerons
les deux prochains chapitres \`a ces g\'en\'eralisations.
\end{itemize}

\chapter[Alg\`ebre de Heisenberg $q-$d\'eform\'ee et $q-$oscillateurs]
{\center{ Alg\`ebre de Heisenberg $q-$d\'eform\'ee et  $q-$oscillateurs}}

%\stepcounter{chapter}

 Etant donn\'e que les particules \'el\'ementaires apparaissent dans des 
repr\'esentations appropri\'ees de groupes de sym\'etrie, 
la classification de tels groupes  va dans le sens d'une
compr\'e\-hension des interactions fondamentales ainsi que des propri\'et\'es
des particules. La premi\`ere classification, pour les alg\`ebres de Lie 
compactes, a \'et\'e effectu\'ee 
par Cartan.  Depuis  l'\'emer\-gence 
de la supersym\'etrie \cite{susy}, on s'est rendu compte que cette nouvelle
sym\'etrie, d\'epassant le cadre des alg\`ebres de Lie,
 permettait de combler un certain nombre de lacunes que le mod\`ele 
standard ne r\'esolvait pas (hi\'erarchie, constante cosmologique, etc.).  
D'un point de vue math\'ematique, la classification des alg\`ebres 
supersym\'etriques (super-alg\`ebres) a \'et\'e men\'ee \`a bien dans 
\cite{class}.

D'apr\`es le th\'eor\`eme de Noether les  sym\'etries fondamentales 
sont engendr\'ees, apr\`es la seconde quantification, par des courants 
conserv\'es d\'efinis \`a partir des champs. Suivant  le caract\`ere
bosonique ou fermionique, les champs seront quantifi\'es par des relations de
commutation ou d'anticommutation. On s'attend donc \`a ce que toutes les
(super-)alg\`ebres de Lie puissent se d\'efinir  en termes d'oscillateurs.
Si on introduit une s\'erie de $n$ oscillateurs bosoni\-ques 
$(a_i,a^\dagger_i)$,
 il est facile de se convaincre que les 
\'el\'ements $M_{ij}=a_i^\dagger a_j$ engendrent 
l'alg\`ebre de Lie $GL(n,\CC)$ 
\cite{br}
 
\beqa
&&\left[a_i^\dagger,a^\dagger_j\right]=0,~~  
\left[a_i,a_j\right]=0,~~
\left[a_i,a^\dagger_j\right]=\delta_{ij} \nonumber \\
&&~~~~~~~~
\left[M_{ij},M_{kl}\right]= M_{il} \delta_{jk}-M_{kj} \delta_{il}. \nonumber
\eeqa 

\noindent
Ce r\'esultat s'\'etend  \`a toutes les alg\`ebres de Lie compactes
\cite{br,now}. Il peut m\^eme 
\^etre g\'en\'eralis\'e aux alg\`ebres de Lie non-compactes \cite{now}. 
On peut \'egalement faire la m\^eme  construction avec une s\'erie de $n$ 
oscillateurs fermioniques et envisager une description analogue, avec
des oscillateurs bosoniques et fermioniques, pour les super-alg\`ebres de Lie 
\cite{now} (ainsi que les r\'ef\'erences contenues dans \cite{now}).  

Pour les oscillateurs bosoniques, les op\'erateurs de cr\'eation et
d'annihilation
sont associ\'es \`a l'alg\`ebre de Heisenberg engendr\'ee par un op\'erateur
position et son moment conjugu\'e. En ce qui concerne les fermions,
c'est une alg\`ebre de Clifford qui  correspond aux oscillateurs et les
variables sont auto-conjugu\'ees. Ces deux alg\`ebres ont des propri\'et\'es
distinctes, les premi\`eres n'admettent pas de repr\'esentation de dimension
finie alors que les secondes, si. Ce distinguo est bien \'evidemment li\'e
aux statistiques diff\'erentes des deux syst\`emes.  Ces deux structures
alg\'ebriques sont \`a la base d'un grand nombre de r\'ealisations concr\`etes
en physique des particules. 

Nous avons, lors du premier chapitre, \'etudi\'e des extensions
de l'alg\`ebre de Clifford et de Grassmann, extensions engendr\'ees par
un polyn\^ome de degr\'e $n$  admettant une sym\'etrie $\ZZ_n$. 
On peut donc l\'egitimement se demander si tout ce qui a pu \^etre fait
avec succ\`es pour les fermions, c'est-\`a-dire  les alg\`ebres
de Clifford et de Grassmann, est envisageable pour les alg\`ebres de Clifford
g\'en\'eralis\'ees. Etant donn\'e que l'alg\`ebre de Clifford est une 
quantification de l'alg\`ebre de Grassmann (on reviendra sur ce point dans
le chapitre 3), il semble  qu'un bon point de d\'epart consiste en l'\'etude de
$\GG_n^1$, l'alg\`ebre de Grassmann g\'en\'eralis\'ee. La premi\`ere
\'etape va donc consister \`a d\'efinir une variable conjugu\'ee pour
$\theta \in \GG_n^1$, c'est-\`a-dire  construire une d\'erivation.
Ce faisant, nous obtiendrons une alg\`ebre de 
Heisenberg $q-$deform\'ee $\H_{q,n}$
($q=\exp{(2i\pi /n)})$ \cite{bf,fik}, r\'esultat qui sera r\'eobtenu 
dans le chapitre suivant par 
quantification de $\GG_n^1$. On verra ensuite que $\H_{q,n}$  est connect\'ee
aux $q-$oscillateurs introduits ind\'ependamment par Biedenharn et
Macfarlane \cite{bm}. Nous \'etudierons en d\'etail cette structure 
alg\'ebrique. Ensuite, nous verrons comment \'etendre les r\'esultats obtenus
pour une variable lorsque  plusieurs variables sont consid\'er\'ees.
A la diff\'erence des polyn\^omes quadratiques, suivant le comportement
des diff\'erentes familles d'oscillateurs, nous aurons une structure distincte.

Tout comme l'alg\`ebre de Clifford s'av\`ere \^etre l'outil de
base pour la construction des th\'eories supersym\'etriques, certaines
alg\`ebres, obtenues par quantification de $\GG_n^p$, seront les \'el\'ements
permettant de construire
des th\'eories allant au-del\`a de la supersym\'etrie, extension
dont nous parlerons dans le dernier chapitre
\cite{abl,dur1,mm,am,fr,prs1,prs2,fsusy1d,fsusy2d,fsusy3d}. 

Toutefois, une approche diff\'erente peut \^etre envisag\'ee \`a partir
des alg\`ebres de Heisenberg $q-$d\'eform\'ees ou des $q-$oscillateurs, 
qui consiste \`a introduire des d\'eformations des sym\'etries 
(groupes quantiques \cite{gq} et alg\`ebres d\'eform\'ees \cite{qa}). Il 
transpara\^{\i}t une certaine similitude entre cette approche, que nous
n'aborderons pas, et les structures que nous consi\-d\'erons. On peut s'en
rendre compte en analysant les 
relations (\ref{eq:GCA}) et (\ref{eq:qtheta}) qui traduisent que les variables
de Clifford et de Grassmann g\'en\'eralis\'ees sont dans une repr\'esentation
du groupe des tresses. A cet \'egard, on verra que $\C^p_n$ et $\GG^p_n$
admettent une $R-$matrice (c.f. eq.(\ref{eq:R})).
Face \`a la litt\'erature abondante sur les groupes
quantiques et les $q-$oscillateurs, nous renverrons le lecteur 
\`a la r\'ef\'erence de J. Fuchs
\cite{fuchs}, nous nous bornerons par la suite \`a \'evoquer certaines
applications  des groupes quantiques ou des alg\`ebres 
$q-$d\'eform\'ees ayant une relation plus ou moins directe avec nos 
r\'esultats. Certains des r\'esultats que nous allons pr\'esenter ont tr\`es
probablement des applications dans ce domaine, mais nous n'avons
pas exploit\'e cette voie. 

Notons enfin que les structures que nous consid\'erons
ont un champ d'application relativement
vaste dans le cadre des alg\`ebres de Lie g\'en\'eralis\'ees 
 \cite{gla} et ouvrent des perspectives  \`a d'autres extensions possibles
de la supersym\'etrie, que nous mentionnerons au chapitre suivant. 

\mysection{Alg\`ebre de Heisenberg $q-$d\'eform\'ee}

\subsection{D\'erivation}

Nous allons d'abord consid\'erer l'alg\`ebre $\GG_n^1$, engendr\'ee
par une variable $\theta$, telle que $\theta^n=0$. Une telle alg\`ebre
est donc constitu\'ee des polyn\^omes de degr\'e $(n-1)$ en $\theta$,

\beq
\label{eq:G1n}
\sum \limits_{i=0}^{n-1} f_i \theta^i.
\eeq

Une d\'erivation sur  $\GG_n^1$ peut \^etre vue comme un automorphisme
(infinit\'esimal) 
de $\GG_n^1$ sur elle-m\^eme. Une telle approche pour munir $\GG_n^1$ d'une 
d\'erivation a \'et\'e suivie par Filipov, Isaev et Kurdikov \cite{fik} et 
d\'ebouche alors sur une r\`egle de Leibniz adapt\'ee, comme nous
allons le voir. Consid\'erons donc une d\'erivation
$\partial$, \'egalement nilpotente ($\partial^{n}=0$)
 qui, par d\'efinition, donne

\beqa
\label{eq:partial}
&&\partial(1) = 0 \nonumber \\
&&\partial(\theta^a) \sim \theta^{a-1},
\eeqa 

\noindent
pour conna\^{\i}tre pr\'ecis\'ement le second membre de 
l'\'equation (\ref{eq:partial}), on utilise la r\`egle de Leibniz adapt\'ee

\beq
\partial \left(\theta^a\right)= 
\partial \left(\theta\right) \theta^{a-1} +
\alpha \theta\partial \left(\theta\right) \theta^{a-2} +
 \cdots +
\alpha^{a-1} \theta^{a-1} \partial \left(\theta\right).
\eeq  
\noindent
Si $a=n$,   
$\partial : 0=\theta^n \longrightarrow 0$,
et donc, $1+\alpha+\cdots+\alpha^{n-1}=0$. De ce fait, $\alpha$ est une
racine  $n-$i\`eme de l'unit\'e. Il y a donc 
 $(n-1)$ d\'eriv\'ees distinctes ($\alpha=q^i$) qui sont solutions, et

\beq
\label{eq:deri}
\partial_i \left(\theta^a\right)= { 1 -q^{ia} \over 1 - q^i}
\theta^{a-1}=\left\{a\right\}_i \theta^{a-1} .
\eeq
\noindent
Il est facile, \`a partir de (\ref{eq:deri}), d'obtenir la relation 
constituant la d\'efinition de l'alg\`ebre de Heisenberg $q^i-$d\'eform\'ee
$\H_{q^i,n}$

\beq
\label{eq:hq}
\partial_i \theta - q^i \theta \partial_i=1.
\eeq

Une telle relation a \'et\'e mise en \'evidence pour la premi\`ere fois
dans \cite{abl}, puis utilis\'ee par les auteurs des R\'efs. \cite{dur1,mm}.
Cette structure alg\'ebrique a  \'et\'e red\'ecouverte 
ind\'ependamment par Baulieu et Floratos \cite{bf} ansi que dans 
\cite{ch,ama}. 
Elle peut \^etre  reli\'ee aux $q-$oscillateurs \cite{bm}.
Notons enfin que,
si $n$ n'est pas un nombre premier ($n=n_1n_2$), les $(n-1)$ d\'eriv\'ees
obtenues ne sont pas toutes \'equivalentes. 
En effet, un calcul direct nous indique
que $\partial_1$ est nilpotente d'ordre $n$ alors que $\partial_{n_1}$ est 
nilpotente d'ordre $n_2$. Quoi qu'il en soit, ind\'ependamment de $n$, il
existe au moins deux d\'eriv\'ees $\partial_1 \equiv \p$ et $\partial_{n-1}
\equiv \d$ nilpotentes d'ordre $n$. On peut remarquer que les relations de
d\'efinition des $\partial_i$ contiennent les nombres $q^i-$d\'eform\'es
$\{a\}_i = { 1 - q^{ai} \over 1 -q^i}$ d\'ej\`a introduits par Euler, et ces
nombres redonnent $a$ dans la limite $q \to 1$. 
Si on se limite aux alg\`ebres de Grassmann, il n'est pas utile d'introduire
de tels nombres, du fait qu'ils
sont cach\'es. Il s'ensuit 
donc que  la seule relation qui nous int\'eresse est
$\partial \theta =1$, \'etant donn\'e que $\theta^2=0$. 

La structure diff\'erentielle exhib\'ee (\ref{eq:hq}) n'est pas la seule 
envisageable et pour une \'etude syst\'ematique des diverses possibilit\'es,
on peut consulter la r\'ef\'erence \cite{fik}. Cependant, c'est la 
structure (\ref{eq:hq}) qui s'av\`erera \^etre
la base de la g\'en\'eralisation de la supersym\'etrie
que nous consid\'ererons au cours du chapitre 3.

\subsection{Repr\'esentations}

Il est bien connu que les variables de Grassmann  admettent une 
repr\'esentation de dimension ${\mathbf 2}$ conduisant \`a la description
des fermions. Il peut donc s'av\'erer utile de mettre
en \'evidence les repr\'esentations de l'alg\`ebre de Heinsenberg 
$q-$d\'eform\'ee. Une repr\'esentation matricielle a \'et\'e 
construite  dans les R\'ef. \cite{fik,dur2}.
Nous avons montr\'e \cite{rdt}
que  les repr\'esentations de $\H_{q,n}$ sont de 
dimension $kn$ et seule, celle de dimension
minimale, est irr\'eductible. Pour aboutir \`a ce r\'esultat, on peut 
proc\'eder de trois mani\`eres diff\'erentes. Soit montrer, par un calcul
direct mais tr\`es fastidieux, que $\left(x \partial + y \theta^{n-1}\right)^n
\sim x^{n-1} y$ car $\sum \limits_{a=0}^{n-1} \partial^a \theta^{n-1}
\partial^{n-a-1} \sim 1$ \cite{dur2}, 
soit utiliser les r\'esultats des extensions
de la supersym\'etrie \cite{abl,dur1} bas\'ees sur les 
variables $\theta$ et $\partial$
qui conduisent au m\^eme r\'esultat, soit enfin, utiliser directement
la repr\'esentation matricielle \cite{fik,dur2,rdt}. Celle-ci peut \^etre 
obtenue \`a partir de l'expression donn\'ee pour $\theta$ au chapitre
sur les alg\`ebres de Grassmann g\'en\'eralis\'ees (\ref{eq:thetamat}), puis
en r\'esolvant l'\'equation (\ref{eq:hq}) pour $i=1$

\beq
\label{eq:mhq}
\theta=\pmatrix{0&\cdots&0&0 \cr
                1&\cdots&0&0 \cr
                \vdots&\ddots&\ddots&\vdots \cr
                0&\cdots&1&0}~~
\p=\pmatrix{0&\{1\}_1&0&\cdots&0 \cr
                  0&0&\{2\}_1&\cdots&0 \cr
                  \vdots&&&\ddots&0 \cr
                  0&\cdots&&\ddots&\{n-1\}_1\cr
                  0&\cdots&&\cdots&0}.
\eeq 
Il est amusant de constater que l'on peut relier les deux matrices
(\ref{eq:mhq}) par une transformation de Fourier finie \cite{rdt}
$\p= {\cal F}\theta {\cal F}^{-1}$ avec ${\cal F}_{ij}={1 \over \{i-1\}_1!}
\delta_{i+j-n-1,0}$.
On voit donc, \`a  partir des matrices (\ref{eq:mhq}), que
$\left(x \partial + y \theta^{n-1}\right)^n=\{n-1\}_1!x^{n-1}y~~
(\{a\}_1!=\{1\}_1 \cdots \{a\}_1).$ Autrement dit,
$\theta$ et $\p$ engendrent une repr\'esentation de l'alg\`ebre de
Clifford du polyn\^ome $x^{n-1}y$. Ce faisant, $\H_{q,n}$ est une 
repr\'esentation
de $\C_{x^{n-1}y}$. Arguant  que (\ref{eq:mhq}) est une repr\'esentation
fid\`ele de $\H_{q,n}$, on en d\'eduit que toutes 
les autres repr\'esentations 
de $\H_{q,n}$ seront de dimension $kn$. On peut alors 
montrer que celles-ci sont
r\'eductibles et consistent en $k$ copies de celle 
obtenue ci-dessus \cite{rdt}.

On munit alors $\H_{q,n}$ de la graduation naturelle 
$\g(\theta)=1,~\g(\p)=-1$ 
(mod. $n$). Un calcul peu int\'eressant, utilisant par exemple la 
repr\'esentation matricielle (\ref{eq:mhq}), permet de montrer que les 
matrices $H_{ii}$, de graduation $0$ (voir section (1.1.2.1.)), sont 
engendr\'ees par les $n$ \'el\'ements
$(1, \p \theta, \cdots, \p^{n-1} \theta^{n-1})$. Bien entendu, les matrices
$H_{i+ai}$, de graduation $a$, sont obtenues \'egalement \`a partir
des \'el\'ements de graduation $a$ de $\H_{q,n}$.  Les autres 
d\'eriv\'ees $\partial_{i\neq1}$ (de graduation $-1$) s'expriment donc
\`a partir de $(\theta^{n-1}, \p, \p^2 \theta,\cdots,\p^{n-1} \theta^{n-2})$ 
et l'alg\`ebre
est compl\`etement sp\'ecifi\'ee par la donn\'ee des trois \'el\'ements
$q,\theta$ et $\p$.  Il s'ensuit donc  

\beq
\label{eq:hincl}
\H_{q,n}  ~~~\left\{  
\begin{array}{ll}
  \sim \H_{q^i,n} &{ \mathrm{ si ~}} n {\mathrm{~est~ premier}} \\
  \supset \H_{q^i,n} & {\mathrm{ si ~}} n {\mathrm{~n'est~ pas~  premier
 ~et~}} i{\mathrm{~divise~}}n.
\end{array}
\right.
\eeq

\subsection{Inclusions}

On s'int\'eresse maintenant \`a une variable de Grassmann g\'en\'eralis\'ee 
d'ordre $n=n_1n_2$, avec $n_1$ et $n_2$ deux nombres premiers. Nous avons
montr\'e \cite{prs2} qu' il est
facile de construire, \`a partir de $\theta$ et $\p$,
$\theta_1=\theta^{n_2} \in \H_{q^{n_2},n_1} \equiv \H_{q_1,n_1} 
\subset \H_{q,n}$,
ainsi que la d\'eriv\'ee $\partial_{\theta_1}$, de graduation $-n_1$ 
(c.f. la remarque de la section pr\'ec\'edente).
On peut donc obtenir l'alg\`ebre $\H_{q_1,n_1}$
comme une sous-alg\`ebre de  $\H_{q,n}$. On  
consid\`ere  l'application 

\beqa
\label{eq:inclus}
f_2 : \H_{q,n}&& \longrightarrow \H_{q,n} \nonumber \\
           \theta && \longmapsto \theta_1 = \theta^{n_2} \nonumber \\
           \p    &&   \longmapsto \partial_{\theta_1} \\
           q     &&   \longmapsto q_1=q^{n_2}, \nonumber
\eeqa
on peut facilement se rendre compte que $f_2$ est un homomorphisme d'alg\`ebre,
et donc que $\H_{q_1,n_1}$ est isomorphe \`a $\H_{q,n}/{ \mathrm{Ker}}(f_2).$
 Il est 
\'egalement possible de d\'efinir un autre isomorphisme: $ \partial_\theta
\to \left(\partial_\theta\right)^{n_2}; \theta \to \theta_1$.
Cette propri\'et\'e sera fort utile dans le chapitre suivant.

\mysection{q-oscillateurs}

\subsection{Alg\`ebre des $q-$oscillateurs}
Tout comme l'alg\`ebre de Heisenberg est reli\'ee
 naturellement aux oscillateurs
bosoniques, il est possible de construire, \`a partir de $\H_{q,n}$, une
alg\`ebre d'oscillateurs $q-$d\'eform\'es \cite{bm} (quand $q=-1$, on
a une co\"{\i}ncidence triviale entre ces deux structures alg\'ebriques).
 Notons tout d'abord que
la repr\'esentation matricielle nous permet de mettre en \'evidence
un op\'erateur nombre \cite{dur2}

\beqa
\label{eq:nonb}
&&{\cal N} = \sum \limits_{i=1}^{n-1} {(1-q) \over (1-q^i)}^i 
\theta^i \p^i \\
&&\left[{\cal N}, \theta \right] = \theta,~~ 
\left[{\cal N}, \p \right] =-\p. \nonumber
\eeqa

\noindent
On peut alors montrer (en utilisant la repr\'esentation matricielle par
exemple) que l'on a 
\beq
\label{eq:qh}
q^{\cal N} =\sigma_3,~~
\sigma_3 \theta = q \theta \sigma_3,~~\sigma_3 \p = q^{-1} \p \sigma_3. 
\eeq

\noindent
Enfin, \`a partir de (\ref{eq:hq}), on peut \'etablir le m\^eme r\'esultat,
\`a savoir que $(\p \theta - \theta \p) \theta = q \theta
(\p \theta - \theta \p)$ et $(\p \theta - \theta \p) \p = q^{-1} \p
(\p \theta - \theta \p)$ \cite{fik}, c'est-\`a-dire que 
$\p \theta - \theta \p = q^{{\cal N}}$.

\noindent
Pour introduire les $q-$oscillateurs, \`a partir de $\H_{q,n}$,
on pose $a^\dag=\theta \sigma_3^{-1/4}, a=
\sigma_3^{-1/4} \p$.  On observe alors que

\beq
\label{eq:hq-os}  
a a^\dag = \sigma_3^{-1/2} \p \theta = \p \theta 
\sigma_3^{-1/2},~~
a^\dag a =q^{1/2} \theta \p \sigma_3^{-1/2},
\eeq

\noindent
et  on obtient donc l'alg\`ebre des $q$-oscillateurs

\beq
\label{eq:qosc}
\left\{
\begin{array}{ll}
a a^\dag - q^{1/2} a^\dag a &= q^{-{\cal N}/2} \\
a a^\dag - q^{-1/2} a^\dag a &= q^{-{\cal N}/2} \sigma_3=q^{{\cal N}/2}.
\end{array}
\right.
\nonumber
\eeq

\noindent
En fait, la d\'emonstration 
produite ici n'est rien d'autre que l'inverse de la 
d\'emonstration primitive faite par Biedenharn \cite{bm}, permettant de
passer du syst\`eme $a, a^\dag$ au syst\`eme $\theta,\p$. L'\'equation
(\ref{eq:qosc}), d\'efinissant le syst\`eme de $q$-oscillateurs, est d'ailleurs
vraie pour toutes les valeurs de $q$. 
%Cependant, uniquement lorsque 
%$q$ est un nombre r\'eel, $a^\dag$ et $a$ sont conjugu\'es hermitiques l'un
%de l'autre. Par contre, si $q$ est une racine primitive de l'unit\'e, les
%conjugu\'es hermitiques de $(a, a^\dag)$ sont 
%$(\left(a^\dag\right)^\star,a^\star) $, $\star$ \'etant la conjugaison 
%complexe.  Il est alors facile de se rendre compte que
%$(a^\star, \left(a^\dag\right)^\star)$ satisfont l'\'equation (\ref{eq:qosc}).
Cependant, seulement lorsque $q$ est une racine primitive de
l'unit\'e, l'alg\`ebre (\ref{eq:qosc}) est de dimension finie.
L'espace de Fock est donc constitu\'e des \'etats $\left|k \right>, k=0,1
\cdots$, orthonorm\'es \cite{bm} 

\beqa
\label{eq:qfock}
a^\dag \left|k\right>&=& \sqrt{[k+1]} \left|k\right> \nonumber \\
a \left|k\right>&=& \sqrt{[k]} \left|k-1\right> \\
{\cal N}  \left|k\right> &=& k \left|k\right>, \nonumber
\eeqa

\noindent
avec $  [x] = { q^{x/2}-  q^{-x/2} \over  q^{1/2}-  q^{-1/2}}$.
La repr\'esentation de l'alg\`ebre des $q-$oscillateurs est donc construite
\`a partir d'un vide annihil\'e par $a$.  Etant donn\'e que les relations de
d\'efinition de $\H$ font intervenir $q^{1/2}$ au lieu de $q$, il vient
que (quand $q$ est une racine primitive de l'unit\'e)
$a$ et $a^\dag$ sont bien conjugu\'es hermitiques l'un de l'autre ($[k]$
est toujours positif). Lorsque $q$ est r\'eel c'est \'egalement le cas. 
En fait  en \'etudiant l'alg\`ebre (\ref{eq:qosc}) on voit que les \'equations
sont stables par conjugaisons uniquement dans ces deux cas (pour
$q=\exp(2i \pi /n)$ \'etant donn\'e que la seconde \'equation
manquait dans les premiers articles, cette propri\'et\'e n'avait pas 
\'et\'e observ\'ee).

%Les \'etats  conjugu\'es
%$\left<k\right|$ sont bien s\^ur obtenus avec $a^\star$ et du vide annihil\'e
%par $ \left(a^\dag\right)^\star, \left<0\right| { \left(a^\star\right)^k\over
%\sqrt{[k] !}^\star} = \left<k\right|$.
%Mais du fait des proprit\'es des op\'erateurs $a, a^\dag$, on peut \'egalement
%d\'efinir $\left<k\right|$  en utilisant $a, a^\dag$.

On voit, \`a partir de ces relations, que si $q$ est une racine
primitive de l'unit\'e, 
on a $[n]=0$. Le caract\`ere fini de l'espace de Hilbert est directement
induit par le fait que les  $q$-oscillateurs sont nilpotents. Les
vecteurs d'\'etats $\left|k\right>$ apparaissant dans l'\'equation
(\ref{eq:qfock}) sont aussi construits \`a partir de $\theta$ et $\p$,
$\theta \left|k\right> = \sqrt{\left\{k+1\right\}_1} \left|k+1\right>,
\p  \left|k\right> = \sqrt{\left\{k\right\}_1} \left|k-1\right>$, 
le passage aux  matrices (\ref{eq:mhq}) se faisant par un changement
de base.

Toujours lorsque $q^n=1$, on peut trouver 
une autre repr\'esentation dite cyclique, pour laquelle  $a\left|0\right>=0$
et $a^\dag\left|n\right> \sim  \left|0\right>$ \cite{flo}. Il est int\'eressant
de constater que  celle-ci admet comme point de d\'epart
l'alg\`ebre $\C^2_n$ au lieu de l'alg\`ebre de Grassmann g\'en\'eralis\'ee
associ\'ee. C'est d'ailleurs en identifiant 
$\sigma_1 \equiv \exp( 2i \pi /n P), \sigma_3 \equiv \exp(i Q)$, que, dans
la limite $n \to \infty$, on voit que $P,Q$ engendre l'alg\`ebre de Heisenberg 
\cite{fqm,flo}.

Il existe enfin une troisi\`eme repr\'esentation, construite
\`a partir d'op\'erateurs boso\-ni\-ques $(b^\dag,b, {\cal N})$ \cite{kd}

$$ a =\left({ [{\cal N}+1] \over {\cal N}+1 }\right)^{1/2} b, 
a^\dag =b^\dag \left({ [{\cal N}+1] \over {\cal N}+1 }\right)^{1/2}.$$

\noindent
Si  $q$ est une racine primitive de l'unit\'e, celle-ci 
consiste en secteurs disconnect\'es  les uns des autres. 

\subsection{Base des \'etats coh\'erents}
Il est \'egalement possible, pour $q$ arbitraire, de d\'efinir une
base des \'etats coh\'erents. Introduisons deux variables $\theta$ et 
$\bar \theta$ satisfaisant $\theta \bar \theta = q \bar  \theta \theta$,
commutant avec  $a, a^\dag$   et,
lorsque $q$ est une racine de l'unit\'e, v\'erifiant 
$\theta^n=\bar \theta^n=0$ 
 \cite{bm,gn,kd}. On \'etablit alors simplement que

\beqa
\label{eq:coh}
\left| \theta \right>& =& \sum \limits_k {\theta^k \over \sqrt{[k]!} }
\left|k\right> \equiv \exp_q{\left(\theta a^\dag\right)}  \left| 0 \right>
\nonumber \\
a \left| \theta \right>& =& \theta \left| \theta \right>   \\
\left<p|\theta\right>&=& {\theta^p \over \sqrt{[p]!}}. \nonumber
\eeqa

\noindent
Si on d\'efinit ($q^n=1$) $\left< \theta \right| = 
\left<0\right| \exp_q\left(a \bar \theta \right)$, alors ce vecteur sera \'etat
propre de $a^\dag ~:~ \left< \theta \right| a^\dag =  \left< \theta \right|
\bar \theta$.
On a donc $\left<\theta|p\right> = {\bar \theta^p \over \sqrt{[p]!}}$.

Bien entendu, la somme est finie si $q$ est une racine primitive de l'unit\'e.
La fonction $ \exp_q{x}= \sum \limits_k {x^k \over [k]!}$, constituant une
g\'en\'eralisation de la fonction exponentielle,  joue un r\^ole central
dans ce type de structures alg\'ebriques; elle est d\'enomm\'ee exponentielle
$q-$d\'eform\'ee. Elle est constitu\'ee de $n$ termes quand $q^n=1$.
C'est cette constation, nous permettant d'associer deux variables 
``classiques''($\theta, \bar \theta$) au syst\`eme $(a, a^\dag)$ ou \`a
($\theta,\p$), 
qui constitue une seconde d\'emonstration
de l'assertion {\it  la quantification des variables} $\theta$ {\it et }
$\bar \theta$ {\it conduit \`a l'alg\`ebre} $\H_{q,n}$ {\it ou \`a celui des
q-oscillateurs}; ce proc\'ed\'e de quantification a \'et\'e suivi
dans \cite{bf}. Nous verrons une autre justification au chapitre suivant.
L'introduction d'une base des \'etats coh\'erents a \'et\'e red\'ecouverte
dans les r\'ef\'erences \cite{bf,ch}. 

L'alg\`ebre (\ref{eq:qosc}) peut \^etre vue comme une d\'eformation des 
relations de commutation de l'oscillateur bosonique, et, de ce fait,
($a,a^\dag$) a  \'et\'e appel\'e
 $q-$boson. Il est \'egalement possible de d\'efinir une
d\'eformation de l'oscillateur fermionique \cite{ck,pvbd} en introduisant
un signe dans la relation de commutation de l'\'equation (\ref{eq:qosc}).
A partir de familles de $q-$bosons, 
commutantes les unes par rapport aux autres,
ou v\'erifiant des relations plus compliqu\'ees, similaires \`a celles
des variables de Grassmann  et de Clifford g\'en\'eralis\'ees 
(\ref{eq:GCA},\ref{eq:qtheta}),
sui\-vant la m\'ethode de Jordan-Wigner, une construction explicite
des g\'en\'erateurs de certaines alg\`ebres d\'eform\'ees a \'et\'e obtenue
\cite{bm,kd,ckl,fz,flo}. Il est \'egalement possible de construire des
extensions $q-$d\'eform\'ees des alg\`ebres conformes \cite{cilpp,flo} ou
des superalg\`ebres \cite{ckl,fsv}.

\mysection{Famille de $q-$oscillateurs}
Un certain nombre de raisons nous donnent envie d'aller au-del\`a du syst\`eme
constitu\'e d'un  oscillateur unique: ainsi par exemple, les r\'esultats 
mentionn\'es au paragraphe pr\'ec\'edent, et surtout le fait qu'un syst\`eme
physique n'est jamais constitu\'e d'un seul oscillateur. A la diff\'erence
de ce qui se passe pour les syst\`emes bosoniques et fermioniques, o\`u le
passage d'un oscillateur \`a un nombre arbitraire d'oscillateurs est
unique (modulo une transformation de Klein pour les fermions), d\`es que
l'on consid\`ere les $q-$oscillateurs, ou les alg\`ebres de Heisenberg
$q-$d\'eform\'ees, ce n'est plus vrai. Suivant les postulats de d\'epart,
on aura une structure avec des propri\'et\'es distinctes et des covariances
diff\'erentes, c'est-\`a-dire des groupes de stabilit\'e distincts. 

\subsection{Syst\`emes de $q-$oscillateurs}
La fa\c con naturelle de consid\'erer un syst\`eme multi-oscillateurs
consiste \`a introduire une famille de  $p$ $q-$bos\-ons ind\'ependants ou,
ce qui revient au m\^eme, $p$ alg\`ebres de Heisenberg $q-$d\'eform\'ees
commutantes $\otimes^p \H_{q,n}$ ($(\xi_i,\partial_{\xi_i}, {\cal N}_i) 
i=1,\cdots p$). Ensuite, on d\'efinit une transformation de
Klein adapt\'ee permettant, \`a partir des oscillateurs $\xi, \partial_\xi$,
de construire un autre syst\`eme. Bien \'evidemment, toutes les alg\`ebres
construites par ce proc\'ed\'e h\'eriteront des propri\'et\'es de $\H_{n,q}$;
elles admettront  en particulier une repr\'esentation de dimension finie
lorsque $q^n=1$.\\

\noindent
{\large {\bf 2.3.1.1. Quantification de l'alg\`ebre de Grassmann
g\'en\'eralis\'ee. }}\\
\addcontentsline{toc}{subsection}{\numberline{}  
2.3.1.1. Quantification de l'alg\`ebre de Grassmann g\'en\'eralis\'ee }

A partir de $p$ variables commutantes, il est facile de d\'efinir
$p$ variables $q-$mutantes, et  ainsi de d\'efinir une alg\`ebre
de Grassmann g\'en\'eralis\'ee v\'erifiant les relations ({\ref{eq:qtheta}).
Si en outre on veut pr\'eserver la propri\'et\'e de d\'erivation (\ref{eq:hq}),
c'est-\`a-dire munir l'alg\`ebre $\GG_n^p$ d'une structure diff\'erentielle,
nous avons montr\'e qu'il n'y a qu'un choix unique  pour exprimer 
$\theta_i, \partial_{\theta_i}$  en fonction de $\xi_i, \partial_{\xi_i}$
\cite{prs1}

\beqa
\label{eq:klein}
\xi_i \longrightarrow &&\theta_i= \left(\prod_{j<i} q^{{\cal N}_j}\right)
 \xi_i \\
\partial_{\xi_i} \longrightarrow &&\partial_{\theta_i} = 
\left(\prod_{j<i} q^{-{\cal N}_j}\right)
\partial_{\xi_i}. \nonumber
\eeqa

\noindent
On peut noter que la premi\`ere expression, du fait de la propri\'et\'e
$q^{{\cal N}} = \sigma_3$, traduit les relations (\ref{eq:qtheta}).
Les couples $\theta_i, \partial_{\theta_i}$ v\'erifient alors les relations
de $q$-mutations suivantes

\beq
\label{eq:qtp}
\left\{
\begin{array}{ll}
\theta_i \theta_j = q \theta_j \theta_i,& i<j \\
\partial_{\theta_i} \partial_{\theta_j} = q
\partial_{\theta_j} \partial_{\theta_i},& i<j  \\
\partial_{\theta_i} \theta_i -q \theta_i \partial_{\theta_i}=1& \\
\partial_{\theta_i} \theta_j = q^{-1} \theta_j \partial_{\theta_i},&
i<j  \\
\partial_{\theta_j} \theta_i = q \theta_i \partial_{\theta_j},& i<j.
\end{array}
\right.
\eeq

\noindent
Une telle transformation est une adaptation de la transformation de Klein,
engendrant \`a partir d' un syst\`eme de $p$ fermions commutant, 
un syst\`eme de fermions anticommutant, aux syst\`emes $q-$mutant. 
Ces relations ont \'egalement
\'et\'e mises en \'evidence dans \cite{chd}. Ces auteurs sont partis de la
premi\`ere \'equation des relations  (\ref{eq:qtp}) et, en imposant une
covariance  sous le groupe Euclidien $q-$d\'eform\'e, ont obtenu le reste
des relations.

En outre, on peut montrer que l'espace ainsi construit admet une $R-$matrice
\cite{wz,pw}

\beq
\label{eq:R}
R^{i_1i_2}_{j_1j_2} = \delta^{i_2}_{j_1} \delta^{i_1}_{j_2}
\left(1+(q-1)\delta^{i_1i_2}\right) + (q-1/q) \delta^{i_1}_{j_1}
\delta^{i_2}_{j_2} \Theta(i_2-i_2).
\eeq

\noindent 
En interpr\'etant $R$ comme une application de $\H_{q,n} \otimes \H_{q,n}$
dans elle-m\^eme, et en indi\c cant par $1,2$ les \'el\'ements du
premier (second) espace, on peut r\'e\'ecrire (\ref{eq:qtheta}) sous la forme
$ (R_{12}-q)\theta_1 \theta_2=0$, avec $\Theta(i_1-i_2)=1 $ si $i_2 >i_1$ et 
z\'ero sinon.  Si maintenant, on consid\`ere le produit de trois espaces
$\H_{q,n} \otimes \H_{q,n} \otimes \H_{q,n}$, 
un calcul direct montre que $R$ satisfait l'\'equation
de Yang-Baxter $R_{12}R_{23}R_{12} =  R_{23}R_{12}R_{23}$ \cite{yb}.\\

\noindent
{\large {\bf 2.3.1.2. L'hyperplan quantique. }}\\ 
\addcontentsline{toc}{subsection}{\numberline{}  
2.3.1.2. L'hyperplan quantique }

Une autre possibilit\'e pour construire un syst\`eme de $q-$oscillateurs 
est de faire agir les op\'erateurs nombres sym\'etriquement pour $\theta$ et
$\p$. Suivant le principe de la transformation de Klein, introduite au
paragraphe pr\'ec\'edent, d\'efinissons

\beqa
\label{eq:qpalne1}
\xi_i \longrightarrow && \eta_i = \xi_i \prod 
\limits_{j >i} q^{{\cal N}_j/2} \\
\partial_{xi_i}  \longrightarrow &&
\partial_{ \eta_i} = \partial_{\xi_i} \prod \limits_{j >i} q^{{\cal N}_j/2},
\nonumber
\eeqa

\noindent
les variables $\eta, \partial_{\eta}$ satisfont alors

\beq
\label{eq:qplane2}
\left\{
\begin{array}{ll}
\eta_i \eta_j = q^{1/2} \eta_j \eta_i,& i<j \\
\partial_{\eta_i} \partial_{\eta_j} = q^{-1/2}
\partial_{\eta_j} \partial_{\eta_i},& i<j  \\
\partial_{\eta_i} \eta_j = q^{1/2} \eta_j \partial_{\eta_i},& i\ne j \\
\partial_{\eta_i} \eta_i -q \eta_i \partial_{\eta_i}=1
+(q-1) \sum \limits_{j >i} \eta_j \partial_{\eta_j}.& \\
\end{array}
\right.
\eeq

\noindent
Une telle structure diff\'erentielle a \'et\'e obtenue par Wess et Zumino
\cite{wz} ainsi que Pusz et Woronowicz \cite{pw} en imposant une covariance
sous le groupe quantique $GL_q(p)$. La dissym\'etrie entre les variables
se r\'esume dans la propri\'et\'e $\partial_{\xi_j} \xi_j - 
\xi_j\partial_{\xi_j}= q^{{\cal N}_j}=1+(q-1)\xi_j \partial_{\xi_j}$.

\vskip1truecm
Bien que les relations de d\'efinition de (\ref{eq:qtp}) et de 
(\ref{eq:qplane2})
soient diff\'erentes, elle pr\'esentent une certaine analogie. Ces relations,
vraies pour toutes les valeurs de $q$, se singularisent lorsque $q^n=1$.
En effet, pour les racines de l'unit\'e, on peut trouver des repr\'esentations
matricielles pour $\theta_i,\partial_{\theta_i}$ ainsi que 
$\eta_i,\partial_{\eta_i}$ (voir \'equations (\ref{eq:thetamat})).
Avec les notations que nous avons choisies, 
pour pouvoir comparer les alg\`ebres
apparaissant dans les deux sous-sections pr\'ec\'edentes, il faut faire
la substitution $q \to q^{1/2}$. Si  $q=-1$ pour la premi\`ere 
alg\`ebre (c'est-\`a-dire $q^{1/2}=-1$ pour la seconde), les deux structures
construites sont identiques et correspondent \`a un syst\`eme de $p$
oscillateurs fermioniques. La diff\'erence dans ces deux extensions r\'eside
dans leur groupe de stabilit\'e; il est a posteriori \'evident que
(\ref{eq:qtp}) et (\ref{eq:qplane2}) n'admettent pas le m\^eme groupe de 
covariance. Car, si on adoptait l'hypoth\`ese contraire,
 on pourrait passer des premi\`eres
relations aux secondes. Pour les  fermions, ces deux s\'eries
d'oscillateurs peuvent \^etre construites \`a partir de l'alg\`ebre de 
Clifford (\ref{eq:theta}). Cette propri\'et\'e traduit simplement le fait 
que les matrices $\gamma$ engendrent la repr\'esentation spinorielle de
$SO(2p)$. Maintenant, si $q$ est une racine primitive $n-$i\`eme
de l'unit\'e, on peut bien s\^ur construire les variables $\theta_i$ \`a partir
des \'el\'ements de l'alg\`ebre de Clifford g\'en\'eralis\'ee (en prenant
par exemple $l=0$ dans l'\'equation (\ref{eq:theta})); mais nous ne sommes
pas en mesure de construire une d\'erivation  en consid\'erant les
variables $\theta_i^{(l)}$. C'est encore une limitation des polyn\^omes
de degr\'e sup\'erieur \`a $2$ par rapport aux quadratiques.\\

\noindent
{\large {\bf 2.3.1.3. Extensions alternatives. }}\\
\addcontentsline{toc}{subsection}{\numberline{}  
2.3.1.3. Extensions alternatives. }

A partir de $p$ s\'eries de $q-$oscillateurs, en introduisant des 
transformations bien choisies, nous avons \'et\'e capables de construire
deux syst\`emes distincts. Il existe une autre 
extension possible, appel\'ee  quons \cite{cgm,gm}, et n'ayant 
pas de rapport avec les oscillateurs $q-$d\'eform\'es. 
La diff\'erence essentielle entre ce syst\`eme et les pr\'ec\'edents est
que nulle relation de commutation entre les diff\'erentes familles
d'oscillateurs n'est connue, except\'e

\beq
\label{eq:quon}
\partial_i \theta_j - q \theta_j \partial_i = \delta_{ij}.
\eeq

D'autres possibilit\'es peuvent encore \^etre introduites. Les arguments
conduisant \`a ces extensions sont \'evidemment diff\'erents
de ceux consid\'er\'es jusqu'\`a pr\'esent. Ainsi, dans le chapitre suivant,
d'autres possibilit\'es, justifi\'ees par la coh\'erence 
du syst\`eme que nous
utiliserons, seront \`a la base d'extensions diff\'erentes.

\vskip1truecm
Formellement, toutes ces $q-$g\'en\'eralisations peuvent \^etre introduites
quelle que soit la dimension de l'espace-temps, mais d'apr\`es le th\'eor\`eme
spin-statistique, les seules extensions envisageables des fermions et des
bosons sont les parafermions et les parabosons \cite{g,ko}. Cependant,
on peut profiter de l'existence de  dimensions ``exceptionnelles'' pour
introduire de telles s\'eries d'oscillateurs sans violation du
th\'eor\`eme spin-statisque. C'est ce que nous ferons dans le chapitre
suivant. 

\subsection{Int\'egration}

Pour les variables de Grassmann, Berezin a montr\'e que l'on pouvait d\'efinir
une int\'egration adapt\'ee. En utilisant
des arguments similaires, nous avons \'etabli qu' il 
est possible d'introduire une int\'egration
pour les variables de Grassmann g\'en\'eralis\'ees \cite{frint}.
En imposant l'invariance par translation, on obtient

\beq
\label{eq:inttheta}
\int d \theta \theta^a = \{n-1\}_1! \delta_{a,n-1},
\eeq

\noindent
on  identifie donc  
$\int d \theta \equiv \left({d \over d \theta}\right)^{n-1}$. La
normalisation apparaissant dans l'\'equation (\ref{eq:inttheta}) \'etant
parfaitement arbitraire, on aurait pu  \'egalement d\'efinir l'int\'egration
\`a partir d'une autre d\'eriv\'ee.
Une telle formule d'int\'egration a aussi \'et\'e obtenue par un certain
nombre d'auteurs \cite{bf,ama,gn}. La relation (\ref{eq:inttheta}) est
d\'efinie pour les racines de l'unit\'e; 
cependant, rien ne restreint $q$ de la sorte. 
Il est \'egalement possible  de consid\'erer une d\'erivation et une 
int\'egration lorsque $q$ est quelconque \cite{gn,bf}. 

L'\'etape ult\'erieure consiste \`a consid\'erer une int\'egration quand
plusieurs variables de Grassmann g\'en\'eralis\'ees sont introduites.
Nous nous limiterons \`a une int\'egration associ\'ee au syst\`eme
(\ref{eq:qtp}). Ce qui revient \`a dire que

\beq
\label{eq:qint}
\left\{
\begin{array}{ll}
\theta_i \theta_j = q \theta_j \theta_i,& i<j \\
\int d {\theta_i} \int d {\theta_j} = 
q\int d {\theta_j} \int d {\theta_i},& i<j  \\
\int d {\theta_i} \theta_j = q \theta_j \int d {\theta_i}, &i<j  \\
\int d {\theta_j} \theta_i = q^{-1} \theta_i \int d {\theta_j},& i<j.
\end{array}
\right.
\eeq   

\noindent
Ceci \'equivaut \`a  imposer que les diff\'erentes 
int\'egrations soient ``emboit\'ees'' les unes dans les autres, 
c'est-\`a-dire, que pour une fonction

$$f(\theta_1,\cdots,\theta_p)=\sum \limits_{ \{a_i\}=0,\cdots, n-1}
\theta_1^{a_1} \cdots \theta_p^{a_p} f_{a_1 \cdots a_p},$$

\noindent
on a
\beq
\label{eq:qint2}
\int \prod \limits_{i=p}^1 d \theta_i f(\theta_1,\cdots,\theta_p) =
\int d^n \theta f(\theta_1,\cdots,\theta_p) = \left(\{n-1\}_1!\right)^p
f_{n-1 \cdots n-1},
\eeq

\noindent
apr\`es int\'egration successive sur les variables $\theta_1,\cdots,
\theta_p$.\\

\noindent
{\large {\bf 2.3.2.1. Int\'egration et base des \'etats coh\'erents.}}\\
\addcontentsline{toc}{subsection}{\numberline{}  
2.3.2.1. Int\'egration et base des \'etats coh\'erents }

A partir des formules d'int\'egration, il 
devient possible de d\'eduire les diverses grandeurs en fonction  de 
$\theta$ et $\bar \theta$ et de la base des \'etats coh\'erents.

En notant que
la relation de fermeture peut se r\'e\'ecrire \cite{bf}

\beq
\label{eq:ferm}
\sum \limits_{p=0}^{n-1} \left|p\right> \left<p\right| =
\int{ d \bar \theta d \theta \over [n-1] !}
 \exp_q{(-\bar \theta \theta)} 
\left|\theta\right>\big< \theta|=1,
\eeq
\noindent 
il est alors possible de r\'eexprimer un vecteur d'\'etat \cite{bf}

\beq
\label{eq:etat}
\left\{
\begin{array}{ll}
\left|\psi\right>=\sum \limits_{p=0}^{n-1} \psi_p \left|p\right>,&
\left<\chi\right| = \sum \limits_{p=0}^{n-1} \left<p\right| \chi_p^\star 
 \\
\psi(\bar \theta)= \left<\theta | \psi\right> =
\sum \limits_{p=0}^{n-1} \psi_p {\bar \theta^p \over \sqrt{[p]!}},&
\bar \chi(\theta) = \left<\chi| \theta\right> = 
\sum \limits_{p=0}^{n-1} \chi_p^\star { \theta^p \over \sqrt{[p]!}},
\end{array}
\right.
\eeq

\noindent
ainsi qu'un op\'erateur (en introduisant deux couples $\theta,\bar \theta$
et $\theta^\prime,\bar \theta^\prime$ commutants) \cite{bf}

\beqa
\label{eq:oper}
A& =&\sum \limits_{p,q=0}^{n-1} A_{pq} \left|p\right>\left<q\right| \\
A(\bar \theta,\theta^\prime)&=& 
\left<\theta |A|\theta^\prime\right> = 
\sum \limits_{p,q=0}^{n-1} A_{pq} {\bar \theta^q \over \sqrt{[q]!}} \nonumber
{\theta^{p\prime}  \over \sqrt{[p]!}}. 
\eeqa 

Ainsi, en utilisant la relation de fermeture, on obtient
$\left<\chi |\psi\right> = \int {d \bar \theta d \theta \over [n-1]!}
 \chi^\star(\theta) \psi(\bar \theta)$, et de ce fait, les vecteurs
${\theta^p \over \sqrt{[p]!}}$ d\'efinissent une base orthonorm\'ee 
\cite{bf,flo}. 

Ce faisant, la base des \'etats coh\'erents permet, par l'interm\'ediaire
d'une int\'egration de Berezin, adapt\'ee aux variables de Grassmann 
g\'en\'eralis\'ees, d'obtenir une repr\'esentation de l'espace des 
\'etats de l'oscillateur $q-$d\'eform\'e (ou bien de $\H_{q,n}$) \`a
partir d'un jeu de variables $\theta, \bar \theta$ \cite{bf,flo}. 
Les propri\'et\'es d'un syt\`eme d\'ecrit 
par une alg\`ebre $\H_{q,n}$ pourront donc
\^etre d\'eduites de variables classiques en utilisant une
int\'egrale de chemin adapt\'ee. \\ 

{\large {\bf 2.3.2.2. Changement de variables }}\\
\addcontentsline{toc}{subsection}{\numberline{}  
 2.3.2.2. Changement de variables }

En supersym\'etrie, la construction d'actions invariantes passe par
la consid\'eration d'int\'e\-grales sur des variables 
bosoniques et fermioniques.
Nous serons amen\'es, lors de l'\'etude de sym\'etries au-del\`a de la 
supersym\'etrie, \`a d\'efinir des int\'egrations o\`u les variables de 
Grassmann g\'en\'eralis\'ees et de type bosonique seront consid\'er\'ees
simultan\'ement. Nous avons \'eta\-bli, quand $n=3$  \cite{fr}, des
relations analogues \`a celle du superd\'eterminant. Ces formules de 
changement de base se font en suivant mot pour mot la d\'emonstration 
de Berezin \cite{ber}. Soient $k$ variables bosoniques $x_a$ et $p$
variables de Grassmann g\'en\'eralis\'ees $\theta_i$. On consid\`ere le
changement de base

\beqa
\label{eq:cgt}
x^\prime&=&  x A+  \theta B \\
\theta^\prime &=&  x C+  \theta D, \nonumber
\eeqa

\noindent
o\`u $A$ est une matrice $k \times k$ de graduation $0$, $B$ une matrice
$k \times p$ de graduation $-1$, $C$ une matrice $p \times k$ de graduation
$1$ et enfin $D$ une matrice $p \times p$ de graduation $0$. On obtient
alors

\beq
\label{eq:sdet}
\int d x^{\prime k} d \theta^{\prime p} = \det( A - BD^{-1}C) 
\left(\det D\right)^{-(n-1)}
\int d x^{k} d \theta^{p}.
\eeq

\noindent
Cette relation est une g\'en\'eralisation directe du superd\'eterminant
apparaissant dans les th\'eories supersym\'etriques \cite{ber}.\\

\noindent
{\large {\bf 2.3.2.3. Int\'egrale de chemin}}\\
\addcontentsline{toc}{subsection}{\numberline{}  
2.3.2.3. Int\'egrale de chemin }

Les relations d'int\'egration et de changement de base introduites ci-avant
vont \^etre utiles pour \'etablir la construction d'actions invariantes.
Au niveau quantique, nous avons besoin de d\'efinir des int\'egrales
de chemin. Le passage d'une int\'egration de dimension infinie se fait
comme d'habitude en consid\'erant, dans un premier temps, une int\'egration
de dimension finie, puis en prenant la limite infinie. 
A cet \'egard, introduisons deux s\'eries de variables de Grassmann 
g\'en\'eralis\'ees $\theta^i_\ell$ $\ell=1, \cdots N$ et $i=1,2$ de graduation
respective ${\mathrm gr}( \theta^1_\ell) =1$ et 
${\mathrm gr} (\theta^2_\ell) =-1$.
Ces deux variables peuvent \^etre comprises comme la version classique
des op\'erateurs $\theta_\ell,\partial_{\theta_\ell}$. On peut supposer que
l'on a une relation d'ordre $(i,\ell) < (j,k)$ si $i < j$ ou $i=j$ et 
$\ell < k$,
permettant de d\'efinir des relations de $q-$mutation entre les $\theta$,
ou tout simplement que les diff\'erents couples commutent. Cette derni\`ere 
alternative est pr\'ef\'erable \'etant donn\'e que dans la limite continue,
$\theta^i_\ell$ et $\theta^i_{\ell+1}$ commutent.
Nous avons  montr\'e que l'on a \cite{prs1}

\beq
\label{eq:path}
\int d^N \theta^1 d^N \theta^2 e^{\theta^1 \Delta \theta^2}= 
{\cal N} \left(\det \Delta \right)^{n-1}.
\eeq

\noindent
En effet, la forme quadratique $\Delta$ peut \^etre diagonalis\'ee en utilisant
deux transformations unitaires $J_1, J_2$: diag=$J_1 \Delta J_2$. Ensuite,
en utilisant les relations de changement de base (\ref{eq:sdet}) et les
lois d'int\'egration sur les variables de Grassmann g\'en\'eralis\'ees 
(\ref{eq:qint2}), on obtient (\ref{eq:path}). Le coefficient ${\cal N}$ est
calculable, il provient d'une part des int\'egrales sur les $\theta$ et
d'autre part, des commutations des diff\'erentes variables. Il peut
donc \^etre r\'eabsorb\'e dans la mesure d'int\'egration. Si d'autres relations
de $q-$mutation entre les couples  de variables $\theta$ sont
d\'efinies, en choisissant une normalisation
appropri\'ee dans la mesure d'int\'egration, le r\'esultat  reste
inchang\'e car les propri\'et\'es primordiales sont 
$\left(\theta_\ell^1\right)^n=0$, et la relation (\ref{eq:inttheta}). Des
r\'esultats similaires pour les variables de Grassmann g\'en\'eralis\'ees
d'ordre $3$ ont \'egalement \'et\'e obtenus par Matheus-Valle {\it et al}
dans \cite{fsusy2d}.

\vskip 1truecm
Dans cette partie, nous nous sommes concentr\'es sur les propri\'et\'es 
des $q-$oscillateurs, ou de $\H_{q,n}$ 
l'alg\`ebre de Heisenberg $q-$d\'eform\'ee.
Nous avons  mentionn\'e leurs applications \`a l'\'etude des
repr\'esentations des alg\`ebres $q-$d\'eform\'ees. C'est d'ailleurs la raison
pour laquelle Macfarlane et Biedenharn \cite{bm} avaient consid\'er\'e
de telles d\'eformations (ils voulaient obtenir des repr\'esentations de
$SU_q(2)$).

La raison pour laquelle nous nous sommes int\'eress\'es \`a ces structures
est diff\'erente. Nous verrons dans le chapitre suivant que les variables
$\theta$ et $\p$ vont \^etre \`a la base d'une sym\'etrie g\'en\'eralisant
la supersym\'etrie. 

%Avant de  clore ce chapitre, metionnons,  parmi les nombreuses 
%applications de $\H_q$, un dernier exemple.
%Le syst\`eme $\theta$, $\p$, peut \^etre repr\'esent\'e par la variable
%bosonique $x$ et la d\'eriv\'ee de Jackson \cite{jack}
%$$ D f(x) = {f(qx)-f(x) \over (q-1)x}.$$
%On peut alors d\'efinir une \'equation diff\'erentielle o\`u la d\'eriv\'ee
%usuelle est remplac\'ee par son symbole de Jackson associ\'e. Ainsi,
%aux solutions classiques, exprim\'ees en termes de fonctions sp\'eciales,
%se substitue les fonctions $q-$d\'eform\'ees correspondantes et 
%au groupe de sym\'etrie le groupe $q-$ d\'eform\'e {\it ad hoc}. 
\noindent

\chapter[Extension de la supersym\'etrie: Supersym\'etrie  fractionnaire]
{\center{ Extension de la supersym\'etrie: Supersym\'etrie fractionnaire}}

%\stepcounter{chapter}

La formulation des th\'eories d\'ecrivant les particules \'el\'ementaires
par exemple est intimement corr\'el\'ee aux sym\'etries sous-jacentes. 
De ce fait, les particules  vont \^etre class\'ees en 
repr\'esentations irr\'eductibles du groupe de  sym\'etries fondamentales
de l'espace-temps, c'est-\`a-dire du groupe de Poincar\'e. Par ailleurs,
on sait que d'une part, une fois ce groupe jaug\'e (ou plus exactement le
sous-groupe des translations), on obtient une description de l'interaction
gravitationnelle et que d'autre part, la notion de groupe de Lie 
compact permet de d\'ecrire les interactions fondamentales ou de jauge. 
En outre, d'apr\`es le th\'eor\`eme de Noether, 
on peut connecter les g\'en\'erateurs
des groupes de sym\'etries aux quantit\'es conserv\'ees, cons\-truites 
directement \`a partir de la densit\'e lagrangienne.  Les g\'en\'erateurs
ainsi obtenus agiront sur l'espace de Hilbert d\'ecrivant les \'etats 
accessibles du syst\`eme consid\'er\'e.

En th\'eorie quantique des champs relativistes,  les processus
d'interaction entre particules (diffusion, cr\'eation, etc)  sont d\'ecrits par
l'interm\'ediaire d'une quantit\'e fondamentale, la matrice $S$. Celle-ci
permet de relier  les \'etats  entrants aux \'etats sortants. Un
certain nombre de contraintes physiques  restreint consid\'erablement les 
sym\'etries de la th\'eorie. En effet, si on impose l'unitarit\'e de la matrice
$S$, ainsi qu'un spectre de particules massives discret et un intervalle 
d'\'energie non-nul entre les \'etats \`a une particule et 
le vide, deux th\'eor\`emes
limitent fortement les sym\'etries envisageables. Si on recherche une th\'eorie
d\'ecrite dans le cadre d'alg\`ebres de Lie, le th\'eor\`eme de Coleman
et Mandula \cite{cm} implique une extension triviale de l'alg\`ebre de 
Poincar\'e. Ceci revient \`a dire que l'on obtient une structure du type
$P \otimes G$, o\`u $P$ est le groupe de Poincar\'e et $G$ un groupe de 
sym\'etries internes, dont les g\'en\'erateurs sont des scalaires de Lorentz,
susceptibles de d\'ecrire les interactions de jauge. Si maintenant on
s'int\'eresse au cas non-massif, on sort des conditions de validit\'e du
th\'eor\`eme de Coleman et Mandula,  
le groupe des sym\'etries fondamentales n'est
plus le groupe de Poincar\'e, mais  le groupe conforme. Si on
va au-del\`a des alg\`ebres de Lie de fa\c con \`a  permettre 
la prise en compte de
super-alg\`ebres de Lie, c'est-\`a-dire que l'on introduise des 
g\'en\'erateurs
dans une repr\'esentation fermionique du groupe de Poincar\'e, 
le th\'eror\`eme de Haag, Lopuszanski et Sohnius \cite{hls} impose
que les g\'en\'erateurs suppl\'ementaires soient dans la repr\'esentation 
spinorielle du groupe de Lorentz. Une telle extension est bien \'evidemment
non-triviale. D'apr\`es le th\'eor\`eme spin-statistique, il semble donc qu'il
existe une {\it unique} extension non-triviale de l'alg\`ebre de Poincar\'e,
au nombre de supercharges pr\`es.

Ainsi, il appara\^{\i}t que  ces deux th\'eor\`emes mettent un point final
aux types de sym\'etries envisageables.  Mais, si l'on regarde de
plus pr\`es les conditions de validit\'e des th\'eor\`emes
 de Coleman \& Mandula
et de Haag, Lopuszanski \& Sohnius, il devient possible, sous certaines 
conditions, d'envisager des sym\'etries allant au-del\`a de 
la supersym\'etrie. On peut classer les extensions permises en deux grandes
cat\'egories: Pour les premi\`eres, on introduit une s\'erie de 
g\'en\'erateurs sans
faire appel aux repr\'esentations du groupe de Lorentz alors que pour les
secondes, on impose que les nouveaux g\'en\'erateurs soient dans une 
repr\'esentation appropri\'ee du groupe de Lorentz.

En ce qui concerne les premiers types d'extensions r\'ealisables,
une premi\`ere  possibilit\'e passe par les
groupes quantiques \cite{fuchs} et  l'espace-temps se trouve alors
d\'efini par une  structure non-commutative et le groupe de sym\'etrie devient
le groupe de Poincar\'e $q-$d\'eform\'e \cite{wzm}.
On peut \'egalement supposer que les g\'en\'erateurs du groupe
de Poincar\'e eux-m\^emes admettent, outre leurs indices d'espace-temps, des
indices internes. Il est alors possible, en utilisant le formalisme des
alg\`ebres de Lie g\'en\'eralis\'ees \cite{gla}, de construire une extension
de l'alg\`ebre de Poincar\'e $\ZZ_n-$gradu\'ee \cite{wt}.  
Une troisi\`eme alternative consiste \`a introduire des alg\`ebres $\ZZ_n$
gradu\'ees dont les relations de fermeture se font via des relations
d'ordre $n$ associ\'ees \`a une sym\'etrie interne,  au lieu de relations 
quadratiques \cite{kern}.
Quant aux se\-conds types d'extensions envisageables,
on admet comme point de d\'epart que les nouveaux
g\'en\'erateurs sont dans une repr\'esentation du groupe de Lorentz qui
n'est ni bosonique, ni fermionique. La premi\`ere possibilit\'e passe
par les parafermions \cite{g,ko}. Ce faisant, on va consid\'erer des 
alg\`ebres admettant une structure trilin\'eaire et conduisant
\`a la parasupersym\'etrie \cite{psusy1,psusy2}; 
appliqu\'ee aux sym\'etries de
l'espace-temps, on peut alors construire une extension parasupersym\'etrique 
(PSUSY) du groupe de Poincar\'e \cite{psusy3}. La derni\`ere possibilit\'e, 
celle que nous avons
suivie, r\'eside dans l'existence de dimensions 
exceptionnelles, pour lesquelles
il est possible d'envisager des champs ni bosoniques ou fermioniques, ni
parabosoniques ou parafermioniques. Une telle sym\'etrie, appel\'ee
supersym\'etrie fractionnaire (FSUSY), a \'et\'e consid\'er\'ee
par un certain nombre d'auteurs \cite{abl,dur1,mm,am,fr,prs1,prs2,dur2,
fsusy1d,fsusy2d,fsusy3d}.  En fait, \`a l'origine,
la PSUSY \cite{psusy1} et la  FSUSY \cite{fsusy1d, fsm} 
avaient \'et\'e introduites dans le but de d\'ecrire 
un syst\`eme quantique pr\'esentant une d\'eg\'en\'erescence du spectre, 
c'est-\`a-dire une sym\'etrie cach\'ee; de telles sym\'etries
sont directement inspir\'ees
de la m\'ecanique quantique supersym\'etrique consid\'er\'ee par Witten
\cite{sqm}.

Dans ce chapitre, nous allons nous int\'eresser \`a la supersym\'etrie
fractionnaire et nous allons \'etudier ses implications en th\'eorie des
champs. Nous allons donc rechercher des dimensions de l'espace-temps pour
lesquelles il est possible d'envisager d'autres types de champs. Il est
int\'eressant de souligner  que tout le formalisme que nous avons
d\'eve\-lopp\'e jusqu'\`a pr\'esent trouve une application naturelle dans le 
cadre de la FSUSY. Pour r\'esumer sommairement, on peut affirmer que
les extensions des alg\`ebres de Clifford et de Grassmann, d\'evelopp\'ees
dans le premier chapitre, conduisent aux $q-$oscillateurs ou aux alg\`ebres
de Heisenberg $q-$d\'eform\'ees. Nous allons mettre en \'evidence le fait que,
gr\^ace \`a la structure math\'ematique
inh\'erente aux $q-$oscillateurs, on peut construire une th\'eorie des
champs ayant une structure $\ZZ_n-$gradu\'ee. On va donc introduire 
des g\'en\'erateurs g\'en\'eralisant les g\'en\'erateurs  
bosoniques/fermioniques, ou g\'en\'erateurs $\ZZ_2-$gradu\'es (les bosons,
fermions \'etant respectivement pairs, impairs). De prime abord, la
prise en compte de tels types de g\'en\'erateurs semble \^etre en opposition
avec les r\'esultats de la th\'eorie des repr\'esentations du groupe de
Lorentz o\`u seuls des champs de spin entier/demi-entier existent. 
C'est \`a ce stade que la prise en compte des dimensions exceptionnelles
prend corps (de fa\c con identique, ce n'est pas de l'\'etude g\'en\'erale
de $SO(n)$ que l'on peut obtenir les propri\'et\'es de trialit\'e de
$SO(8)$). 

Lorsque la dimension de l'espace-temps est  $1$, le groupe des rotations est 
inexistant, le seul g\'en\'erateur r\'eside alors dans les translations
sur l'axe du temps. La supersym\'etrie fractionnaire d'ordre $F$ 
consiste alors en la prise en consid\'eration d'un g\'en\'erateur $Q$ 
s'interpr\'etant comme la racine $F-$i\`eme du Hamiltonien ($Q^F=\partial_t$).
Nous avons appliqu\'e une telle sym\'etrie dans le formalisme de la ligne 
d'univers o\`u, dans un langage issu des cordes, les propri\'et\'es
spatio-temporelles peuvent \^etre d\'eduites de celles sur la  ligne d'univers
\cite{fr}, r\'esultats que nous allons d\'evelopper dans une premi\`ere
section apr\`es avoir construit une version locale de la supersym\'etrie
fractionnaire, que nous avons appel\'ee supergravit\'e fractionnaire. En 
suivant le formalisme de la quantification avec contraintes d\'evelopp\'e par
Dirac, nous avons d\'eriv\'e une \'equation pour laquelle l'op\'erateur 
de base est une racine cubique de l'op\'erateur de Klein-Gordon ($F=3$).

Dans les espaces de dimension $2$, si on se limite aux \'etats de masse nulle,
on peut tirer parti de l'invariance conforme et de l'alg\`ebre de Virasoro.
On peut donc profiter de l'existence de champs de spin  (plus
exactement de poids conforme) arbitraire et introduire en plus du tenseur
\'energie-impulsion de poids conformes $2$, un g\'en\'erateur de 
poids conformes $1+1/F$ \cite{prs1,prs2}.
Une telle extension contiendra des champs
primaires de poids conformes $(0,1/F,\cdots, (F-1)/F)$. Nous allons nous
attacher \`a d\'efinir une telle sym\'etrie dans la seconde section, o\`u
nous construirons explicitement l'alg\`ebre (OPE)
et nous verrons que la coh\'erence impose les relations de commutation
entre les diff\'erents oscillateurs ($q-$oscillateurs) apparaissant dans
la th\'eorie, ainsi qu'un th\'eor\`eme de Wick adapt\'e.  

Enfin, le dernier cas particulier est celui des espaces de dimension $1+2$,
o\`u des \'etats de spin arbitraire existent. Les anyons ont \'et\'e mis
en  \'evidence pour la premi\`ere fois par Leinaas et Myrheim \cite{lm}.
On peut comprendre l'existence de telles particules de deux fa\c cons 
diff\'erentes. Premi\`erement, en dimension $3$ un syst\`eme de plusieurs
particules est dans une repr\'esentation du groupe des tresses, et non des
permutations, et le syst\`eme peut donc prendre une phase arbitraire 
(diff\'erente de $1$ ou $-1$) quand on permute les particules.
Secundo, le petit groupe, associ\'e aux \'etats  massifs est 
$\overline{SO(2)}$ (le groupe de recouvrement universel de $S0(2)$). 
Un tel groupe n'\'etant pas quantifi\'e,
la fonction d'onde, de dimension $1$, peut prendre une phase arbitraire 
apr\`es une rotation de $2 \pi$. Nous avons alors, en consid\'erant des
op\'erateurs dans une repr\'esentation ``anyonique'' de $SO(1,2)$, pu
d\'efinir une extension non-triviale de la supersym\'etrie. Cette sym\'etrie 
sera une sym\'etrie entre les diff\'erents champs anyoniques \cite{fsusy3d}. 
Nous  pr\'esenterons nos r\'esultats dans la troisi\`eme section.

Le point commun \`a $D=1,2,3$ est que si $F=2$, on  retrouve la supersym\'etrie
usuelle. De plus, tout comme en supersym\'etrie, la FSUSY est une sym\'etrie
agissant sur des champs de graduations $(0,1,\cdots F-1)$, g\'en\'eralisant
le concept de boson ou de fermion. Ensuite, on observe que le formalisme du
superespace d\'evelopp\'e dans le cadre de la supersym\'etrie trouve une
extension naturelle en termes de $\H_{q,F}$, l'alg\`ebre de Heisenberg
$q-$d\'eform\'ee. Notons enfin qu'une classification de cette famille de 
th\'eories appara\^{\i}t. Ainsi, si $F=F_1F_2$, une th\'eorie FSUSY d'ordre
$F$ sera automatiquement FSUSY d'ordre $F_1$ ou $F_2$.

Bien que le point de d\'epart des alg\`ebres FSUSY et celui consid\'er\'e
dans \cite{kern} soit diff\'erent, on peut relever un certain nombre 
de points communs.
En effet, dans ces r\'ef\'erences, une alg\`ebre admettant une structure
ternaire a \'et\'e construite (qui correspond  \`a $F=3$). Les structures
fondamentales qui transparaissent lors de la construction de cette alg\`ebre
sont une repr\'esentation non-fid\`ele de l'alg\`ebre $3-$ext\'erieure et
de l'alg\`ebre de Clifford du polyn\^ome $x_1^3 + \cdots$, repr\'esentation
diff\'erente des alg\`ebres de Grassmann et de Clifford g\'en\'eralis\'ees
apparaissant dans les th\'eories FSUSY.  
Cette approche d\'ebouche \'egalement sur une extension de la 
supersym\'etrie pour laquelle les g\'en\'erateurs des translations sont
obtenus \`a partir de relations d'ordre $3$. 

R\'ecemment, il a \'et\'e montr\'e que la supersym\'etrie fractionnaire
en dimension $1$, tout comme la supersym\'etrie, pouvait \^etre obtenue comme
une certaine limite de la ligne ``tres\-s\'ee'' (braided line) \cite{bl}.
Cette interpr\'etation
g\'eom\'etrique de la FSUSY permet \'egalement  de montrer que l'approche
des groupes quantiques et les extensions supersym\'etriques ou
supersym\'etriques fractionnaires sont corr\'el\'ees, 
du moins en dimension $1$.
Evidemment, en utilisant un tel formalisme, on peut se poser la question de la
pertinence de 
l'extension de ces r\'esultats  en dimension $2$ et $3$, et se demander si, 
quand la dimension de l'espace-temps est sup\'erieure ou \'egale \`a $4$, 
on peut montrer que la supersym\'etrie est la sym\'etrie
la plus g\'en\'erale que l'on puisse consid\'erer.
 
\mysection{Supersym\'etrie fractionnaire en dimension 1} 

Comme nous l'avons mentionn\'e ci-avant, la FSUSY avait \'et\'e introduite
pour la description de syst\`emes quantiques pr\'esentant 
une sym\'etrie cach\'ee.
Cependant, on peut en  donner une autre interpr\'etation. 
Parall\`element aux th\'eories de cordes s'est d\'evelopp\'e
le formalisme de la ligne d'univers. En th\'eorie des (super-)cordes,
les propri\'et\'es d'espace-temps sont induites par celles de la surface
d'univers de la corde. Les repr\'esentations du groupe de jauge sont
contr\^ol\'ees par l'alg\`ebre de Kac-Moody \cite{km} et celles du
groupe de Poincar\'e par l'invariance super-conforme \cite{gsw}. 
Un tel formalisme
a en fait \'et\'e inspir\'e par les r\'esultats obtenus \`a une dimension.
En effet, par une description des sym\'etries sur la ligne d'univers,
on peut obtenir une formulation des champs relativistes et m\^eme
une th\'eorie quantique des champs \cite{wl1,wl2,wl3,wlg,wlf,wlfey}.
L'essence d'une telle approche r\'eside dans l'introduction de variables
de Grassmann pour la prise en compte des degr\'es de libert\'e de spin
\cite{wl1,wl2,wl3} ou de jauge \cite{wlg}.

Etant donn\'ees les propri\'et\'es de la  ligne d'univers, les variables de
Grassmann, associ\'ees aux particules de spin $1/2$, ne sont
pas les plus g\'en\'erales que l'on puisse d\'efinir. 

\subsection{Groupe d'invariance de la ligne d'univers}

On consid\`ere une vari\'et\'e Riemanienne ${\cal R}_1$ de dimension $1$
d\'ecrite par une seule  coordonn\'ee $\tau$ et  repr\'esentant la trajectoire
d'une particule ponctuelle. On note $g_{--}$ sa m\'etrique,
les symboles ($--$) permettent de garder la trace du caract\`ere tensoriel du
tenseur m\'etrique. En effet, de prime  
abord, les  propri\'et\'es g\'eom\'etriques de ${\cal R}_1$ semblent triviales 
du fait que tous les tenseurs paraissent identiques. La vari\'et\'e 
${\cal R}_1$
est invariante sous le groupe   des diff\'eomorphismes $\tau^\prime \to \tau 
- f(\tau)$. Une particularit\'e des diff\'eomorphismes en dimension $1$
est que, par d\'efinition, ils conservent la m\'etrique \`a un facteur
d'\'echelle pr\`es et donc co\"{\i}ncident avec les transformations
conformes. Dans le formalisme des actions effectives ou lors des corrections 
quantiques \`a une boucle,
la ligne
d'univers a la topologie d'un cercle et il devient possible de d\'evelopper
la fonction $f$ en s\'erie de Laurent: le groupe des diff\'eomorphismes
du cercle s'identifie alors avec l'alg\`ebre de Virasoro sans charge
centrale \cite{kr} $(\ell_n = {i T \over 2 \pi} e^{{2 i \pi n \tau \over T}}
{ d \over d \tau}$), $T$ repr\'esentant la circonf\'erence du cercle. 
C'est cette propri\'et\'e remarquable qui permet de 
consid\'erer un champ ``de poids conforme'' arbitraire. Un tel
champ $\Phi_h(\tau)$, de poids conforme $h$, admet la loi de transformation

\beq
\label{eq:conf1}
\Phi^\prime_h(\tau^\prime) = \left( {d \tau \over d \tau^\prime }\right)^h
\Phi_h(\tau),
\eeq

\noindent
pour un tenseur d'ordre $(p,q),~h=p-q$.
Pour \'eviter les probl\`emes de racine carr\'ee de nombres n\'egatifs,
on impose  la fonction $f$  croissante.

Lorsque la dimension est $1$, le tenseur de courbure est nul et 
il n'y a pas de connexion de spin au sens usuel du terme. Cependant, il
est possible de d\'efinir une d\'eriv\'ee covariante. En remarquant
que le symbole de Christoffel vaut $\Gamma = \dot e / e$, on
introduit la d\'eriv\'ee covariante

\beq
\label{eq:cov}
\nabla \Phi_h(\tau) = \dot \Phi_h(\tau)  - h { \dot e \over e}
\Phi_h(\tau),
\eeq

\noindent
et on peut r\'e\'ecrire la loi de transformation  (\ref{eq:conf1})
au niveau infinit\'esimal, en notant que $f$, tout comme $\tau$,
est de poids conforme $-1$.

\beqa
\label{eq:conf2}
\delta_f \Phi_h(\tau) &=& \Phi_h^\prime(\tau)- \Phi_h(\tau)  \nonumber \\
&=& f(\tau) \dot \Phi_h(\tau) + h \dot f(\tau) 
\Phi_h(\tau) \\  
&=& f(\tau) \nabla  \Phi_h(\tau) + h \nabla f(\tau)  \Phi_h(\tau). \nonumber
\eeqa 

Nous avons utilis\'e cette description pour interpr\'eter 
les r\'esultats connus
sur la supergravit\'e en dimension $1$ et leur relation avec la description
des champs relativistes de spin $1/2$ \cite{rss}. Nous allons maintenant,
\`a partir de ce formalisme,  donner un nouvel \'eclairage sur ce  que
nous avons obtenu sur la supersym\'etrie fractionnaire en dimension 1 
\cite{fr}.

\subsection{ De la ligne d'univers \`a la super-ligne d'univers}

La supersym\'etrie correspond \`a l'adjonction, au c\^ot\'e  de $H=\partial_t$,
d'un g\'en\'erateur $Q$ v\'erifiant la relation fondamentale $Q^2=H$. 
Une mani\`ere rapide de construire une th\'eorie
supersym\'etrique passe par l'introduction de variables de Grassmann $\theta$,
vues comme les partenaires supersym\'etriques des composantes d'espace-temps.
Le concept de superespace $(t,\theta$) ainsi d\'efini 
peut \^etre introduit plus
formellement en utilisant la notion d'espace quotient \cite{susy}.

La FSUSY \`a une dimension est donc engendr\'ee par deux g\'en\'erateurs
$Q$ et $H$ satisfaisant les relations

\beqa
\label{eq:fsusy1}
\left[Q, H\right] &=& 0 \\
Q^F &=& H. \nonumber
\eeqa

\noindent
Le Hamiltonien engendre les  translations dans le temps et $Q$ les 
transformations FSUSY.
Il est important de noter que (\ref{eq:fsusy1}) ne d\'efinit ni une alg\`ebre
de Lie ni une super-alg\`ebre de Lie \'etant donn\'e que les relations
de fermeture ne sont pas quadratiques mais d'ordre $F$. Une question qui
sera r\'ecurrente tout au long de ce chapitre concerne le statut de
$(Q^2,\cdots,Q^{F-1})$. Parmi les puissances de $Q$, peut-on exhiber d'autres
g\'en\'erateurs associ\'es \`a une sym\'etrie plus large? On verra que si
$F$ n'est pas un nombre premier, et si $p$ divise $F$, alors $Q^p$ sera
un tel g\'en\'erateur. Cette propri\'et\'e, ind\'ependante de la dimension
de l'espace temps, a \'et\'e d\'emontr\'ee en dimension $2$ \cite{prs2}
et $3$ \cite{fsusy3d}. 

Avec comme alg\`ebre de  base (\ref{eq:fsusy1}), en utilisant un formalisme
inspir\'e des superespaces, nous allons construire
une th\'eorie invariante par transformation FSUSY \cite{fr}.  
On consid\`ere $\theta$, une variable de Grassmann 
g\'en\'eralis\'ee d'ordre $F$, r\'eelle, et $\p,\d$ 
ses d\'eriv\'ees associ\'ees. 
En reprenant la terminologie
du chapitre pr\'ec\'edent, les variables $t,\theta$ seront de graduation
$0,1$ dans $\ZZ_F$.

Comme en SUSY, on introduit alors la notion
de superespace fractionnaire $(t,\theta)$, avec $\theta$ une variable
de Grassmann g\'en\'eralis\'ee r\'eelle, par sa param\'etrisation

\beq 
\label{eq:fse}
\exp_{gr} \left(t H + \theta Q\right) \equiv \exp\left(tH\right)
\exp_q\left(\theta Q \right).
\eeq

\noindent
L'exponentielle $q-$d\'eform\'ee consid\'er\'ee dans ce chapitre est
$\exp_q(x)= \sum \limits_{n=0}^{F-1} {x^n\over \{n\}_1!}$
avec $\{n\}_1 \equiv \{n\}$ d\'efini dans le chapitre 2. On remarque alors
que lorsque $F=2$ on a $\exp_{q=-1}(x) = \exp(x)$ si $x^2=0$, c'est la
raison pour laquelle en supersym\'etrie, point n'est besoin de d\'efinir
des exponentielles $-1$ d\'eform\'ees.

On consid\`ere alors les deux sym\'etries globales engendr\'ees respectivement
par $H$ et $Q$. On note $f$ le param\`etre de la transformation associ\'e 
\`a $H$ et $\epsilon$ celui  associ\'e \`a $Q$. On obtient alors,
en utilisant les relations de $q-$mutation
 
\beq
\label{eq:qfmut}
Q \theta = q^{-1}  \theta Q, ~ Q \epsilon = q^{-1} \epsilon Q, ~ 
\theta \epsilon= q \epsilon \theta, 
\eeq

\noindent
ainsi que le fait que la somme
$\exp_q$ est finie (elle s'arr\^ete \`a l'ordre $F$)

\beqa
\label{eq:transfo}
\exp_{gr} \Big(t^\prime H + \theta^\prime Q\Big)
&=&\exp_{gr} \Big(-f H + \epsilon Q\Big) \exp_{gr} 
\Big(t H + \theta Q\Big) \\
&=&\exp\Big(\big\{ t- f + q^{-F(F-1)/2}\sum 
\limits_{a=0}^{F-1} \left[{\epsilon^a 
\theta^{F-a} \over \{a\}_1! \{F-a\}_1!} 
\right] \big\}H \Big)  \exp_q\Big( \left[\theta + \epsilon\right] Q \Big).
\nonumber
\eeqa

\noindent
En fait, nous avons explicitement d\'emontr\'e une telle loi de transformation
pour $F=3$  \cite{fr} et nous avons utilis\'e les r\'esultats \'etablis dans 
\cite{am,bl} pour $F$ quelconque.  Dans \cite{am}, il est montr\'e qu'une telle
transformation admet une structure de groupe.
Les relations de $q-$mutation sont en fait 
impos\'ees, leurs justifications sont donn\'ees dans l'appendice de \cite{fr}. 
On peut montrer, du fait des relations de $q-$mutation (\ref{eq:qfmut}), que 
$t^\prime$ est  r\'eel (lorsque $F$ pair on doit alors effectuer un
changement de normalisation, par exemple en SUSY $t^\prime=i\epsilon \theta$). 
En anticipant sur la suite, on peut trouver trois autres 
justifications  des relations de $q-$mutation propos\'ees. Elle permettent
d'assurer que $\epsilon^F=\theta^F=0$ conduisent bien \`a $(\epsilon+
\theta)^F=0$ (voir chap.1 sur les alg\`ebres de Grassmann g\'en\'eralis\'ees).
On peut \'egalement d\'eduire de (\ref{eq:qfmut}) que $\epsilon Q$ commute
avec la d\'eriv\'ee covariante et v\'erifie la r\`egle de Leibniz  \cite{dur2}
(voir ci-apr\`es).

\subsection{Action invariante par transformations FSUSY}

\noindent
{\large {\bf 3.1.3.1. $F$ est un nombre premier}}\\
\addcontentsline{toc}{subsection}{\numberline{}  
 3.1.3.1. $F$ est un nombre premier }

On consid\`ere un champ scalaire r\'eel d\'efini sur la superligne 
fractionnaire et on se restreint au cas o\`u $F$ est un nombre
premier 

\beq
\label{eq:fsc1}
\phi(t,\theta) = x(t) + \sum \limits_{n=1}^{F-1}
q^{n^2/2} \theta^n \psi_n(t).
\eeq

\noindent
Les composantes de $\phi$ g\'en\'eralisent le 
concept de fermion et de boson. Elles 
satisfont $\psi_n^F=0, n \neq 0$ et sont de graduation $(-n)$ 
($x=\psi_0$ est de graduation $0$)
dans $\ZZ_F$ et de ce fait $\psi_{F-n}$ a les m\^emes propri\'et\'es que 
$\theta^n$.  
Dans le formalisme d\'evelopp\'e dans la section (3.1.1),
en notant que $\theta$ est de poids conforme $ -1/F$, on voit que les
champs $\psi_n$ sont de poids conforme $n/F$.
On postule alors les relations de $q-$mutation 

\beqa
\label{eq:qmutchp}
\theta x(t) &=& x(t) \theta \nonumber \\
\theta \psi_n(t) &=& q^{-n} \psi_n(t) \theta    \\
\psi_n(t) \psi_{F-n}(t) &=& q^n  \psi_{F-n}(t) \psi_n(t)  \nonumber \\
\psi_n(t) \psi_p(t) &=& \psi_p(t) \psi_n(t),~~ p \ne F-n. \nonumber
\eeqa

\noindent 
Les deux premi\`eres  traduisent que $\theta$ est un compteur
de graduation  et la troisi\`eme prendra tout son sens lorsque l'on
montrera que la quantification du syst\`eme $(\psi_n,\psi_{F-n})$ conduit
\`a une alg\`ebre de Heisenberg $q^n-$d\'eform\'ee. Si on choisit les 
composantes $\psi_n$ r\'eelles, les relations de $q-$mutation 
(\ref{eq:qmutchp})
et la normalisation (\ref{eq:fsc1}) nous indiquent que le superchamp
$\phi$ est r\'eel. On indique \'egalement que les champs
$\psi_n$ et $\psi_p$ ont les m\^emes relations de 
$q-$mutation que $\psi_n$ et $\dot \psi_p$ \cite{dur2,fr}.

Si dans la loi de transformation (\ref{eq:transfo}), on ne tient pas compte
du facteur $q^{-F(F-1)/2}$, de fa\c con \`a avoir des notations 
coh\'erentes avec
celles de
la litt\'erature (en rempla\c cant $\epsilon \to q^{-F(F-1)/2} \epsilon$), 
et que l'on consid\`ere $t^\prime$ et $\theta^\prime$  au niveau 
infinit\'esimal, on obtient 

\beqa
\label{eq:tfsusy1}
\delta_\epsilon x(t) &=& q^{1/2} \epsilon \psi_1(t) \nonumber \\
\delta_\epsilon \psi_n(t) &=& q^{1/2} \{n+1\}_1  \epsilon \psi_{n+1}(t),~~
a=1,\cdots, F-2 \\
\delta_\epsilon \psi_{F-1}(t) &=& (-1)^F q^{1/2} {(1 - q) \over F}^F 
\partial_t x(t).
\nonumber
\eeqa

\noindent
On constate que les deux membres des \'equations ont, comme il se doit,
la m\^eme graduation. Tout comme en supersym\'etrie, le param\`etre de la
transformation $\epsilon$ n'est pas de type bosonique.
En notant que $\epsilon$ satisfait les m\^emes relations de $q-$mutation
avec les champs que $\theta$, on observe que $\psi_n^\prime(t)=
\psi_n(t) + \delta_\epsilon \psi_n(t)$ ($\psi_0 \equiv x$) ne satisfont
pas les relations de $q-$mutation (\ref{eq:qmutchp}). Pour rem\'edier \`a ce
probl\`eme, Matheus Valle {\it et al} ont introduit des cocycles pour corriger
la statistique. Nonobstant, il n'est pas n\'ecessaire d'introduire de tels 
cocycles.  En effet,
ce n'est pas $\psi_n^\prime(t)$ qui doit se comporter vis-\`a-vis
de $\theta$ comme $\psi_n(t)$,  mais $\psi_n(t^\prime)$. On peut v\'erifier
que ces derniers ont le bon comportement. Le fait que  $\psi_n^\prime(t)$
ne v\'erifie pas  (\ref{eq:qmutchp}) n'est pas \'etonnant car on est
pass\'e de $\psi_n(t^\prime)$ \`a $\psi_n^\prime(t)$ en effectuant un
d\'eveloppement de Taylor, brisant explicitement la sym\'etrie dans le
superespace.

Les relations (\ref{eq:tfsusy1}) nous indiquent que l'on peut d\'efinir
l' op\'erateur $Q=\p + {(1-q) \over F}^{F-1}$ $ \theta^{F-1} \partial_t$.
En utilisant les r\'esultats \'etablis dans le chapitre $2$, on  sait que 
$(\theta,\p)$ engendre l'alg\`ebre de Clifford du polyn\^ome $x^{F-1}y$ 
(\ref{eq:hq}), ce qui revient \`a dire que $Q^F=\partial_t$. Ce r\'esultat
n'est pas surprenant car $Q$ n'est rien d'autre que l'op\'erateur
agissant sur $\phi$ et  engendrant la transformation FSUSY

\beq
\label{eq:fieldt}
\delta_\epsilon \phi(t,\theta)= \phi(t^\prime,\theta^\prime) -\phi(t,\theta)
\nonumber =\epsilon Q  \phi(t,\theta).
\eeq

Pour construire une action invariante sous de telles transformations
(\ref{eq:tfsusy1}), il est n\'eces\-saire de construire une d\'eriv\'ee
covariante commutant avec $\epsilon Q$. C'est \`a ce stade que la seconde
d\'eriv\'ee $\d$ devient importante. En remarquant que $\p \d = q \d \p$,
on montre que 

\beq
\label{eq:covar}
D=\d + {(1 - q^{-1}) \over F}^{F-1} \theta^{F-1} \partial_t,
\eeq

\noindent
satisfait les relations

\beqa
\label{eq:QD}
&&QD=q DQ \\
&&D^F=\partial_t, \nonumber
\eeqa

\noindent
et donc $D$ commute avec la transformation FSUSY $\epsilon Q$. 
Si $F=2$, il n'y a qu'une d\'eriv\'ee ($\p=\d$) et $D=\d - \theta \partial_t$ 
($D^2=-\partial_t$). En outre, en utilisant les relations de $q-$mutation
(\ref{eq:qfmut}), on observe, pour deux superchamps $A$ et $B$,
que $\epsilon Q(AB)= \epsilon Q(A) B +
A \epsilon Q(B)$ (le produit de deux superchamps est un superchamp),
$\epsilon Q(D A)= D(\epsilon Q A)$ (la d\'eriv\'ee covariante d'un superchamp
est un superchamp) \cite{dur2}.
 
De plus, d'apr\`es les lois de transformation (\ref{eq:tfsusy1}), on
observe que la composante de  $\phi$ suivant $\theta^{F-1}$ se
transforme comme une d\'eriv\'ee totale. Ce faisant, en utilisant les
propri\'et\'es d'int\'egration sur les 
variables de Grassman g\'en\'eralis\'ees 
(\ref{eq:inttheta}), on construit l'action invariante 

\vfill \eject
\beqa
\label{eq:action1} 
S&=&{1 \over 2} \left({1-q \over 1-q^{-1}}\right)^{F-1}\int dt d \theta 
\left\{ D  \phi(t,\theta) \dot \phi(t,\theta)  \right\} \nonumber \\
&=&  \int dt \left\{ {1 \over 2}\dot x^2(t) + {1 \over 2}
{F \over (q^{-1}-1)^F}
\sum \limits_{n=1}^{F-1} \left(q^{-n} -1\right)
\psi_{F-n}(t)~ \dot \psi_n(t \right\}.
\eeqa

Un calcul direct nous indique que le lagrangien est r\'eel.
Une telle action a \'et\'e obtenue pour $F=3$ par divers d'auteurs
\cite{fsusy1d,am,fr} et pour $F$ quelconque, par Durand \cite{dur1}.

On peut alors calculer les charges conserv\'ees,
 associ\'ees aux transformations
FSUSY et aux translations \cite{dur2,prs2}

\beqa
\label{eq:charg}
H(t)&=& {1 \over 2} \dot x^2(t) \\
G(t)&=& -\dot x(t) \psi_1(t) + {1 \over 2} { F \over (q-1)^{F-1}}
\sum \limits_{n=1}^{F-1} \{n+1\} \{-n\} \psi_{F-n}(t)  \psi_{n}(t). 
\nonumber
\eeqa

\noindent
Le fait que $H$,  l'hamiltonien du syst\`eme, ne s'exprime
qu'en fonction de $x$ est symptomatique d'un syst\`eme contraint, comme
nous allons le voir. Les lois de transformation (\ref{eq:tfsusy1}) et
(\ref{eq:fieldt}) nous montrent explicitement que la FSUSY est une
g\'en\'eralisation directe de la SUSY et transforme des champs de diff\'erentes
graduations ou poids conformes.\\

\noindent
{\large {\bf 3.1.3.2. FSUSY d'ordre $F$ non-premier}}\\
\addcontentsline{toc}{subsection}{\numberline{}  
 3.1.3.2. FSUSY d'ordre $F$ non-premier}

Il est possible, \`a partir  d'une action invariante par 
transformation FSUSY d'ordre $F$ ($F$ \'etant un nombre premier),
de construire une action invariante par transformation
FSUSY d'ordre  $fF$. Nous avons \'etabli un tel r\'esultat en dimension
$2$ \cite{prs2}, mais il reste vrai en dimension $1$. Nous allons reproduire
les \'etapes de la d\'emonstration, sans toutefois nous pr\'eoccuper
des normalisations pr\'ecises. On peut d'abord noter que les r\'esultats 
\'etablis ci-avant restent vrais pour tout $F$ \cite{dur2}. On va donc
reproduire l'action (\ref{eq:action1}) en d\'ecomposant l'ordre de la
supersym\'etrie fractionnaire en ses produits de facteurs premiers.

Le point de d\'epart consiste \`a introduire un supermultiplet scalaire d'ordre
$fF$ et de le d\'ecomposer en supermultiplets d'ordre $F$

\beq
\label{eq:fmult}
\Phi_0^{(fF)} = \bigoplus \limits_{a=0}^{f-1}
\Phi_{ {a\over fF}}^{(F)},
\eeq

\noindent
o\`u   $\Phi_{ {a\over fF}}^{(F)} \sim \sum \limits_{n=0}^{F-1}
\theta_F^a\psi_{nf+a}$ est un supermultiplet d'ordre $F$ de poids
conforme ${a\over fF}$ ($\psi_{nf+a}$ est de poids conforme ${nf+a \over fF}$).
Les r\`egles de $q-$mutation des diff\'erentes composantes avec $\theta$ sont

\beq
\label{eq:multiplet}
\theta \psi_{nf+a} = q^{-(nf+a)}\psi_{nf+a} \theta
 ,~~ \psi_{nf+a}^{fF}=0, {\mathrm{~quand~}}
a \neq 0, \psi_{nf}^F=0, q=\exp({2i \pi \over fF}),
\eeq
\noindent
montrant explicitement que $\psi_{nf}$ est une variable de Grassmann 
g\'en\'eralis\'ee d'ordre $F$ et non d'ordre $fF$.
En utilisant l'isomorphisme (\ref{eq:inclus})
d\'emontr\'e lors du chapitre pr\'ec\'edent, on construit
 
\beqa
\label{eq:iclusf}
Q_F&=& \left(Q_{fF}\right)^f
=\partial^f_{\theta_{fF}} + {(1-q) \over fF}^{fF-1}\left[
\theta_{fF}^{fF-1} \partial_{\theta_{fF}} + \cdots 
\partial_{\theta_{fF}} \theta_{fF}^{fF-1}\right]\partial_t \\
&\equiv&
\partial_{\theta_{F}} + {(1-q^f) \over F}^{F-1} \theta_{F}^{F-1}\partial_t,
\nonumber
\eeqa

\noindent
avec $\theta_{fF},\theta_{F}$ deux variables de Grassmann d'ordre
$fF$ et $F$. En utilisant les multiplets (\ref{eq:fmult}), on peut
reproduire l'action (\ref{eq:action1}).

\beq
\label{eq:action11}
S \sim \int dt d \theta_F\left\{ D_F \Phi_0^{(F)}(t,\theta_F) 
\dot \Phi_0^{(F)}(t,\theta_F) + \sum_{n=1}^{f-1} 
\Phi_{ {f-a\over fF}}^{(F)}(t,\theta_F) 
\dot \Phi_{ {a\over fF}}^{(F)}(t,\theta_F) 
\right\}. 
\eeq  

On peut montrer  que l'action (\ref{eq:action11}) co\"{\i}ncide avec
(\ref{eq:action1}) \cite{prs2}. Ceci nous permet de conclure que  si nous avons
un syst\`eme FSUSY d'ordre $fF$ ($fF-$SUSY), il sera automatiquement FSUSY
d'ordre $F$ ($F-$SUSY  (ou $f$, $f-$SUSY). Le passage d'une action invariante 
par $fF-$SUSY \`a un syst\`eme pr\'esentant la sym\'etrie $F-$SUSY se fait
en d\'ecomposant le supermultiplet scalaire d'ordre $fF$ en supermultiplets
d'ordre $F$. R\'eciproquement, si on a un supermultiplet scalaire
d'ordre $F$, la consid\'eration de supermultiplets de poids conformes
$a/fF, a=1,\cdots f-1$ permet de construire une action invariante d'ordre
$fF$. 

En conclusion, on peut \'enoncer que si on a une th\'eorie invariante
par $F-$SUSY, elle sera $f-$SUSY invariante si $f$ divise $F$. Le r\'esultat
est que l'on peut mettre en \'evidence des sous-syst\`emes pr\'esentant des 
sous-sym\'etries.\\

\noindent
{\large {\bf 3.1.3.3. Structure de l'alg\`ebre FSUSY}}\\
\addcontentsline{toc}{subsection}{\numberline{}  
3.1.3.3. Structure de l'alg\`ebre FSUSY }

L'existence de
sous-sym\'etries, obtenues par d\'ecomposition de l'ordre
de la FSUSY en ses produits de facteurs premiers, peut \^etre interpr\'et\'ee
par l'isomorphisme (\ref{eq:inclus}). Par ailleurs, celui-ci
nous indique que, outre $H$ et $Q$, si $f$ divise $F$, l'ordre de la 
supersym\'etrie, alors $Q^f$ est le g\'en\'erateur d'une sym\'etrie.
Qu'en est-il  de $Q^a$ lorsque  $a$ ne divise pas $F$? En fait, dans
cette approche, on peut montrer que $Q^a$ {\it n'est pas le g\'en\'erateur
d'une sym\'etrie}. Rappelons\--en bri\`evement les raisons  \cite{prs1}.

Tout superchamp fractionnaire $\phi$ est dans une repr\'esentation 
appropri\'ee de $(t, \theta)$, la superligne fractionnaire. De ce fait, 
l'action des sym\'etries de $(t, \theta)$ sera induite par les
op\'erateurs $\partial_t,\p,\d$. A priori, on peut donc construire des
g\'en\'erateurs de graduation $(0,1, \cdots, F-1)$. Etant donn\'e l'action
(\ref{eq:action1}), $\phi$ et $D \phi$ admettent les m\^emes lois de 
transformation et donc les divers g\'en\'erateurs doivent commuter avec
$D$. Les seules solutions envisageables sont $Q, Q^2, \cdots, Q^{F-1}, H$.
Enfin, comme le produit de deux superchamps doit se transformer sui\-vant
la r\`egle de Leibniz, les op\'erateurs diff\'erentiels apparaissant
sont imp\'erativement du premier ordre. Comme $Q^a$ s'exprime en fonction
de $\p^a$, utilisant l'isomorphisme (\ref{eq:inclus}) uniquement quand
$a$ divise $F$, on peut r\'einterpr\'eter $\p^a$ comme une  d\'eriv\'ee
du premier ordre. Ainsi et ce, seulement si $a$ divise $F$, $Q^a$ engendre une
sym\'etrie. Une analyse bas\'ee sur la r\`egle de Leibniz a conduit \`a 
des r\'esultats similaires, mis \`a part le cas de $Q^a$ quand
$a$ divise $F$ \cite{dur2}. Il a  \'egalement \'et\'e montr\'e explicitement 
(pour $F=3$) que les transformations
associ\'ees \`a $Q^2$ n'engendraient pas de sym\'etrie \cite{mm}. Enfin, 
notons qu'un probl\`eme analogue appara\^{\i}t dans le 
troisi\`eme article de \cite{kern}.

On voit donc que la raison profonde pour laquelle on ne peut pas consid\'erer
les autres puissances de $Q$ comme des g\'en\'erateurs r\'eside dans la 
structure non-lin\'eaire de l'alg\`ebre sous-jacente: nous n'obtenons pas une
alg\`ebre de Lie. Pour rem\'edier \`a cette dissym\'etrie, une solution
envisageable consisterait \`a lin\'eariser l'alg\`ebre et \`a enrichir la
structure par des variables interm\'ediaires $\theta \equiv \theta_1,
\theta^2 \equiv \theta_2, \cdots, \theta^{F-1} \equiv \theta_{F-1}$
comme cela a \'et\'e propos\'e par Saidi {\it et al} dans \cite{fsusy2d}.
La structure ainsi consid\'er\'ee ``ressemblerait'' \`a une supersym\'etrie
\'etendue. Il est bien connu que pour les th\'eories  supersym\'etriques
\'etendues, la notion de superespace na\"{\i}f n'est pas adapt\'ee, avec comme
incidence le fait que les champs auxiliaires ne sont pas connus. Dans le
contexte des th\'eories FSUSY, il ne para\^{\i}t pas trivial que l'on puisse
d\'efinir une action pr\'esentant de telles invariances. Cette voie
demande donc une analyse plus pr\'ecise et, soit la recherche 
d'un superespace adapt\'e, soit la construction d'une action invariante
sans la notion de superespace. A cet \'egard, on peut noter que, 
historiquement, le concept
de superespace, en supersym\'etrie, a \'et\'e introduit apr\`es la construction
d'actions invariantes.

\subsection{Invariance locale}

Lorsque $F=2$, c'est-\`a-dire pour une th\'eorie supersym\'e\-trique, la 
consid\'eration d'une inva\-riance locale conduit \`a une description 
relativiste d'une particule de spin $1/2$ \cite{wl1,wl2,wl3}. 
Nous avons \'etendu ces r\'esultats pour $F=3$ \cite{fr}. 
Suivant la proc\'edure de Noether, nous voulons jauger la sym\'etrie. Pour ce 
faire, au moins deux champs doivent \^etre introduits: l'un coupl\'e
au courant conserv\'e $H$  contr\^olerait ainsi les diff\'eomorphismes,
et l'autre  au supercourant $G$, serait associ\'e aux
transformations de supersym\'etrie fractionnaire locale (supergra\-vit\'e
fractionnaire FSUGRA). Le premier champ de jauge $e$ est appel\'e einbein
et le second $\chi$ fractino, par analogie au gravitino.

En introduisant les moments conjugu\'es

\beqa
\label{eq:mt}
\pi_x(t)&=& {\delta S\over  \delta \dot x(t)}= \dot x(t)   \nonumber \\
\pi_2(x)&=& {\delta S \over  \delta \dot \psi_2(t)}= {q^2 \over 2} \psi_1(t) \\
\pi_1(x)&=& {\delta S \over  \delta \dot \psi_1(t)}= -{q \over 2} \psi_2(t), 
\nonumber
\eeqa

\noindent
l'action devient

\beqa
\label{eq:sugra}
L&=&\pi_x(t) \dot x(t) + \dot \psi_2(t) \pi_1(t) +  \dot \psi_1(t) \pi_2(t) \\
&-&{1 \over 2} e \pi_x(t)^2 + {q^2 \over 4} \chi \left( \pi_x(t) \psi_1(t)
+{1 \over 2} \psi_2^2(t)\right) + {q \over 4}
\left( \psi_1(t) \pi_x(t)+{1 \over 2} \psi_2^2(t)\right)\chi \nonumber,
\eeqa

\noindent
avec $H= \pi_x^2(t)$ et $G=-\pi_x(t) \psi_1(t)-{1 \over 2} \psi_2^2(t)$. 
Comme le fractino $\chi$ v\'erifie les m\^emes relations de commutation 
que $\theta$ avec
les champs, il est n\'ecessaire d'introduire le couplage au fractino, 
suppos\'e r\'eel,  sous la forme $q^2 \chi G + q G^\dag \chi$ pour 
que l'action
soit r\'eelle. Ensuite, si  l'on fait varier $L$ par rapport \`a $\pi_x$, on
obtient

\beq
\label{eq:pix}
\pi_x(t)= {\dot x(t) \over e} + {q^2 \over 2} {\chi \over e} \psi_1(t),
\eeq

\noindent
et le lagrangien devient

\beq
\label{eq:fsugra1}
L= {1 \over 2e} \dot x^2(t) + {q^2 \over 2} \dot \psi_2(t) \psi_1(t) 
- {q \over 2} \dot \psi_1(t) \psi_2(t) + {q^2 \over 2} \chi
\left({\dot x(t) \over e} \psi_1(t) +{1 \over 2} \psi_2^2(t)\right)
-{q^2 \over 4} { \chi^2 \over e} \psi_1^2(t).
\eeq

Une fois  l'action (\ref{eq:fsugra1}) obtenue, nous n'avons pas pu d\'eriver
les transformations de supergravit\'e fractionnaire la laissant invariante.
Ceci est peut-\^etre la cons\'equence de la structure non-lin\'eaire, qui met
en d\'efaut la proc\'edure de Noether. On doit tr\`es probablement introduire
un autre champ de jauge, un second fractino en quelque sorte, couplant
\`a $G_2(t)$ d\'efini par $\delta G(t) \equiv \epsilon G_2(t)$. $G_2$
serait un courant conserv\'e si $Q^2$ \'etait un g\'en\'erateur associ\'e
\`a une sym\'etrie. De ce fait, nous pouvons appeler $G_2$ un 
pseudo-g\'en\'erateur. Deux arguments vont dans le sens de la prise 
en compte de deux 
fractini: un supermultiplet de la FSUGRA d'ordre $3$ doit contenir trois
champs, or dans l'action (\ref{eq:fsugra1}), on n'a consid\'er\'e que deux
champs de graduation $0$ et $1$.  Une autre argumentation passe par le
concept de superespace fractionnaire et courb\'e. 

Le passage d'une action invariante par transformation locale se fait
par l'introduction de d\'eriv\'ees covariantes adapt\'ees. Il faut donc
consid\'erer des d\'eriv\'ees analogues  \`a 
$\partial_t, D \equiv D_\theta$. Cette substitution, tout
comme en supergravit\'e, se fait par l'introduction d'un supereinbein 
fractionnaire. 
On d\'efinit $X^M = (X^\mu \equiv \tau,  X^m \equiv \Theta)$ les composantes
de la superligne courb\'ee et $X^A=(X^\alpha \equiv t, X^a \equiv \theta)$ 
celles 
de la superligne plate tangente.
On introduit le supereinbein fractionnaire  $E_M^A$
et  son inverse $E_A^M$
($E_A^M E_M^B = \delta_A^B,~E_N^A E_A^M = \delta_N^M$)
et  on consid\`ere  la loi de transformation
$X^M \longrightarrow X^M - \xi^M(X)$  lin\'eaire en $X^m$, donc \'equivalente
\`a (\ref{eq:cgt}). On obtient ainsi les lois de transformation, analogues
\`a celles \'etablies en supergravit\'e,

\beqa
\label{eq:tfsugra}
\phi^\prime(X^\prime)&=& \phi(X) \nonumber \\
E^{\prime~M}_A(X^\prime) &=& E^{N}_A(X) 
{\partial X^{\prime~M} \over \partial X^N} \\
E^{\prime~A}_M(X^\prime) &=&  
{\partial X^N \over \partial X^{\prime M}} E^{A}_N(X). \nonumber
\eeqa

\noindent
Tout comme en supergravit\'e, lors du processus de contraction, il faut
faire attention \`a la position des indices: l'indice sup\'erieur est
toujours \`a gauche de l'indice inf\'erieur.

On consid\`ere les d\'eriv\'ees covariantes $D_A= E_A^M \partial_M$
(\`a une dimension, \'etant donn\'e l'absence de connexion affine,
ce sont les d\'eriv\'ees partielles qui interviennent dans la d\'efinition
de $D_A$). $D_\alpha$ remplace $\partial_t$ et $D_a$, $D_\theta$
(les notations suivent de tr\`es pr\`es celles de Brink {\it et al},
lors de l'\'etude de la supergravit\'e sur la ligne d'univers \cite{wl1}).
 D'apr\`es
(\ref{eq:sdet}), on introduit  le ``volume'' invariant
$V_3= \left[E^\alpha_\mu- E^\alpha_m \left(E^a_m\right)^{-1}E^a_\mu \right]
\left(E^a_m\right)^{-2}$  et on obtient une action invariante sous les 
transformations (\ref{eq:tfsugra})

\beq
\label{eq:fsugraact}
S \sim \int V_3 d^2 X  \left\{ D_\alpha \phi(X) D_a \phi(X) \right\}.
\eeq

La construction est enti\`erement analogue \`a celle suivie sur la superligne,
dans le cas supersym\'etrique \cite{wl1} ou dans  
dans la Ref.\cite{wl2} (Henty, Howe et Townsend). L'action
(\ref{eq:fsugraact}) pr\'esente trop de sym\'etries par
rapport aux transformations de FSUGRA et aux diff\'eomor\-phismes, 
tout comme pour $F=2$. Ainsi, un
certain nombre de contraintes doivent \^etre impos\'ees. Par
exemple, on a dans l'espace plat  $D_\theta^3 = \partial_t$, 
mais on n'a aucune raison pour avoir $D_a^3 = D_\alpha$
dans l'espace courb\'e. 
En supersym\'etrie usuelle, cela revient \`a dire que
le tenseur de torsion de l'espace plat ne correspond pas \`a celui de l'espace
courb\'e et donc si on construit l'action avec $D^2_a= E^M_a \partial_M
\left\{E^N_a \partial_N\right\}$ au lieu de $D_\alpha$, il n'y a aucune raison
pour que l'on obtienne le m\^eme r\'esultat.
Pour r\'esoudre cette ambiguit\'e, il faut imposer des contraintes dans le
superespace courb\'e \cite{wl1}, \cite{wl2} (Henty  {\it et al}) 
et \cite{rss} (comme par exemple $D^2_a=D_\alpha$), celles-ci conduisent \`a 
une action invariante par SUGRA. En dimension 
$D \ge 4$, de telles contraintes, outre l'avantage  de r\'esoudre les
d'ambiguit\'es \'enonc\'ees pr\'ec\'edemment,
 permettent d'\'eliminer les champs de spin
plus grands que deux \cite{susy}. 

Une proc\'edure analogue doit pouvoir \^etre \'etablie quand $F=3$.
On doit tr\`es probablement imposer $D_a^3 = D_\alpha$ et rechercher 
un sous-groupe des transformations (\ref{eq:tfsugra}) pr\'eservant les
contraintes.
Quoi qu'il en soit, on voit que si l'on d\'eveloppe le supereinbein $E_M^A$,
on obtient $3$ champs qui vont conduire au einbein et aux deux fractini.
 
Comme l'action  (\ref{eq:fsugraact}) est ind\'ependante de $F$, cette
m\'ethode est transposable  lorsque $F$ est quelconque. Parmi la 
famille $G(t)=G_1(t), \cdots, G_{F-1}(t)$, o\`u $G_{i+1}$ est d\'efini
\`a partir de $G_i  (\delta_\epsilon G_i=\epsilon G_{i+1}$), quand
$i$ divise $F$, $G_i$ sera un courant conserv\'e et quand $i$ ne divise
pas $F$, $G_i$ sera un pseudo-g\'en\'erateur. Corr\'elativement, il faudra
consid\'erer $F-1$ fractini.

\subsection{Quantification de Dirac}

Bien que nous n'ayons pas pu mettre la main sur une action locale compl\`ete,
(\ref{eq:fsugra1}) va nous permettre, dans une certaine limite, d'obtenir
une \'equation relativiste, tout comme la supergravit\'e sur la ligne
d'univers conduit \`a l'\'equation de Dirac \cite{wl1,wl2,wl3}. Le fait
que certains termes soient omis dans l'action nous donne les indications
suivantes: une d\'emarche analogue \`a celle suivie  pour la particule de 
spin $1/2$, conduira \`a une \'equation relativiste associ\'ee au 
syst\`eme consid\'er\'e, pour laquelle  les contraintes induites par le 
second fractino  vont certainement manquer.

Le syst\`eme d\'ecrit par l'action (\ref{eq:fsugra1}) pr\'esente des 
contraintes de seconde et de premi\`ere classe. Les premi\`eres contraintes
sont li\'ees au fait que $\psi_1$ est le moment conjugu\'e de $\psi_2$

\beqa
\label{eq:1c}
\Xi_1 &\equiv& \pi_1 - {q^2 \over 2} \psi_2 = 0 \\
\Xi_2 &\equiv& \pi_2 + {q \over 2} \psi_1 = 0, \nonumber
\eeqa

\noindent
alors que les secondes r\'esultent de l'invariance de jauge

\beqa 
\label{eq:2c}
{\delta S \over \delta e} &=& H = {1 \over 2} \pi_x^2 = 0 \\
{\delta S \over \delta \chi}&=& -G = {q^2 \over 2} \left( \pi_x \psi_1 + 
{1 \over 2} \psi_2^2 \right)=0.
\eeqa

\noindent
On voit  que le einbein et le fractino apparaissent  comme des 
multiplicateurs de Lagrange. La quantification des syst\`emes contraints
a \'et\'e construite par Dirac \cite{dir}.
Pour les contraintes de seconde classe, on doit remplacer le 
crochet de Poisson
par un crochet de Dirac. La premi\`ere \'etape consiste \`a introduire un
crochet de Poisson pour des variables qui $q-$mutent. 
On peut introduire un tel crochet de deux mani\`eres diff\'erentes, soit
\`a partir de la m\'etrique $q-$symplectique 
associ\'ee \`a $\psi_1, \psi_2$ \cite{mal}, soit gr\^ace \`a 
la $R-$matrice (\ref{eq:R}). On montre  alors \cite{fr} que 

\beq
\label{eq:PB}
\left\{A,B\right\}_{PB}={\partial A \over \partial x}
                   {\partial B \over \partial\pi_x} -          
{\partial A \over \partial \pi_x}{\partial B \over \partial  x} +
{\partial A \over \partial \psi_1} {\partial B \over \partial\pi_1} -
q^2 {\partial A \over \partial \pi_1} {\partial B \over \partial\psi_1} +
{\partial A \over \partial \psi_2} {\partial B \over \partial\pi_2} -
q {\partial A \over \partial \pi_2} {\partial B \over \partial\psi_2}.
\eeq

\noindent
On obtient alors l'alg\`ebre des contraintes

\beq
\label{eq:contr}
\left\{\Xi_i,\Xi_j\right\}= C_{ij}= \pmatrix{0&q \cr
                                     -q^2&0}_{ij},
\eeq
\noindent
qui permet de d\'efinir le crochet de Dirac

\beq
\label{eq:DB}
\left\{A,B\right\}_{DB}= \left\{A,B\right\}_{PB} -
\left\{A,\Xi_i\right\}_{PB} C_{ij}^{-1}\left\{\Xi_j,B\right\}_{PB}.
\eeq
\noindent
Si on applique (\ref{eq:DB}) pour les variables $(\psi_1,\psi_2)$,
on obtient $\left\{\psi_1,\psi_2\right\}_{DB}=-q^2$ et donc

\beq
\label{eq:quant}
\psi_1 \psi_2 - q \psi_2 \psi_1 = - q^2.
\eeq

Une telle \'equation  nous indique que la quantification d'une alg\`ebre
de Grassmann g\'en\'era\-li\-s\'ee est 
une alg\`ebre de Heisenberg $q-$d\'eform\'ee.
Ceci constitue une version adapt\'ee des r\'esultats connus pour les
variables de Grassmann. L'\'equation (\ref{eq:quant}) n'est pas 
surprenante \'etant donn\'e
que  $\psi_1$ et $\psi_2$ sont deux variables 
conjugu\'ees ($\psi_1 \sim {\partial \over \partial \psi_2}$). On montre
ensuite que l'on a bien $G^3=H$ \cite{fr}.

Les contraintes de premi\`ere classe sont, elles, impos\'ees sur les
\'etats physiques.  La premi\`ere \'etape consiste \`a supposer que
les variables $x,\psi_1, \psi_2$ sont dans la repr\'esentation vectorielle
du groupe de Poincar\'e. L'espace-temps n'est rien d'autre qu'un espace
cible dans lequel est plong\'ee la ligne d'univers \cite{fr}. On note
$\eta_{\mu \nu}= {\mathrm{diag}}(-1,1,\cdots,1)$, la m\'etrique de Minkowski.
Tous les r\'esultats pr\'ec\'edents sont alors transposables et l'alg\`ebre
impose que

$$G^3=\left(\pi_x^\mu \psi_1^\nu \eta_{\mu \nu} + 
{1 \over 2} \psi_2^\mu \psi_2^\nu \eta_{\mu \nu}\right)^3= 
{1 \over 2} \pi_x^\mu \pi_x^\nu \eta_{\mu \nu}.$$

\noindent 
On voit donc que les variables $(\psi_1^\mu,\psi_2^\mu)$ engendrent l'alg\`ebre
de Clifford du polyn\^ome $x_0^2y^{ }_0 - x_1^2y^{ }_1 - \cdots - 
x_{D-1}^2y^{ }_{D-1}$ (car $\sum_\mu\left(x^\mu \psi_1^\mu  + 
y^\mu\psi_2^\mu \psi_2^\mu \right)^3= x_0^2y^{ }_0  - \cdots -
x_{D-1}^2y^{ }_{D-1}$).
En choisissant une repr\'esentation de cette alg\`ebre, on peut construire
l'espace de Fock $\left|\lambda_{\mathrm{phys}} \right>$ et introduire
la fonction d'onde $\left<x|\lambda_{\mathrm{phys}} \right> = \lambda(x)$.
Les contraintes des premi\`ere classe se r\'esument \`a

\beqa
\label{eq:3rac}
\partial_\mu \partial^\mu  \lambda(x)&=&0 \\
\left(i \psi_1^\mu \partial_\mu + {1 \over 2} \psi_2^\mu \psi_{2~\mu} \right)
\lambda(x)&=&0. \nonumber
\eeqa
\noindent
De part la structure de l'alg\`ebre, on obtient 
$\left(i \psi_1^\mu \partial_\mu + {1 \over 2} \psi_2^\mu \psi_{2~\mu} 
\right)^3=
{1 \over 2}\partial_\mu \partial^\mu $; une telle \'equation constitue donc
une g\'en\'eralisation de l'\'equation de Dirac, et est induite par les
sym\'etries sur la ligne d'univers. Les variables $\psi_1$ et $\psi_2$ 
d\'ecrivent des degr\'es de libert\'e internes, analogues au spin.

Si maintenant, on suit la m\^eme proc\'edure pour $F$ quelconque,
il y a des chances pour que l'on obtienne un op\'erateur dont
la puissance $F-$i\`eme s'identifie \`a l'op\'erateur de Klein-Gordon.

Dans \cite{kern}, une \'equation correspondant \`a la racine cubique
de l'\'equation de Dirac est propos\'ee, mais une telle \'equation diff\`ere
de celles obtenues par la FSUGRA.\\

\subsection{ Perspectives et questions ouvertes}

%\vskip .5truecm
%\underline{ \bf Perspectives et questions ouvertes}
%\vskip .5truecm
La premi\`ere \'equation de (\ref{eq:3rac}) nous indique que nous avons
affaire \`a une particule non massive. Quant \`a la seconde, son
interpr\'etation reste un probl\`eme  ouvert.

On sait d'apr\`es le premier chapitre que l'alg\`ebre engendr\'ee par
les $\psi$ est de dimension infinie. Deux possibilit\'es s'offrent \`a nous

\begin{enumerate}
\item
soit construire une repr\'esentation de dimension finie, comme nous 
l'avons propos\'e dans \cite{fr};
\item
soit construire un espace de Fock de dimension infinie tout en pr\'eservant
la relation $\psi_1^\mu \psi_2^\mu - q \psi_2^\mu \psi_1^\mu= -q^2$.
\end{enumerate}

\noindent
Le comportement des variables pour deux indices diff\'erents 
distingue les deux possibilit\'es.

Il est possible que ce soit le second choix qui conduise \`a des r\'esultats
int\'eressants.  En effet, comme on le verra dans la section $3$, le groupe
de Lorentz en dimension $1+2$ admet des repr\'esentations de dimension
infinie. Mais pour pouvoir identifier ces repr\'esentations \`a celles
construites pr\'ec\'edemment, il faut d'une part \'etendre le formalisme pour
une particule de masse non-nulle (par l'introduction d'un champ
auxiliaire par exemple) et d'autre part, calculer les g\'en\'erateurs 
du groupe
de Lorentz pour pouvoir conclure dans quelle repr\'esentation de
$SO(1,2)$ se trouve $\lambda$. Une direction envisageable consisterait 
\`a utiliser les travaux 
 de Strassler \cite{wl3}. En effet, celui-ci  a montr\'e que l'action 
effective d'un champ de Dirac conduisait \`a une action pr\'esentant
l'invariance de la supergravit\'e \`a une dimension. On pourrait donc refaire
la d\'emonstration de Strassler en prenant comme point de d\'epart
l'action (\ref{eq:fsugra1}) et en essayant de construire une action 
effective en $3D$. Enfin, nous ne sommes pas parvenus \`a
contraindre la dimension de l'espace-temps.

\vfill \eject
\mysection{Supersym\'etrie fractionnaire en dimension $2$ et invariance 
conforme}

En dimension deux, \`a la diff\'erence des autres dimensions, le groupe
des transformations conformes est de dimension infinie \cite{bpz}. Cette
propri\'et\'e remarquable a \'et\'e mise \`a profit lors de l'\'etude
des th\'eories de cordes \cite{gsw}, ou dans la description des syst\`emes
int\'egrables \cite{id}. 

La raison essentielle qui conduit \`a un groupe infini
r\'eside dans la structure math\'ematique des transformations conformes, qui
ne m\'elangent
pas les composantes holomorphes et anti-holomorphes. Les cons\'equences
d'un tel comportement sont multiples. Ainsi, le groupe de sym\'etrie devient
infini (il est engendr\'e par deux copies de l'alg\`ebre de Virasoro)
et on peut d\'efinir des champs de poids conforme arbitraire. Une th\'eorie
conforme est alors sp\'ecifi\'ee par deux nombres, la charge centrale $c$ 
et le poids conforme $h$. Si on impose une repr\'esentation unitaire
et  de plus bas poids (construite \`a partir d'un vecteur primitif), 
on contraint la s\'erie
$(h,c)$ \cite{fqs}. D'autres th\'eories sont envisageables si on enrichit
les sym\'etries bi-dimensionnelles. A cet \'egard, la supersym\'etrie permet,
par l'introduction de fermions, de consid\'erer d'autres syst\`emes \cite{fqs}.
Mais comme nous l'avons dit, les fermions ne sont plus les champs les
plus g\'en\'eraux que l'on puisse introduire \`a deux dimensions; ainsi,
les parafermions\footnote{Ces parafermions n'ont pas de rapport avec les
parafermions introduits pour les parastatistiques.} \cite{fzz}
permettent de d\'efinir
d'autres s\'eries unitaires \cite{fsvir,fstr}
pour lesquelles des courants
de spin fractionnaire sont pr\'esents. Toutes ces extensions ont comme point
commun le fait qu'elles peuvent \^etre d\'efinies \`a partir de la construction
dite GKO \cite{gko}. Nous n'avons pas pu exhiber de construction GKO 
conduisant \`a notre syst\`eme.

En ce qui nous concerne, nous avons suivi une voie diff\'erente pour \'etendre
l'alg\`ebre de Virasoro \`a savoir par la construction d'une th\'eorie
FSUSY en dimension $2$ \cite{prs1,prs2}. Une telle extension, tout
comme celles consid\'er\'ees dans \cite{fsvir}, contient, outre le tenseur
\'energie-impulsion, un courant conserv\'e de poids conforme $1+1/F$ et
des champs primaires de poids conformes $(0,1/F,\cdots,(F-1)/F)$.

\subsection{Superespace fractionnaire et extension de l'alg\`ebre de Virasoro}

Pour construire une action invariante sous les transformations FSUSY, nous
allons proc\'eder de mani\`ere diff\'erente par rapport au 
paragraphe pr\'ec\'edent.
Nous allons directement \'etendre le plan complexe $(z,\bar z)$ et mettre
en avant une sym\'etrie plus large que celle engendr\'ee par l'alg\`ebre
de Virasoro.

La premi\`ere \'etape va donc consister \`a associer aux variables $z$
et $\bar z$ leur partenaire FSUSY. A ce stade, il est crucial de noter
que l'alg\`ebre de Grassmann g\'en\'eralis\'ee (\ref{eq:qtheta}) 
est incompatible avec l'introduction d'une structure complexe \cite{prs1}.
Cela revient \`a dire que les partenaires de $z$ et $\bar z$ {\it
ne pourront pas \^etre complexes conjugu\'es l'un de l'autre}. On va donc
\'etendre $z,\bar z$ de fa\c con h\'et\'erotique, c'est-\`a-dire que nous
associons \`a $(z, \bar z)$ deux variables de Grassmann g\'en\'eralis\'ees
r\'eelles $\tl,\tr$ et ainsi, la construction se fait ind\'ependamment
sur les modes $L$ et $R$. En utilisant (\ref{eq:qtp}), on obtient les
relations qui m\'elangent les modes $L$ et $R$ (on note
$d=\partial_\theta,\delta_\theta$)

\beq
\label{eq:LR}
\begin{array}{ll}
\tl^F=\tr^F=0& d_l^F=d_R^F=0 \\
\tl \tr = q \tr \tl & d_L d_R= q d_r d_L \\
d_L \tr= q^{-1} \tr d_L& d_R \tl= q \tl d_R.
\end{array}
\eeq

Il est important de noter que les relations (\ref{eq:LR}) ne sont stables ni  
sous la conjugaison complexe $\star$ ni sous la permutation $\sigma$
des deux variables $L$ et $R$. Cependant, elles sont 
stables sous la composition $\sigma \circ \star$ des deux. Sous une
telle transformation, $(z,\tl,\partial_{\tl}, \delta_{\tl})$ a pour
image $(\bar z,\tr,\delta_{\tr}, \partial_{\tr})$ et
il est important de noter que $\star \circ \sigma$ \'echange les 
r\^oles de $\partial_{\tl}$ et  $\delta_{\tr}$. Un tel automorphisme connecte
donc les modes $L$ et $R$. De ce fait, on peut d\'efinir la
d\'eriv\'ee covariante et le g\'en\'erateur des transformations FSUSY des modes
$R$ \`a partir de ceux des modes $L$; et on obtient alors \cite{prs1}

\beqa
\label{eq:fsusy2}
Q_L&=& \partial_L + {(1-q) \over F}^{F-1} \theta_L^{F-1} \partial_z 
\nonumber \\
D_L&=& \delta_L + {(1-q^{-1})\over F}^{F-1}  \theta_L^{F-1} \partial_z 
\nonumber \\
Q_R&=& \left(Q_L\right)^{\star \circ \sigma}=
 \delta_L + {(1-q^{-1})\over F}^{F-1}  \theta_R^{F-1} \partial_{\bar z} 
\\ 
D_R&=& \left(D_L\right)^{\star \circ \sigma}= 
\partial_L + {(1-q)\over F}^{F-1}  \theta_R^{F-1} \partial_{\bar z},
\nonumber
\eeqa

\noindent
satisfaisant l'alg\`ebre (${\cal Q}=Q,D$)

\beqa
\label{eq:fsusyb} 
Q_L^F=D_L^F=\partial_z~~~~~&&Q_R^F=D_R^F=\partial_{\bar z}\nonumber \\
Q_L D_L=q^{-1}D_L Q_L &&Q_R D_R=q D_R Q_R \\
{\cal Q}_L {\cal Q}_R = q {\cal Q}_R {\cal Q}_L. ~~&& \nonumber
\eeqa

On peut  remarquer que les g\'en\'erateurs
(\ref{eq:fsusy2}) peuvent \^etre \'etendus (on ne consid\`ere que les modes
gauches et on note $\tl$, $\theta$)

\beqa
\label{eq:fv1}
L_n&=&z^{1-n}\partial_z -{1 \over F}(n-1) z^{-n} {\cal N},
n \in \ZZ  \\
G_r&=&z^{{1 \over F} -r}(\partial_{\theta} +{(1-q) \over F}^{F-1}
 \theta^{F-1} \partial_z) - {(1-q) \over F}^{F-1}(r-{1 \over F})z^{{1 \over F}
-r-1}\theta^{F-1} {\cal N}, r \in \ZZ + {1 \over F} \nonumber.
\eeqa

\noindent
On peut montrer que de tels g\'en\'erateurs
engendrent une extension de l'alg\`ebre de Virasoro sans charge centrale
\cite{dur1}

\beqa
\label{eq:fv2}
\left[L_n,L_m\right]& =& (n-m)L_{m+n} \nonumber \\
\left[L_n,G_r\right] &=& ({n \over F}-r)G_{n+r} \\
\left\{G_{r_1},\cdots,G_{r_F}\right\}&=& L_{r_1 + \cdots + r_F}, \nonumber
\eeqa 

\noindent
avec ${\cal N}$ l'op\'erateur nombre (\ref{eq:nonb}) introduit au chapitre 2, 
et $\left\{ \cdots \right\}$ le produit sym\'etrique 
(\ref{eq:g}) d\'efini au cours du chapitre 1.
La seconde \'equation traduit le fait que les g\'en\'erateurs $G_i$,
compl\'etant l'alg\`ebre de Virasoro, d\'efinissent un champ de poids conforme 
$1+1/F$. A la diff\'erence de $F=2$ o\`u  on peut extraire la sous-alg\`ebre
$OSp(1|2)$ engendr\'ee par $(L_{-1},L_0,L_1)$ et ($G_{-1/2}, G_{1/2}$), 
d\`es que $F > 2$, en notant que $\left[L_1,G_{1/F} \right]=0$,
on construit une repr\'esentation de $SL(2,\RR)$ \`a partir
de $G_{1/F}$ (cette repr\'esentation appartient aux s\'eries
infinies de $SL(2,\RR)$ voir section suivante): 
$\left[L_{-1}, \cdots \left[L_{-1},G_{1/F}\right ]\cdots \right] \sim 
G_{1/F-n}$ ($L_{-1}$ a \'et\'e appliqu\'e $n$ fois). 
On construit de la sorte l'extension de $OSp(1|2)$ engendr\'ee
par $(L_{-1}, L_0, L_1)$ et $(G_{1/F - n}, n=0,1,2, \cdots)$. 

\subsection{Action}
Forts de cette extension alg\'ebrique, nous sommes en mesure de construire une
action invariante. Etant donn\'e que le proc\'ed\'e est identique \`a celui
d\'evelopp\'e sur la ligne d'univers, nous insisterons uniquement sur les
sp\'ecificit\'es de la dimension 2. Un superchamp fractionnaire admet le
d\'eveloppement  

\beq
\label{eq:fsf2}
\phi(z,\tl,\bar z,\tr) \sim \sum \limits_{a,b=0}^{F-1}
\tl^a \tr^b \psi_{a,b}(z,\bar z).
\eeq

\noindent
En anticipant sur les r\'esultats, on peut d\'ecomposer  $\phi$ de la fa\c con
suivante

$$\vbox{\offinterlineskip \halign{
\tv#  &\cc{#}&\tv# &\cc{#}  &\tv# &\cc{#} &\tv# &\cc{#} &\tv#  \cr
\noalign{\hrule}
&\cc{champs}&&\cc{poids conforme}&&\cc{sur couche}
&&\cc{hors couche}& \cr
&\cc{}&&\cc{}&&\cc{de masse}
&&\cc{de masse}& \cr
\noalign{\hrule}
&$X(z,\bar z)$&&$(0,0)$&&$1 + 1$ &&$1 + 1$& \cr
\noalign{\hrule} 
&$\psi_{a,0}(z,\bar z)$&&$(a/F,0)$&&$1 + 0$ &&$1 + 1$& \cr
\noalign{\hrule} 
&$\psi_{0,b}(z,\bar z)$&&$(0,b/F)$&&$0 + 1$&&$1 + 1$& \cr
\noalign{\hrule} 
&$\psi_{a,b}(z,\bar z)$&&$(a/F,b/F)$&&$0 + 0$&&$1 + 1$& \cr 
\noalign{\hrule} 
}}$$

\noindent
($a+b$ indique que le champ contient $a$ composante(s) holomorphe(s) et
$b$ anti-holomorphe(s)),
montrant explicitement, que ce soit sur ou hors couche de masse, que
l'on a autant de degr\'es de libert\'e de poids conforme $(0,1/F,\cdots,
(F-1)/F)$. Les champs $\psi_{a,0}$ sont holomorphes alors que 
$\psi_{0,b}$ anti-holomorphes. Enfin, $\psi_{a,b}$ d\'efinissent des champs
auxiliaires dont l'\'equation du mouvement est $\psi_{a,b}=0$. 

Les relations de $q-$mutation des composantes de $\phi$ sont identiques \`a
(\ref{eq:qmutchp}) 

\beqa
\label{eq:qmutchp2}
&&\tl \psi_{a,b}= q^{-(a+b)} \psi_{a,b} \tl,~~
\tr \psi_{a,b}= q^{-(a+b)} \psi_{a,b} \tr  \\
&&\psi_{a,0} \psi_{F-a,0} = q^a \psi_{F-a,0} \psi_{a,0},~~
\left(\psi_{a,b}\right)^F=0, \nonumber
\eeqa

\noindent
(une \'equation analogue \`a la troisi\`eme doit \^etre postul\'ee pour
les champs anti-holomorphes). 
Si $F$ n'est pas un nombre premier, aux relations (\ref{eq:qmutchp2}), il
faudra substituer des relations analogues \`a (\ref{eq:multiplet}).

L'action se r\'eduit donc \`a

\beqa
\label{eq:fact2}
S &\sim& \int d z d \bar z d \tl d \tr 
\left[\overrightarrow{D_L} \phi\right] \left[\phi \overleftarrow{D_R}\right] \\
\nonumber
&\sim&\int d z d \bar z \left\{
 \partial_z X(z,\bar z)  \partial_{\bar z} X(z,\bar z) +
\sum \limits_{a=1}^{F-1} 
 \psi_{a,0}(z,\bar z)\partial_{\bar z}\psi_{F-a,0}(z,\bar z) \right. \\ 
&+& \left. 
\sum \limits_{b=1}^{F-1} 
\partial_{ z} \psi_{0,b}(z,\bar z)\psi_{0,F-b}(z,\bar z) +
\sum \limits_{a,b=1}^{F-1} \psi_{a,b}(z, \bar z) \psi_{F-a,F-b}(z, \bar z)
\right\}.
\nonumber
\eeqa 

\noindent
Dans le calcul de l'action (\ref{eq:fact2}), $D_L$ agit par la gauche et 
$D_R$ par la droite. Ceci est une cons\'equence de l'invariance $\star \circ
\sigma$ \cite{prs1}. On observe en outre, en tenant compte des diff\'erents
champs, que le poids conforme de l'action est bien nul.
Dans la r\'ef\'erence \cite{prs1} nous avons, pour $F=3$,
calcul\'e pr\'ecis\'ement tous les termes, alors que pour $F$ quelconque,
nous nous sommes int\'eress\'es uniquement \`a la partie holomorphe \cite{prs2}
(en ne conservant que les termes $\psi_{a,0} \equiv \psi_a$ du d\'eveloppement
de $\phi=X(z,\bar z) + \sum \limits_{a=1}^{F-1} q^{a^2/2} \tl^a \psi_a(z)$)

\beqa
\label{eq:fachol}
S &=&  \left({1-q \over 1-q^{-1}}\right)^{F-1}
\int  d z d \bar zd \tl \partial_{\bar z} \phi D_L \phi \\
&=& \int d z d \bar z \left\{
\partial_z X(z,\bar z ) \partial_{\bar z}X(z, \bar z)+ 
{ F \over (1 -q ^{-1})^F} \sum \limits_{a=1}^{F-1}
\left(q^{-a}-1 \right)  \psi_{{a }}(z, \bar z)
\partial_{\bar z}\psi_{{F-a}}(z, \bar z) \right\}\nonumber.
\eeqa

\noindent
Cette action correspond \`a celle propos\'ee par Matheus-Valle et Monteiro 
\cite{fsusy2d} (ils ne consi\-d\'eraient que les modes gauches et se limitaient
\`a $F=3$). Il ressort d'une telle action que les champs $\psi_{a \ne 0}$
satisfont des \'equations du mouvement $\partial_{\bar z} \psi_{a \ne 0} =0$,
identiques \`a celles d'un fermion, cependant leurs propri\'et\'es physiques
sont diff\'erentes comme nous allons le voir.

On peut noter que (\ref{eq:fachol}) co\"{\i}ncide avec l'action 
(\ref{eq:action1}) obtenue en dimension 1 .

\subsection{Invariance}

Une fois l'action construite, en proc\'edant tout comme dans la section 
pr\'ec\'edente, il est facile de montrer que celle-ci est invariante sous les
transformations conformes et FSUSY. Les lois de transformation des champs
holomorphes sont identiques \`a celles obtenues en dimension 1 
$\delta_{\epsilon_L} \phi = \epsilon_L Q_L \phi$
(\ref{eq:tfsusy1}) \cite{prs2}. Les lois de transformation, incluant les 
champs anti-holomorphes et auxiliaires, ont \'et\'e d\'eriv\'ees dans 
\cite{prs1} 
$\delta_{\epsilon} \phi = \epsilon_L Q_L \phi + \phi Q_R \epsilon_R$
(la structure de  l'alg\`ebre ainsi que l'invariance $\star \circ 
\sigma$ permettent 
d'obtenir les relations de $q-$mutation entre les param\`etres
de la transformation et les champs \cite{prs1}).

Si $F$ n'est pas un nombre premier, les r\'esultats de la ligne d'univers 
restent valables \cite{prs2}, et conduisent \`a des 
sous-syst\`emes qui pr\'esentent
des sous-sym\'etries. Par exemple, une action invariante sous la FSUSY  
d'ordre pair sera  automatiquement invariante sous les transformations
supersym\'etriques \cite{prs2}. 

Nous allons interpr\'eter ces r\'esultats dans le langage de l'invariance
conforme et, sauf mention contraire, nous nous int\'eresserons uniquement
\`a la partie holomorphe de l'action.

\subsection{Int\'egrale de chemin et fonctions de Green}

A partir des r\`egles d'int\'egration (\ref{eq:inttheta}), il est possible
de d\'efinir la fonction de partition

\beq
\label{eq:part}
Z[0]= \int {\cal D}X  \left( \prod \limits_{a=1} ^{F-1}{\cal D} \psi_a \right)
 e^{-S[X,\psi]}.
\eeq

\noindent
Etant donn\'e que la d\'erivation des r\'esultats ne d\'epend pas de $F$, 
nous allons nous concentrer sur $F=3$, l'extension \`a $F$ arbitraire \'etant
imm\'ediate. En utilisant (\ref{eq:path}) et en notant
$\Delta_{1,2}(z-w)= \pmatrix{0&-q^2 \partial_{\bar z} \cr
                    q \partial_{\bar z}&0}^2 \delta(z-w) \delta(\bar z-\bar w)$
l'op\'erateur cin\'etique agissant sur $\psi_1,\psi_2$ 
($S_{\psi_1,\psi_2}=-\int d^2z d^2w \left\{ \pmatrix{\psi_2(w)&\psi_1(w)} 
\Delta_{1,2}(z-w) \pmatrix{\psi_2(z) \cr 
\psi_1(z)} \right\}$, $F=3$),
l'int\'egration sur les champs $\psi$  (pour $F=3$) conduit \`a \cite{prs1}

\beq
\label{eq:z2}
Z[0]= \det \pmatrix{0&-q^2 \partial_{\bar z} \cr
                    q \partial_{\bar z}&0}^2.
\eeq

\noindent
Pour $F$ arbitraire, mais premier, en regroupant deux par deux les champs 
($\psi_a$ et $\psi_{F-a}$),  on obtient la puissance $F-$i\`eme du
produit des  op\'erateurs cin\'etiques agissant sur les divers couples. 
Si $F$ n'est pas premier, le r\'esultat est analogue, mais plus compliqu\'e.

On  d\'efinit alors le
propagateur (en toute rigueur, on arrive \`a un tel r\'esultat lors du 
processus de discr\'etisation, et ensuite par passage \`a la limite)
\vfill \eject

\beqa  
\label{eq:prop1}
\left< \pmatrix{ \psi_2(z) \cr
                 \psi_1(z)} \pmatrix{\psi_2(w)&\psi_1(w)} \right>& =&
{\delta \over \delta \Delta_{1,2}(z-w)} Z[0] \nonumber \\
& =& \pmatrix{0&q^2 \cr
              -q&0} { 1 \over z-w}.
\eeqa
             
\noindent
Lors de la d\'erivation de (\ref{eq:prop1}),  une normalisation appropri\'ee
dans la mesure  d'int\'egration permet de s'affranchir des facteurs 
ind\'esirables. 
La fonction de Green \`a deux points peut \'egalement \^etre d\'eduite par
les m\'ethodes standard, en introduisant des sources (Matheus-Valle {\it
et al} \cite{fsusy2d}).

Pour $F$ quelconque, on d\'etermine \'egalement les fonctions de Green
\`a deux points. Les seules non-nulles sont \cite{prs2}

\beqa
\label{eq:green}
<\psi_{{F-a}}(z) \psi_{{a}}(w)>&=& {(q-1)\over F}^F {1\over
q^{-a}-1}
~{1\over z-w}  \nonumber\\
<\psi_{{a}}(z) \psi_{{F-a}}(w)>&=& {(q-1)\over F}^F {1\over
q^{a}-1}
~{1\over z-w} \\
<X(z) X(w)> &=& -\ln(z-w). \nonumber
\eeqa

\subsection{Oscillateurs, ordre normal et th\'eor\`eme de Wick}

A la place d'une description quantique 
par l'int\'egrale de chemin, il est, dans certains
cas, souhaitable d'avoir une description en termes d'oscillateurs. Pour des
raisons de commodit\'e, nous n'\'etabli\-rons les r\'esultats que 
pour $F=3$ o\`u
nous avons deux champs primaires conjugu\'es $\psi_1$ et $\psi_2$. Lorsque
$F$ est arbitraire, on a $F-1$  champs et les r\'esultats  
s'\'etendent moyennant les  substitutions $(\psi_1,\psi_2) \longrightarrow
(\psi_a,\psi_{F-a})$, et $\exp(2i\pi/3) \longrightarrow \exp(2ia \pi/F)$,
$a=1,\cdots E(F/2)$.

Le fait que les champs soient holomorphes nous permet d'effectuer les
d\'eveloppements 

\beqa
\label{eq:modes}
\psi_1(z) &=& \sum \limits_{r \in \ZZ + a/3} \psi_{1,r} z^{-r-1/3} \\
\psi_2(z) &=& \sum \limits_{s \in \ZZ+ 2a/3} \psi_{2,s} z^{-s-2/3}. \nonumber
\eeqa

\noindent
Tout comme en th\'eorie des cordes, nous pouvons d\'efinir une version
adapt\'ee des secteurs de Ramond-Neveu-Schwarz en fonction des conditions
aux limites des champs ($z \to \exp(2i\pi) z$), sp\'ecifi\'ees par
la valeur de $a$:

$$\vbox{\offinterlineskip \halign{
\tv#  &\cc{#}&\tv# &\cc{#}  &\tv# &\cc{#} &\tv# &\cc{#} &\tv#  \cr
\noalign{\hrule}
&\cc{champs}&&\cc{a=0} &&\cc{a=1}&&\cc{a=2}& \cr
\noalign{\hrule}
&$\psi_1$&&$q^2$&&$q$ &&$1$& \cr
\noalign{\hrule}
&$\psi_2$&&$q$&&$q^2$&&$1$& \cr
\noalign{\hrule}
}}$$

On peut construire une  repr\'esentation de l'alg\`ebre en d\'efinissant un
vide $\left|0\right>$ annihil\'e par $\psi_{1,r}, r>0$ et  $\psi_{2,s}, s>0$. 
On remarque que, tout comme pour les secteurs de Neveu-Schwarz, 
les modes correspondant \`a $a=0$ vont engendrer un vide non-trivial
(des modes z\'ero); nous n'exploiterons pas cette possibilit\'e.  

La structure de l'alg\`ebre sous-tendue par $\psi$ est tellement contrainte
que le comportement des diff\'erents oscillateurs, et donc
des champs $\psi$, est impos\'e. On observe que les champs $\psi_1$ et 
$\psi_2$ sont conjugu\'es; de ce fait, en introduisant deux
s\'eries de variables de Grassmann g\'en\'eralis\'ees $a$ et $b$,
 on peut r\'e\'ecrire $\psi_{1,r} \equiv a_r,\psi_{2,s} \equiv b_s$ avec 
$r,s>0$ ainsi que $\psi_{1,-s} \equiv b^{\dag}_s,\psi_{2,-r} 
\equiv a^{\dag}_r$, avec  $r,s>0$ 
%\footnote{Si on applique la transformation
%(\ref{eq:hq-os}), on obtient l'alg\`ebre de $q-$oscillateurs et donc un
%espace de Hilbert d\'efini positif}

\beqa
\label{eq:champs}
\psi_{{1}}(z) &=& \sum \limits_{s >0} \left\{ b_s^{\dag} z^{s - 1/3}+
a_s z^{-s-1/3} \right\} \equiv \psi_{ 1/3 <}(z) +  \psi_{ 1/3 >}(z) \\
\psi_{{2}}(z) &=& \sum \limits_{r >0} \left\{ a_r^{\dag} z^{r - 2/3}+
b_r z^{-r-2/3} \right\} \equiv \psi_{ 2/3 <}(z) +  \psi_{ 2/3 >}(z). \nonumber
\eeqa

Ceci \'etant pos\'e, le comportement des divers oscillateurs les
uns par rapport aux autres n'est pas libre. La structure alg\'ebrique va
induire des relations de commutation ne cor\-respondant pas 
\`a celles mentionn\'ees au chapitre pr\'ec\'edent.

Si on se concentre sur les champs $\psi_1$ et $\psi_2$, la coh\'erence implique
que
\begin{enumerate}
\item les  fonctions de corr\'elation \`a deux points redonnent
(\ref{eq:green})
\beqa
%\label{eq:green2}
\left<0\right| \psi_1(z) \psi_2(w)\left|0\right> &=&
\sum_{r,s} \left<0\right| a_s a^\dag_r\left|0\right> z^{-s-1/3} w^{r-2/3}
\sim { -q \over z-w} \nonumber \\
\left<0\right| \psi_2(z) \psi_1(w)\left|0\right> &=&
\sum_{r,s} \left<0\right| b_s b^\dag_r\left|0\right> z^{-s-2/3} w^{r-1/3}
\sim { q^2 \over z-w}. \nonumber \\
\left<0\right| \psi_1(z) \psi_1(w)\left|0\right> &=&
\sum_{r,s} \left<0\right| a_s b^\dag_r\left|0\right> z^{-s-1/3} w^{r-1/3}
=0 \nonumber \\
\left<0\right| \psi_2(z) \psi_2(w)\left|0\right> &=&
\sum_{r,s} \left<0\right| b_s a^\dag_r\left|0\right> z^{-s-2/3} w^{r-2/3}
=0 \nonumber
\eeqa
\item
Avant quantification, nous avions $\psi_1(z) \psi_2(z)= 
q \psi_2(z) \psi_1(z)$, ce qui nous sugg\`ere de d\'efinir une prescription 
d'ordre normal adapt\'ee conduisant \`a 
$$:\psi_1(z) \psi_2(z): \sim q :\psi_2(z) \psi_1(z):~.$$
\end{enumerate}

\noindent
Nous avons r\'esolu les contraintes $(1)$ et $(2)$ \cite{prs1,prs2}.
Les \'equations (1) conduisent \`a des relations quadratiques entre les 
op\'erateurs de cr\'eation $A^\dag=a^\dag_r, b^\dag_s$
et ceux  d'annihilation $A=a_r, b_s$~: 
$A^\dag A -q^a A A^\dag \sim 1$; mais \`a ce stade, la puissance de $q$
apparaissant dans les relations de $q-$mutation  n'est 
pas
fix\'ee. Si maintenant on impose que que l'alg\`ebre soit ferm\'ee,
ce qui ce traduit dans notre cas par une charge centrale
r\'eelle (voir section (3.2.7)), alors on obtient  

\beqa
\label{eq:qab1}
\begin{array}{ll}
a_r a^{\dag}_s - q a^{\dag}_s a_r = -q\delta_{rs}&
b_r a^\dag_s -q a^\dag_s b_r =0\cr
b_r b^{\dag}_s - q^2 b^{\dag}_s b_r = q^2 \delta_{rs} 
& a_r b^{\dag}_s -q^2 b^{\dag}_s a_r=0.
\end{array}
\eeqa

\noindent
Un calcul direct nous montre que (\ref{eq:qab1}) conduit bien aux
fonctions de Green (\ref{eq:green}). Ensuite, si on d\'efinit une prescription
d'ordre normal adapt\'e (on note $\phi_i= \psi_1, \psi_2$)
\beq
\label{eq:no}
:\phi_{1 }(z) \phi_{2 }(z): \equiv
\lim \limits_{z \to w}\left[\phi_{1 }(z) \phi_{2 }(w)-
<\phi_{1}(z) \phi_{2 }(w)>\right] ,
\eeq
\noindent
on obtient
\beqa
\label{qchamp1}
:\psi_{{1}<}(z)\psi_{{2}>}(z): &=&
q :\psi_{{2}>}(z)\psi_{{1}<}(z):\nonumber\\
:\psi_{{2 }<}(z)\psi_{{1 }>}(z): &=&
q^2 :\psi_{{1}>}(z)\psi_{{2 }<}(z):\\
\label{qchamp2}
:\psi_{{2 }<}(z)\psi_{{2}>}(z): &=&
q :\psi_{{2 }>}(z)\psi_{{2 }<}(z):\nonumber\\
:\psi_{{1 }<}(z)\psi_{{1}>}(z): &=&
q^2 :\psi_{{1 }>}(z)\psi_{{1 }<}(z): \nonumber
\eeqa

\noindent
Les deux premi\`eres relations sont
la traduction quantique de la relation de $q-$mutation entre les
champs $\psi_1(z),\psi_2(z)$ classiques. Il s'av\`ere donc qu'apr\`es 
quantification, les champs ne satisfont plus de relations de monodromie
simples (on ne peut pas exprimer $\psi_1(z) \psi_2(w)$ en fonction de
$\psi_2(w) \psi_1(z)$). En outre, si maintenant on recherche une relation
bilin\'eaire entre deux op\'erateurs de cr\'eation ou d'annihilation
($a_r a_s = q^{\alpha_{rs}} a_s a_r$),
on  peut montrer que l'on d\'ebouche sur une inconsistance dans l'alg\`ebre
des oscillateurs, donc au niveau de l'espace de Hilbert (en calculant
par exemple $a_r a_r^\dag b_s^\dag |0>$ de deux fa\c cons diff\'erentes).
De ce fait, la structure qui \'emerge naturellement apr\`es quantification
de la FSUSY en dimension deux est l'alg\`ebre des quons introduite par
Greenberg et Mohapatra \cite{gm}. 
La cons\'equence de l'alg\`ebre des quons est que nous avons un th\'eor\`eme
de Wick affaibli et nulle relation entre les champs \`a fr\'equence positive
ou n\'egative n'est connue.
Une telle structure avait \'et\'e introduite
pour produire une statistique avec une faible violation du principe de
Pauli. Pour \^etre tout \`a fait pr\'ecis, il faut souligner le fait
que pour Greenberg et Mohapatra nous avons $q \in [0,1[$, alors que
dans notre cas, $q$ est un nombre complexe de module $1$. 
Enfin, pour \^etre
complet dans la structure des oscillateurs, nous devons sp\'ecifier les
relations

$$
(a_r)^3=(a_r^{\dag})^3=(b_r)^3=(b_r^{\dag})^3=.0$$

\subsection{Espace de Hilbert et quons}
La notation du paragraphe pr\'ec\'edent est trompeuse et laisse supposer
que les op\'erateurs $a, b$ et $a^\dag, b^\dag$ sont hermitiques conjugu\'es
les uns des autres alors que ce n'est pas le cas. On pourrait alors vouloir
rechercher une transformation analogue \`a celle \'etablie dans le chapitre
2, et permettant de connecter l'alg\`ebre de Heisenberg $q-$d\'eform\'ee
au $q-$oscillateurs, mais une telle  transformation 
(voir la section (2.2.1)) ne permet pas de construire des op\'erateurs 
conjugu\'es les uns des autres. 
Une transformation un peu plus subtile est certainement n\'ecessaire. 
En outre, vu la structure de l'alg\`ebre des quons, l'op\'erateur nombre
admet une expression complexe \cite{nomb}. Ainsi, \`a la diff\'erence
des $q-$oscillateurs, nous ne sommes pas encore en mesure de prouver que
la repr\'esentation de l'alg\`ebre de la FSUSY en $2D$ est unitaire.
Quoiqu'il en soit, on peut d'ores et d\'ej\`a 
\'etablir quelques propri\'et\'es de l'espace de Hilbert. On
se limite, par souci de simplicit\'e, au cas o\`u nous n'avons qu'une
s\'erie d'op\'erateurs $(a,a^\dag)$ et on
note $\left(a_{r_1}^{\dag}\right)^{k_1} \cdots \left(a_{r_i}^{\dag}\right)^
{k_i} |0> \sim |(k_1)_{r_1};\cdots,(k_i)_{r_i}>$.

\begin{enumerate}
\item 
Il n'y a pas de relation bilin\'eraire entre deux op\'erateurs de cr\'eation.
De ce fait, il faut distinguer les deux \'etats

$$a_r^{\dag} a_s^{\dag} |0> \neq a_s^{\dag} a_r^{\dag} |0> \mathrm{~ si~} 
r \neq s.$$

\noindent
Il s'ensuit que  les \'etats vont se d\'ecomposer en repr\'esentations 
irr\'eductibles
du groupe des permutations (voire du groupe des tresses) tout comme
les parafermions de Green \cite{g,ko}.
\item
On ne peut pas appliquer un op\'erateur de cr\'eation plus de $F$ fois.
Ainsi, si on consid\`ere 
$$ |h>= A_0 \left(a_r^{\dag}\right)^{k_1} A_1 \cdots A_{i-1}
\left(a_r^{\dag}\right)^{k_i} A_{i} |0>,$$
\noindent
avec $A_0, \cdots A_i$ des produits arbitraires d'op\'erateurs de cr\'eation
distincts de $a_r^\dag$, on  a $a_r^\dag |h>=0$ si $k_1  + \dots + k_i = F-1$.
En d'autre termes, on peut d\'ecomposer l'espace de Hilbert 

$$ H=H_0^{(r)} \oplus \cdots \oplus H_{F-1}^{(r)},$$
\noindent 
o\`u $H_i^{(r)}$ est le sous-espace o\`u $a_r^\dag$ a \'et\'e appliqu\'e
$i$ fois. Cette propri\'et\'e est la cons\'equence directe de la relation
$a_r^F=0$.
\item
On peut se convaincre facilement  que $A_0 \cdots A_i |0>$ est annihil\'e par
$a_r$.
\end{enumerate}

\subsection{Alg\`ebre des courants}
Ayant d\'efini pr\'ecis\'ement les diff\'erentes fonctions de Green et un
th\'eor\`eme de Wick adapt\'e aux types de champs rencontr\'es en FSUSY,
nous sommes en mesure de d\'eterminer l'alg\`ebre des courants. 
Dans le calcul des diverses OPE, il faut tenir compte de la prescription
d'ordre normal, ainsi pour calculer $:\psi_{{1}}^2(z):\psi_{{2}}(w)$,
on revient \`a la d\'efinition du th\'eor\`eme de Wick,
$:\psi_{{1 }}^2(z):=\lim\limits_{\epsilon\to 0}\psi_{{1 }}(z+\epsilon)
\psi_{{1}}(z)$, puis on effectue toutes les contractions possibles, et
enfin on prend la limite $\epsilon\to0$. 

En se r\'ef\'erant
aux r\'esultats que nous avons \'etablis \cite{prs1,prs2}, nous obtenons
le tenseur \'energie-impulsion $T$ et le supercourant $G$

\beqa
\label{eq:TG}
T(z) &=& -{1\over 2}:\partial_z X(z)\partial_z X(z):+{F\over (q-1)^F}
\sum\limits_{a=1}^{F-1} \left[{F-a\over 2F}\left(
(q^{-a}-1):\partial_z\psi_{{a\over
F}}(z)\psi_{{F-a\over F}}(z)\right.\right.\nonumber\\
&&\left.\left.+(1-q^{a}):\psi_{{F-a\over F}}(z)\partial_z\psi_{{a\over
F}}(z):\right)\right], \nonumber \\
G(z) &=&q^{1/2}\left[:\partial_zX(z)\psi_{{1\over F}}(z):+
{F\over(q-1)^{F-1} }\sum\limits_{a=1}^{F-2}\{a+1\}\{-a\}{1\over
1+\eta(a)}:\psi_{{F-a\over
F}}(z)\psi_{{1+a\over F}}(z):\right], 
 \nonumber \\
\eeqa
\noindent
o\`u $\eta(a)=q^{{F-1\over 2}}$ si $a={F-1\over 2}$ (si $F$
est un nombre impair) et $\eta(a)=1$ sinon
(une telle normalisation provient du calcul des contractions
avec  $:(\psi_{{F-1 \over 2}})^2:$).
En utilisant les fonctions de Green \`a deux points (\ref{eq:green}), on  
obtient

\beqa
\label{eq:trans}
T(z) X(w) &=& {\partial_w X(w) \over z-w}+\cdots\nonumber\\
T(z) \psi_{{a}}(w)&=& { {a\over F}\psi_{{a}}(w) \over (z-w)^2}
+{\partial_w\psi_{{a}}(w)\over z-w}+\cdots\nonumber\\
G(z) X(w) &=& q^{1/2}{\psi_{{1}}(w)\over z-w}+ \cdots\\
G(z)\psi_{{F-1}}(w)&=&(-1)^Fq^{1/2}{(1-q)\over F}^{F-1}{\partial_wX(w)\over
z-w}+\cdots\nonumber\\
G(z)\psi_{{a}}(w)&=&q^{1/2}\{a+1\}{\psi_{{a+1}}\over
z-w}+\cdots~.\nonumber
\eeqa

\noindent
Ces transformations montrent bien que les champs $X,\psi_a$ sont de poids
conformes $0,a/F$ et que $G$ est le courant conserv\'e associ\'e aux
transformations FSUSY  (voir les \'equations (\ref{eq:tfsusy1})).
On \'etablit \'egalement

\beqa
\label{eq:transTG}
T(z) T(w) &=& {1\over 2} {c_F\over (z-w)^4} + {2 T(w)\over
(z-w)^2}+{\partial_w
T(w)\over z-w}+\dots\\
T(z) G(w) &=& {{F+1\over F}G(w)\over (z-w)^2}+ {\partial_wG(w)\over
z-w}+\dots~,\nonumber
\eeqa
\noindent
o\`u la charge centrale vaut

\beq
\label{eq:cf}
c_F= 1 + ({1 \over 2})+2\sum\limits_{a=1}^{E({F-1 \over 2})}
\cos({2\pi  a\over F}) \left\{ \left( {a \over F} \right)^2  +
\left( {F-a \over F} \right)^2 -4 { a(F-a) \over F^2} \right\},
%1 -24\sum\limits_{a=1}^{{F-1 \over 2}} \cos({2\pi  a\over F})
%{a(F-a)\over F^2}.
\eeq
 
\noindent
$(1/2)$ est pr\'esent uniquement si $F$ est pair, il repr\'esente 
la contribution
d'un fermion \`a la charge centrale.
Uniquement $F=2,3,4$ admettent une charge centrale rationnelle et peuvent donc
avoir un lien avec des syst\`emes int\'egrables. On peut noter que
$F=4$ a la m\^eme charge centrale que $F=2$.

L'extension de l'alg\`ebre de Virasoro consid\'er\'ee est une 
g\'en\'eralisation naturelle de l'inva\-riance superconforme et contient un
courant conserv\'e de poids conforme $1+1/F$. Cette nouvelle sym\'etrie 
connecte les champs primaires de poids conforme $(0,1/F,\cdots,(F-1)/F)$.

Pour  des raisons analogues \`a celles d\'evelopp\'ees
pr\'ec\'edemment, on voit
que l'alg\`ebre FSUSY va impliquer des relations de fermeture d'ordre $F$
et non pas quadratiques. De ce fait, lorsque l'on calcule $G(z) G(w)$,
on engendre le pseudo-courant $G_2$, de poids conforme $1+2/F$. Ce processus
it\'eratif va conduire \`a  $G \equiv G_1,G_2,\cdots,G_{F-1}$, et parmi
ces pseudo-courants, de poids conformes $(1+1/F, 1+2/F,\cdots,1+(F-1)/F)$, 
on peut conclure que $G_i$ engendre une sym\'etrie si $i$ divise $F$.
On retrouve bien \'evidemment une structure alg\'ebrique identique \`a ce 
qui se passe \`a une dimension.  Enfin, $G_1(z) G_{F-1}(w)$ donne le tenseur
\'energie-impulsion $T$, et \'eventuellement  des termes exprim\'es \`a partir
des champs primaires. Ceci est la r\'epercussion, au niveau de l'alg\`ebre des
courants, de $Q_L^F=\partial_z$. Il est facile de se rendre compte que
(\ref{eq:transTG}) conduit aux deux premi\`eres relations de (\ref{eq:fv2})
(avec une charge centrale). Par contre, l'analogue de la troisi\`eme
ne semble pas \'evidente \`a obtenir.

On peut comparer la structure construite \`a l'alg\`ebre de Virasoro 
fractionnaire (FSV) consid\'er\'ee dans \cite{fsvir,fstr}, o\`u, au c\^ot\'e 
du tenseur \'energie-impulsion,   un courant conserv\'e de 
poids conforme $1+1/F$ est introduit.  Ces deux alg\`ebres sont diff\'erentes,
et ceci pour plusieurs raisons. 
FSV implique des relations de fermeture
quadratiques mais non-locales, alors que FSUSY
admet des relations de fermeture d'ordre $F$ mais locales. En outre,
les charges centrales diff\`erent dans les deux th\'eories 
($c_K=3K/(K+2)$,
$K=4, C_K=2$ correspondant \`a $F=3$ etc.) \cite{fsvir,fstr}. Enfin, les champs
$\psi$ que nous avons introduits ne sont pas des parafermions.

\subsection{Associativit\'e}
Nous avons mentionn\'e   que l'alg\`ebre associ\'ee \`a la 
FSUSY induisait des relations de fermeture non-quadratiques,
mais pour qu'une th\'eorie conforme soit coh\'erente, il n'est
pas suffisant qu'elle soit ferm\'ee, elle doit \'egalement \^etre
associative. Ce nouveau type de contraintes a eu une incidence non nulle  sur
la structure introduite dans  \cite{fsvir,fstr}. L'associativit\'e
se manifeste lorsque l'on calcule une fonction de corr\'elation, et impose
que le r\'esultat soit ind\'ependant de la mani\`ere dont on groupe les
diff\'erents champs .

Ainsi, par exemple, si on veut d\'eterminer $\left<\phi_1(z_1) \phi_2(z_2)
\phi_3(z_3) \right>$, nous avons  deux fa\c cons de distribuer les
parenth\`eses :

$$\left<\phi_1(z_1) \Big(\phi_2(z_2) \phi_3(z_3)\Big) \right> {\mathrm{~~ou~~}}
  \left<\Big(\phi_1(z_1) \phi_2(z_2) \Big) \phi_3(z_3) \right>,$$
\noindent
ces deux fa\c cons devant conduire au m\^eme r\'esultat. Et ainsi
de suite pour les fonctions \`a quatre, cinq {\it etc.} points.

Une solution \`a ce probl\`eme d\'elicat a \'et\'e
donn\'ee par l'\'etude des fonctions de Green \`a $4$ points
entre champs primaires, et d\'ebouche
sur l'\'equation de bootstrap \cite{bpz}. Une telle \'equation permet de
d\'eterminer les divers coefficients apparaissant dans les OPE entre les
champs primaires. Dans le cas o\`u $F=3$, on a

\beqa
\label{eq:opepf}
\partial X(z)\partial X(w)&=& {-1\over (z-w)^2}+\sum\limits_{n>0}
C_{00}^{0,(n)}
\partial^n X(w) (z-w)^{n-2}\nonumber\\
\partial X(z)\psi_{{a}}(w)&=&\sum\limits_{n\ge 0} C_{0a}^{a,(n)}
\partial^n  \psi_{{a}}(w)(z-w)^{n-1},~~a=1,2\nonumber\\
\psi_{{1}}(z)\psi_{{1}}(w)&=&\sum\limits_{n\ge 0}
C_{11}^{2,(n)}\partial^n\psi_{{2}}(w)(z-w)^n\\
\psi_{{2}}(z)\psi_{{2}}(w)&=&\sum\limits_{n\ge 0}
C_{22}^{1,(n)}\partial^n\psi_{{1}}(w)(z-w)^{n-1}\nonumber\\
\psi_{{1}}(z)\psi_{{2}}(w)&=&{-q\over z-w}+\sum\limits_{n> 0}
C_{12}^{0,(n)}\ \partial^n X(w)(z-w)^{n-1}\nonumber\\
\psi_{{2}}(z)\psi_{{1}}(w)&=&{q^2\over z-w}+\sum\limits_{n> 0}
C_{21}^{0,(n)}\partial^n X(w)(z-w)^{n-1}\nonumber
\eeqa

Si on calcule 
$$<\phi_1(z_1) \phi_2(z_2)\phi_3(z_3)  \phi_4(z_4)>,$$ 
\noindent 
avec $\phi_i= \psi_1, \psi_2, \partial_z X$ on peut (en utilisant les
\'equations (\ref{eq:opepf}))

\begin{itemize}
\item
soit contracter
$\phi_1$ avec $\phi_2$ et $\phi_3$ avec $\phi_4$ et ensuite calculer les
fonctions \`a deux-points;
\item
soit contracter
$\phi_2$ avec $\phi_3$ et $\phi_1$  avec  $\phi_4$ puis calculer les
fonctions \`a deux points.
\end{itemize}

\noindent
Imposer que les deux fa\c cons de proc\'eder conduisent au m\^eme r\'esultat
fixe les coefficients $C$ introduits dans (\ref{eq:opepf}). Nous n'avons pas
encore effectu\'e de tels calculs qui peuvent \^etre tr\`es lourds.

\subsection{Applications}

Le premier type d'applications envisageable \`a partir de
la FSUSY consiste en l'\'etude des syst\`emes int\'egrables 
ou des mod\`eles minimaux (en particulier
pour $F=2,3,4$ o\`u la charge centrale est un nombre rationnel): ainsi,
une th\'eorie FSUSY d'ordre $4$  est obligatoirement 
supersym\'etrique. 

Tout comme $F=2$ contient un champ primaire fermionique et d\'ebouche
sur une description du mod\`ele d'Ising, on peut se poser la question
du syst\`eme (si syst\`eme il y a) sous-tendu par la th\'eorie $F=3$. On
pourrait penser que les champs $\psi_1$ et $\psi_2$ sont corr\'el\'es au
mod\`ele de Potts \`a trois \'etats. Mais tel n'est pas le cas car la
charge centrale de $\psi_1$ et $\psi_2$ est $1/3$ alors que celle de
Potts \`a trois \'etats est $c=4/5$.\\

En marge (ou reli\'e ?) \`a ce probl\`eme, plusieurs extensions 
sont alors envisageables:

\begin{enumerate}
\item
introduire d'autres superchamps que le superchamp scalaire (\ref{eq:fsf2}),
par exemple des champs de poids conforme non-nul. C'est  ce
principe qui permet de construi\-re une th\'eorie $fF-$SUSY \`a partir
d'une $F-$SUSY. Il a d'ailleurs \'et\'e \'etabli que les repr\'esenta\-tions
de FSUSY sont en bijection avec celles (cycliques) du groupe quantique 
$U_q(sl(2))$
(Saidi {\it el al} \cite{fsusy2d});
\item
on peut \'egalement introduire des termes d'interaction en 
d\'efinissant un superpotentiel (Debergh \cite{fsusy2d}, Saidi {\it et al}
\cite{fsusy2d}) ou encore
un couplage entre diff\'erents superchamps de poids conformes distincts
(Collato {\it et al} \cite{fsusy2d});
\item
enfin, si on modifie les relations de commutation  entre les oscillateurs
de la fa\c con suivante \cite{prs1}
\beq
\label{eq:para}
a_r a^{\dag}_s - q a^{\dag}_s a_r = k_n(\Delta)\delta_{rs},
\eeq
on obtient 
\beq
\label{eq:para1}
\left<0\right| \psi_1(z) \psi_2(w)\left|0\right>=
{1\over {z^\Delta}} \sum\limits_{n\geq 0}
k_n(\Delta) ({w\over z})^{n} \sim {1 \over (z-w)^\Delta},
\eeq
et on peut reproduire des fonctions de Green non-locales. Evidemment, tout
ce que nous avons fait n'est plus valable dans ce cas-l\`a, et on peut se
demander si une telle description a (ou n'a pas) de rapport avec les
parafermions et FSV?
\end{enumerate}

On peut \'egalement envisager  que la FSUSY soit une sym\'etrie sur la surface
d'univers et donc appliquer les r\'esultats en
 th\'eorie des cordes. Si maintenant, on  veut construire une
action pr\'esentant une  invariance locale (tout comme en dimension $1$), les
probl\`emes \'enonc\'es dans la premi\`ere partie 
restent entiers et il faudrait
construire proprement une telle action. D'apr\`es la discussion
envisag\'ee dans la section (3.1.4), il semblerait qu'il faille introduire
$F-1$ fractini, contr\^olant l'invariance par FSUSY, et une m\'etrique
(ou un ``zweibein'') contr\^olant les diff\'eomorphismes. Parmi ces champs
de jauge, seuls le zweibein et les fractini $\chi_i$ couplant aux courants
$G_i$, lorsque $i$ divise $F$, pourront avoir une incidence sur la coh\'erence
de la th\'eorie (seulement ces champs se transformeront par FSUSY
d'ordre $i/F$ comme une d\'eriv\'ee totale) \cite{prs2}. 
C'est par l'interm\'ediaire des
fant\^omes que Polyakov a le premier d\'emontr\'e que les th\'eories primitives
des cordes/supercordes imposaient une dimension de l'espace-temps de
$26$, $10$ \cite{pol1}. Nous allons reproduire les \'etapes du raisonnement 
appliqu\'e \`a une th\'eorie FSUSY. Une extension des th\'eories des cordes
a \'egalement \'et\'e obtenue dans le cadre des th\'eories FSV \cite{fstr}.

\noindent
Lors de la quantification par l'int\'egrale de chemin, le fait de fixer une
jauge n\'ecessite l'introduction de fant\^omes. Sans conna\^{\i}tre 
pr\'ecis\'ement la forme de l'action invariante locale, utilisant les
lois de transformation (\ref{eq:sdet}), on peut \'ecrire (pour 
la FSUSY d'ordre $F/i$)

\beqa 
\label{eq:ghost}
S_i &=&\left( \det J^{(i)}_{\mathrm{ghost}}\right)^{-F/i +1} \\
&=& \int {\cal D} \beta^{(i)}_1
{\cal D} \gamma^{(i)}_1 \cdots {\cal D} \beta^{(i)}_{F/i -1}
{\cal D} \gamma^{(i)}_{F/i-1} 
\exp \left[S_{\beta^{(i)}_1,\gamma^{(i)}_1} + \cdots +
S_{\beta^{(i)}_{F/i -1},\gamma^{(i)}_{F/i -1}}\right], \nonumber
\eeqa 

\noindent
o\`u nous avons introduit $F/i -1$ paires de fant\^omes/anti-fant\^omes, 
champs bosoni\-ques exprimant le d\'eterminant apparaissant dans
(\ref{eq:ghost}), chaque paire de fant\^omes contribuant pour une puissance $1$
dans l'expression du d\'eterminant. Comme le param\`etre de 
la transformation $F/i-$SUSY est
de poids conforme $-i/F$, les fant\^omes $\gamma^{(i)}$,   
anti-fant\^omes $\beta^{(i)}$  sont de poids conformes $-i/F$, $1+i/F$.
Un tel couple aura une charge centrale (Polyakov \cite{wlf}) de 
$c_{\beta^{(i)},\gamma^{(i)}}=2(1+6 i/F(i/F+1))$. 
Les fant\^omes introduits de cette fa\c con sont associ\'es \`a une sym\'etrie
car $G_i$ est un g\'en\'erateur. Etant donn\'e que les repr\'esentations
de la FSUSY font appel \`a des champs de tous les types, il est tr\`es 
probable qu'il faille \'egalement introduire des champs de poids conformes
$-i/F$, $1+i/F$ lorsque $i$ ne divise pas $F$, de telle sorte que l'on puisse
compl\'eter les divers super-multiplets. Suivant la terminologie introduite,
on pourrait appeler pseudo-fant\^omes de tels champs. Du fait qu'ils sont
associ\'es \`a une pseudo-sym\'etrie, ils ne contribueraient pas \`a
l'anomalie.

On r\'esume alors
dans le tableau suivant les diff\'erentes contributions \`a l'anomalie
(on rappelle que la contribution du syst\`eme $c,b$, contr\^olant l'anomalie
des diff\'eomorphismes est toujours $-26$) 

$$\vbox{\offinterlineskip \halign{
\tv#  &\cc{#}&\tv# &\cc{#}  &\tv# &\cc{#} &\tv# &\cc{#} &\tv#  \cr
\noalign{\hrule}
&FSUSY&&\cc{mati\`ere} &&\cc{fant\^omes}&&\cc{dimension critique}& \cr
\noalign{\hrule}
&$F=1$&&$1$&&$-26$&&$26$& \cr
\noalign{\hrule}
&$F=2$&&$1 + {1 \over 2}$&&$-26+11$&&$10$& \cr
\noalign{\hrule}
&$F=3$&&$ 1+ {1 \over 3}  $&&$ -26+ 2 \times { 22 \over 3}$&&${17 \over 2}$& 
\cr
\noalign{\hrule}
&$F=4$&&$1 + {1 \over 2} +0$&&$-26 +11 + 3 \times {23 \over 4}$&&n\'egative& \cr
\noalign{\hrule}
&$F >  4$&& irrationnel && && irrationnel& \cr
\noalign{\hrule}
}}$$

\noindent
Les diverses contributions correspondent:

\noindent
dans la colonne mati\`ere:  aux diff\'erents champs (\ref{eq:cf}); 

\noindent
dans la colonne fant\^omes: aux contributions provenant des diverses 
sym\'etries
(par exemple pour $F=4$, on a la supersym\'etrie, la FSUSY d'ordre $4$ et
les diff\'eomorphismes);

\noindent
la dimension critique est obtenue en prenant $D$ familles de multiplets d'ordre
$F$ de telle sorte que l'anomalie totale est nulle. 

\noindent
Cette \'etude g\'en\'erale montre que la FSUSY n'est pas applicable (du
moins dans sa pr\'esente version) pour la description d'une th\'eorie de corde
car,  pour $F=3$, la dimension n'est pas enti\`ere et pour $F=4$, elle est 
n\'egative. Et a fortiori, d\`es que $F >4$, on ne peut pas construire de 
th\'eorie satisfaisante. 

Dans les Refs.\cite{naki}, une approche tr\`es semblable \`a  celle que nous
avons suivie pour construire la FSUSY a \'et\'e propos\'ee. La diff\'erence
essentielle r\'eside dans les propri\'et\'es
des diff\'erents champs qui commutent ou anticommutent.
Alors, d'une part, la charge centrale de la mati\`ere est toujours un nombre
rationnel (il n'y a pas de $q$ conduisant \`a un cosinus). 
D'autre part, dans cette approche, il  appara\^{\i}t des sym\'etries de tous
les types et les restrictions de type  ``$G_i$ est un g\'en\'erateur 
si $i$ divise $F$''
ne sont pas de mise. De cette fa\c con, 
d'autres dimensions critiques apparaissent.\\

\vskip .5 truecm

Pour clore cette partie, mentionnons que
tous ces r\'esultats militent en faveur d'une modification de la FSUSY
en dimension $2$

\begin{itemize} 
\item
soit par la consid\'eration d'autres multiplets;
\item
soit par une modification pertinente de l'alg\`ebre des oscillateurs;
\item
???
\end{itemize}

\noindent
En effet, la FSUSY en $2D$ pr\'esente des avantages et des inconv\'enients
par rapport aux th\'eories  analogues telles que FSV.\\

{\bf Avantage}: nous avons une structure de superespace explicite 
et des variables
g\'en\'eralisant les variables de Grassmann adapt\'ees pour d\'ecrire une 
telle
extension de l'alg\`ebre de Virasoro.\\

{\bf Inconv\'enients}: la charge centrale est en g\'en\'eral irrationnelle
et de ce fait, la connexion avec une th\'eorie de corde n'est pas envisageable.
Nous n'avons \'egalement pas  de relation \'evidentes avec des syst\`emes 
int\'egrables,
et en particulier  avec les parafermions de Fateev et Zamolodchikov.

\vfill \eject

\mysection{Supersym\'etrie fractionnaire en dimension $1+2$ et anyons}
La derni\`ere dimension qui pr\'esente des propri\'et\'es particuli\`eres et
permette de construire une extension des th\'eories supersym\'etriques se
trouve \^etre $1+2$. Dans de tels espaces, l'existence de repr\'esentations
qui ne sont ni bosoniques (ou parabosoniques) ni fermioniques
(ou parafermioniques) permet de d\'efinir des particules de spin {\it
arbitraire}. Nous allons voir comment, \`a partir de ces repr\'esentations
{\it anyoniques}, il est possible de construire  une sym\'etrie
allant au-del\`a de la supersym\'etrie et constituant une extension naturelle
des th\'eories FSUSY d\'ej\`a d\'efinies en dimension $1$ et $2$.
Nous nous attacherons ensuite \`a \'etudier les repr\'esentations de cette
sym\'etrie.

\subsection{Repr\'esentations de l'alg\`ebre de Poincar\'e}
\textsc{Il} existe plusieurs fa\c cons d'introduire les statistiques 
fractionnaires en
dimension $1+2$. Par exemple, on peut se demander si le fait de permuter
deux particules dans le sens r\'etrograde ou trigonom\'etrique conduit
\`a un m\^eme d\'ephasage ($\exp(i\pi \alpha )=\exp(-i \pi \alpha )$, voir
figure).
\par
%\begin{floatingfigure}{2cm}
\begin{figure}[!h]
\epsfysize = 5.truecm
%\mbox{\epsffile{b.eps}}
%\mbox{
$$\epsfig{file=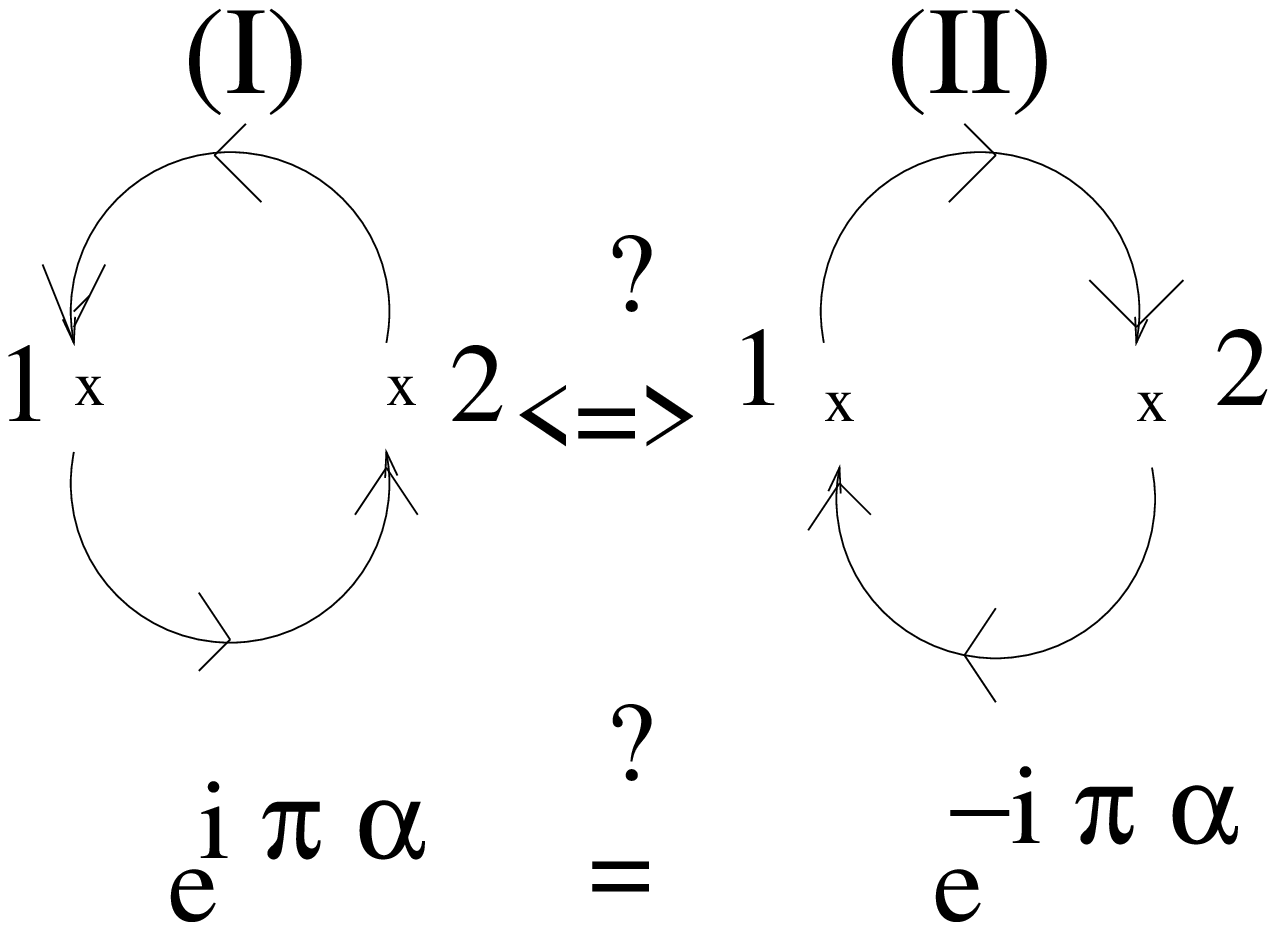}$$
\end{figure}
%}
%  \end{floatingfigure}
\par
%\epsfxsize =5cm
%\epsfysize = 5cm
%$$
%\hskip10truecm
%\epsffile{braided.eps}
%$$
\quad D\`es que la dimension de l'espace-temps est sup\'erieure \`a $4$,
l'existence de dimensions du genre espace perpendiculaires au plan de la
figure impose que les deux mani\`eres d'\'echanger les particules $1$ et $2$ 
soient identiques (on peut faire une rotation, dans un plan contenant
l' axe reliant $1$ \`a $2$ et perpendiculaire au plan de la figure, 
\'echangeant $I$ et $II$) et donc $\alpha=0,1$. Par contre,  si $D=3$, tel 
n'est plus le cas et $\alpha$ est arbitraire. Il s'ensuit,
lorsque la dimension de l'espace-temps est \'egale (sup\'erieure ou \'egale)
\`a $3$, que les particules sont dans une repr\'esentation du groupe
des tresses (permutations). Ainsi, en dimension $1+2$, des statistiques ni
bosoniques ($\alpha =0$) ni fermioniques ($\alpha =1$) sont permises \cite{lm}.

On peut \'egalement introduire les \'etats de spin fractionnaire par la
th\'eorie des repr\'esen\-tations. Etant donn\'e que le premier groupe
d'homotopie est $\pi_1(SO(1,2))=\pi_1(SO(2))=\ZZ$, il existe un groupe
de recouvrement universel permettant d'autres repr\'esentations que celles
bosoniques et fermioniques. 

On d\'efinit les g\'en\'erateurs du groupe
de Poincar\'e $J^\alpha={1 \over 2} \eta^{\alpha \beta}
\epsilon_{\beta \mu \nu }J^{\mu \nu}$ et $P^\alpha$
($\eta_{\alpha \beta} = {\mathrm {Diag}}(-1,$ $1,1)$  
est la m\'etrique de Minkowski et $\epsilon_{\beta \mu \nu}$ le
tenseur de Levi-Civita). L'alg\`ebre de Poincar\'e s'\'ecrit alors

\beqa             
\label{eq:P}      
\left[ P^\alpha ,P^\beta \right]  &=& 0 \nonumber \\
\left[ J^\alpha ,P^\beta \right] &=& i \eta^{\alpha \gamma} \eta^{\beta \delta}
\epsilon_{\gamma \delta \eta} P^\eta \\
\left[ J^\alpha,J^\beta \right] &=&  i \eta^{\alpha \gamma} \eta^{\beta \delta}
\epsilon_{\gamma \delta \eta} J^\eta.    \nonumber          
\eeqa

Les repr\'esentations pour les particules de masse $m \ne 0$ 
sont obtenues \`a partir de l'\'etude du petit groupe laissant invariant
l'impulsion $p^\alpha=(m,0,0)$, calcul\'ee dans un r\'ef\'erentiel o\`u 
la particule
est au repos. Etant donn\'e que ce groupe de stabilit\'e est ab\'elien
($SO(2)$), les repr\'esentations sont de dimension $1$ et non-quantifi\'ees
($J^0 \longrightarrow J^0 + s$ laisse invariantes les relations de structure
de $SO(2)$). Les repr\'esentations sont donc param\'etris\'ees par un 
param\`etre continu $s \in \RR$. 
Une telle substitution montre explicitement que lorsque l'on fait
une rotation de $2 \pi$, on engendre une phase $e^{2 i s \pi }$. 
Il  est remarquable de constater alors, que
pour le groupe $SO(1,2)$, la 
transformation simultan\'ee $J^i \longrightarrow J^i + s { P^i \over P^0 +m}$
pr\'eserve les relations de commutation de $SO(1,2)$. Puisque
pour une transformation propre et orthochrone, le d\'enominateur ne s'annule
jamais, $s$ n'est pas quantifi\'e (ce n'est bien s\^ur pas vrai pour $SO(3)$).
En outre, \`a partir de la m\'ethode des repr\'esentations induites, on peut
montrer que, pour une repr\'esentation $s$, on a 
(les repr\'esentations du groupe
de Poincar\'e ont \'et\'e \'etudi\'ees dans \cite{bin})

\beqa
\label{eq:induit}
J^0_s &=& i \left(p^1 {\partial \over \partial p_2}
 - p^2 {\partial \over \partial p_1}\right)
+ s \nonumber \\
J^1_s &=& - i \left(p^2 {\partial \over \partial p_0}-
p^0 {\partial \over \partial p_1}\right)
+ s { p^1 \over p^0+m} \\
J^2_s &=& - i \left(p^0 {\partial \over \partial p_1}-
p^1 {\partial \over \partial p_0}\right)
+ s { p^2 \over p^0+m}, \nonumber
\eeqa

\noindent
avec $p$ la valeur propre de $P$. Une telle modification des g\'en\'erateurs 
de Lorentz a \'et\'e  propos\'ee dans \cite{sch}.
L'id\'ee de la d\'emonstration est la 
suivante: partant du r\'ef\'erentiel o\`u $p^\alpha=(m,0,0)$, par une 
transformation de Lorentz, on passe dans un rep\`ere quelconque, pour lequel
$p^\alpha=(p^0,p^1,p^2)$ (avec $p^\alpha p_\alpha = m^2$) et on regarde ce
qu'induit un tel changement de base au niveau des g\'en\'erateurs de Lorentz
\cite{jn}.

Une fois  les repr\'esentations  (\ref{eq:induit}) mises en \'evidence,
il a \'et\'e observ\'e \cite{bin,jn,plu}
qu'\`a partir des deux op\'erateurs
de Casimir ($P.P$ et $P.J)$, on pouvait construire une \'equation du 
mouvement pour une  particule de masse $m$ ($P.P=m^2$) et de spin 
$s$ ($P.J=ms$)

%\vfill \eject
\beqa
\label{eq:ms}                                                              
(P^2 - m^2) \Psi &=& 0 \\                                                 
(P.J - sm) \Psi  &=& 0.  \nonumber                                            
\eeqa  

\noindent
Diverses approches  ont \'egalement conduit  \`a des
\'equations analogues \cite{plu}. 
Il est important de noter qu'une telle  description diff\`ere de celle o\`u, 
par  l'interm\'ediaire d'un terme de Chern-Simons,
on transmute un champ scalaire en un champ de spin fractionnaire
\cite{trans}. 

Les \'equations (\ref{eq:ms}) ne sont vraies que sur la couche de masse, et
pour obtenir des \'equations covariantes, il faut aller au-del\`a de la
condition de masse et \'etudier les repr\'esenta\-tions du groupe de Lorentz
en dimension $3$. Ainsi, en partant de ces repr\'esentations, en imposant
une \'equation du mouvement appropri\'ee, on peut retrouver (\ref{eq:ms}) avec
le bon nombre de degr\'e(s) de libert\'e. C'est d'ailleurs ainsi que l'on
passe, \`a quatre dimensions, d'un champ de Dirac \`a $4$ composantes 
complexes, aux degr\'es de libert\'e de l'\'electron par exemple.

Jackiw et Nair  \cite{jn} et Binegar \cite{bin}  ont montr\'e, 
par un calcul explicite, que les repr\'esen\-ta\-tions
non-unitaires de dimension $2s+1$ conduisaient \`a (\ref{eq:ms}) (du moins
pour les repr\'esen\-tations vectorielle et spinorielle). Ces r\'esultats ont
ensuite \'et\'e \'etendus pour $s$ arbitraire \cite{jn,plu} et admettent
comme point de d\'epart les repr\'esentations de dimension infinie de
$SO(1,2)$. Celles-ci ont \'et\'e mises en \'evidence par Wybourne \cite{wyb}
et consistent en les s\'eries discr\`etes 
born\'ees de plus haut ou de plus bas 
poids. C'est \`a partir de ces repr\'esentations que l'on peut obtenir
une \'equation relativiste pour un anyon \cite{jn}. Il existe \'egalement
les s\'eries dites continues \cite{wyb} qui ne sont born\'ees ni par le haut
ni par le bas; une \'equation relativiste a \'egalement \'et\'e propos\'ee
pour de telles repr\'esentations (voir le troisi\`eme article de \cite{plu}), 
mais nous n'en parlerons pas et nous nous  
limiterons aux s\'eries discr\`etes. 
En notant $J_{\pm}=J_1 \mp i J^2$, on obtient 

\beq
\label{eq:r+}
\begin{array}{ll}
&J_s^0 |s_+,n \rangle = (s+n) |s_+,n \rangle \cr
{\cal D}_s^+:&
J_{s,+} |s_+,n \rangle = \sqrt{(2s+n)(n+1)} |s_+,n+1 \rangle \cr
&J_{s,-} |s_+,n \rangle = \sqrt{(2s+n-1)n} |s_+,n-1 \rangle,
\end{array}
\eeq

\noindent 
pour les s\'eries de plus bas poids, et pour celles de plus haut poids,

\beq
\label{eq:r-}
\begin{array}{ll}
&J_s^0 |s_-,n \rangle = -(s+n) |s_-,n \rangle  \cr
{\cal D}_s^-:&J_{s,+} |s_-,n \rangle = - \sqrt{(2s+n-1)n}|s_-,n-1 \rangle \cr 
&J_{s,-} |s_-,n \rangle = - \sqrt{(2s+n)(n+1)} |s_-,n+1 \rangle. 
\end{array}
\eeq

\noindent
Ces deux repr\'esentations admettent $s(s-1)$ comme valeur de l'op\'erateur
de Casimir. Cette approche qui passe par les repr\'esentations du groupe
de recouvrement universel de $SO(1,2)$ diff\'ere de celle o\`u des 
repr\'esentations multi-valu\'ees de $SO(1,2)$ sont consid\'er\'ees 
\cite{fj}.

Pour terminer cette section, mentionnons les r\'esultats essentiels de
Jackiw et Nair \cite{jn}: il est possible de construire une \'equation
relativiste pour un anyon de spin $s$. La solution d'une telle \'equation
s'exprime comme la somme directe d'une particule d'\'energie positive dans
la repr\'esentation ${\cal D}_s^+$ et d'une solution d'\'energie n\'egative
dans la repr\'esentations ${\cal D}_s^-$. Ceci revient \`a dire que  
l'on obtient un \'etat d'\'energie positive et d'h\'elicit\'e $h=s$ et 
un \'etat d'\'energie n\'egative et d'h\'elicit\'e $h=-s$:
$|s \rangle = |h=s, +\rangle \oplus |h=-s,- \rangle$. Ces deux \'etats sont
donc $CP$ conjugu\'es l'un de l'autre et finalement, on obtient une 
repr\'esentation r\'eelle (comme il se doit pour $SO(2)$).

En fait, pour construire des \'equations covariantes
reproduisant (\ref{eq:ms}), on rel\`eve deux m\'ethodes. La premi\`ere
admet comme point de d\'epart  la
repr\'esentation  ${\cal D}_{-1} \oplus {\cal D}_{s+1}$ \cite{jn} et la
seconde ${\cal D}_{-1/2} \oplus {\cal D}_{s+1/2}$  \cite{plu} (troisi\`eme
article); ${\cal D}_{-1},
{\cal D}_{-1/2}$  sont les repr\'esentations vectorielle et  spinorielle. 
Dans les deux cas, (\ref{eq:ms})
ont \'et\'e d\'eriv\'ees \`a partir d'une \'equation lin\'eaire plus un
certain nombre de conditions subsidiaires. Mentionnons que dans les deux 
derni\`eres r\'ef\'erences de \cite{plu}, une \'equation lin\'eaire a \'et\'e
obtenue.

Nous allons, dans la section suivante, consid\'erer le cas $s=-1/F$.
Si on observe les \'equations (\ref{eq:r+}) et (\ref{eq:r-}), il appara\^{\i}t
imm\'ediatement une ambiguit\'e li\'ee \`a $\sqrt{-2/F}$. De ce fait, il y
a {\it a priori} quatre repr\'esentations possibles: deux de plus haut
poids et deux de plus bas poids  li\'ees au choix $\sqrt{-1}=\pm i$.
On note ${\cal D}_{-1/F;\pm}^\pm$ ces quatre repr\'esentations. On 
peut alors \'etablir que celles-ci sont reli\'ees de la mani\`ere suivante

\begin{itemize}
\item la repr\'esentation duale de  ${\cal D}_{-1/F;+}^+$ est d\'efinie
moyennant la substitution $J^a \longrightarrow -\left(J^a\right)^t$:
$\left[{\cal D}^+_{-1/F;+}\right]^*={\cal D}^-_{-1/F;+}$;
\item la repr\'esentation complexe conjugu\'ee de ${\cal D}_{-1/F;+}^+$,
est obtenue en introduisant  $J^a \longrightarrow -\left(J^a\right)^\star$
\footnote{Etant donn\'e que pour les math\'ematiciens il n'y a pas de facteur
$i$ dans la d\'efinition d'une alg\`ebre de Lie, --{\it c.f.} (\ref{eq:P})--
il n'y a pas de signe moins dans la d\'efinition de la repr\'esentation 
complexe conjugu\'ee.}$ ^,$ 
\footnote{On peut noter que pour une matrice complexe $X$,  $X^\star$
repr\'esente la matrice complexe conjugu\'ee, alors que pour un espace 
vectoriel $V$, $V^*$ repr\'esente son espace dual.} (comme,  quelque soit la 
repr\'esentation choisie, par d\'efinition on a $J^\pm = J^1 \mp i J^2$, 
$\left(J^\pm\right)^\star = -\Big( \left(J^1\right)^\star 
\mp i\left(J^2\right)^\star
\Big)$. On obtient alors  $\overline{{\cal D}^+_{-1/F;+}}={\cal D}^-_{-1/F;-}$;
\item finalement 
$\left[\overline{{\cal D}^+_{-1/F;+}}\right]^*={\cal D}^+_{-1/F;-}$.
\end{itemize}

Si maintenant on  consid\`ere 
$\psi_a \in {\cal D}^+_{-1/F;+} ,\psi^a \in {\cal DC}^-_{-1/F;+},
\bar{\psi}_{\dot a} \in {\cal D}^-_{-1/F;-}$ et  
$\bar{\psi^{\dot a}} \in {\cal D}^+_{-1/F;-}$ (les notations suivent de tr\`es
pr\`es celles de $SL(2,\CC)$), nous avons les lois de transformation
suivantes

%\vfill \eject
\beqa
\psi^\prime_a &=& S_a^{~~b} \psi_b \nonumber \\
\psi^{\prime a} &=& \left(S^{-1}\right)_{b}^{~~a} \psi^b  \\
\bar{\psi}^\prime_{\dot a} &=& \left(S^\star\right)_{\dot a}^{~~ \dot b} 
\bar {\psi}_{\dot b} 
\nonumber \\
\bar{\psi}^{\prime {\dot a}}&=& \left((S^\star)^{-1} \right)_{\dot b}^
{~~ \dot a}  \bar{\psi}^{\dot b}. \nonumber
\eeqa

On peut alors montrer que les repr\'esentations de plus haut/bas poids
sont isomorphes (\'etant donn\'e que nous consid\'erons des espaces de
dimension infinie, on a plus exactement $\overline{{\cal D}^+_{-1/F;+}}
\subset \left[{\cal D}^+_{-1/F;+}\right]^*$).
Consid\'erons les matrices $g_{\dot a a}= g^{a \dot a}=
g_{a \dot a}= g^{\dot a a}={\mathrm{diag}} (-1,1,\cdots,1)$ ($g_{\dot a a}=
\left(g^{a \dot a}\right)^{-1}$ et $g_{ a \dot a}=
\left(g^{\dot a  a}\right)^{-1}$ ), on peut alors d\'efinir le produit
scalaire

\beq
\label{eq:ps}
\varphi^a \psi_a = \bar \varphi_{\dot a} \psi_a g^{\dot a a}=
-\bar \varphi_{\dot 0} \psi_0 + \sum \limits_{a>0} 
\bar \varphi_{\dot a} \psi_a,
\eeq 

\noindent
\'etablissant explicitement l'inclusion $\overline{{\cal D}^+_{-1/F;+}}
\subset \left[{{\cal D}^+_{-1/F;+}}\right]^*~:
\left(S^{-1}\right)_b^{~~a}= g^{a \dot a} 
\left( S^\star\right)_{\dot a}^{~~\dot b} g_{\dot b b}$ (pour
$\overline{{\cal D}^+_{-1/F;+}}$ on prend des sommes finies alors que pour
$\left[{{\cal D}^+_{-1/F;+}}\right]^*$ les sommes peuvent \^etre en principe
infinies.

La raison pour laquelle nous avons un produit scalaire pseudo-hermitien est 
li\'ee au fait que nous avons des repr\'esentations non-unitaires d'un
groupe de Lie non-compact.

\subsection{Extension de l'alg\`ebre de Poincar\'e et supersym\'etrie
fractionnaire}

Le point de  d\'epart dans l'\'etude des extensions de l'alg\`ebre de 
Poincar\'e consiste \`a d\'efinir des s\'eries infinies de charges,
dans les repr\'esentations appropri\'ees de $SO(1,2)$. Etant donn\'es les 
r\'esultats des sections consacr\'ees \`a l'\'etude de la FSUSY en une
et deux dimensions, on choisit des charges dans les repr\'esentations
${\cal D}_{-1/F;+}^+$ et  ${\cal D}_{-1/F;-}^-$ not\'ees respectivement
$\left\{Q_{-1/F+n}^+\right\}$ et $\left\{Q_{-1/F+n}^-\right\}$. On a donc
$\left(Q_{-1/F+n}^+\right)^\dag=\left(Q_{-1/F+n}^-\right)$. 
On r\'esoud l'ambiguit\'e li\'ee \`a la racine carr\'ee d'un nombre n\'egatif
en fixant $\sqrt{-1}=i$.
D'apr\`es les \'equations (\ref{eq:r+}) et 
(\ref{eq:r-}), on en d\'eduit  les relations de commutation  \cite{fsusy3d}

\beqa
\label{eq:Q}
\left[J^0,Q^+_{-1/F+n} \right] &=& (-1/F+ n)~~ Q^+_{-1/F+n}  \nonumber \\
\left[J_+,Q^+_{-1/F+n} \right] &=& \sqrt{(-2/F+n)(n+1)}~~ Q^+_{-1/F+n+1} 
\nonumber \\
\left[J_-,Q^+_{-1/F+n} \right] &=& \sqrt{(-2/F+n-1)n}~~ Q^+_{-1/F+n-1} 
\nonumber \\
&& \\
\left[J^0,Q^-_{-1/F+n} \right] &=& - (-1/F+n) ~~Q^-_{-1/F+n}  \nonumber \\
\left[J_+,Q^-_{-1/F+n} \right] &=& - \left(\sqrt{(-2/F+n-1)n}\right)^\star~~ 
Q^-_{-1/F+n-1} 
\nonumber \\
\left[J_-,Q^-_{-1/F+n} \right] &=& - \left(\sqrt{(-2/F+n)(n+1)}\right)^\star~~
Q^-_{-1/F+n+1}, 
\nonumber
\eeqa
\noindent
On obtient donc une infinit\'e de g\'en\'erateurs d'h\'elicit\'e $h=n-1/F$
pour la repr\'esentation ${\cal D}_{-1/F;+}^+$ et $h=-n+1/F$ pour 
${\cal D}_{-1/F;-}^-$; ces deux repr\'esentations ne sont pas \'equivalentes.
Dans le cas particulier o\`u $F=2$, ${\cal D}_{-1/2}^+ \sim {\cal D}_{-1/2}^-$
et la repr\'esentation est de dimension $2$. 

Comme on a introduit des charges dans une repr\'esentation 
anyonique de $SO(1,2)$,
on aimerait pouvoir construire une extension non-triviale de l'alg\`ebre
de Poincar\'e. Nous notons de fa\c con g\'en\'erique ${\cal A}$,
${\cal B}$ les  g\'en\'erateurs anyoniques et bosoniques. On sait
que les op\'erateurs du type ${\cal A}$ induisent des g\'en\'erateurs du
type ${\cal B}$ en dimension un et deux. Par ailleurs, \'etant donn\'e que
${\cal A}$ prend une phase $\exp (-{2 i \pi \over F})$ lors d'une rotation
de $2 \pi$, ses \'el\'ements admettront une graduation $F-1$ dans $\ZZ_F$,
alors que les op\'erateurs bosoniques une graduation $0$. Tout comme en 
dimension $1$ et $2$, on peut se poser la question de l'existence de charges
de graduation $-a/F$, et on montrera que de telles charges peuvent \^etre
d\'efinies si $a$ divise $F$. Nous nous pencherons sur cette alternative
ult\'erieurement. Forts de ces  r\'esultats, on obtient

\beqa
\label{eq:PQ0}
&&\left\{\cal{A},\cdots, \cal{A} \right\}_F \sim \cal{B} \nonumber \\
&&\left[\cal{B},\cal{A}\right]  \sim \cal{A} \\
&&\left[\cal{B},\cal{B}\right]  \sim \cal{B}, \nonumber
\eeqa
\noindent
la justification du produit compl\`etement sym\'etrique 
$\{ \cdots \}_F$ (voir \'equation (\ref{eq:g})) sera donn\'ee par
la suite.

A ce stade deux \'etapes, sont encore n\'ecessaires:  identifier 
pr\'ecis\'ement
les relations entre $\{ {\cal A} \}$ et $\{ {\cal B} \}$  et ensuite
s'assurer de la coh\'erence de l'alg\`ebre ainsi d\'efinie. 
Ces deux contraintes
seront en fait r\'esolues en une seule \'etape, en d\'efinissant des 
identit\'es de Jacobi appropri\'ees

\beqa
\label{eq:Jac}
&&\left[\left[{\cal B}_1,{\cal B}_2\right],{\cal B}_3\right] +
\left[\left[{\cal B}_2,{\cal B}_3\right],{\cal B}_1\right] +
\left[\left[{\cal B}_3,{\cal B}_1\right],{\cal B}_2\right] =0 \nonumber \\
&&\left[\left[{\cal B}_1,{\cal B}_2\right],{\cal A}_3\right] +
\left[\left[{\cal B}_2,{\cal A}_3\right],{\cal B}_1\right] +
\left[\left[{\cal A}_3,{\cal B}_1\right],{\cal B}_2\right]  =0 \nonumber \\
&&\left[{\cal B},\left\{{\cal A}_1,\dots,{\cal A}_F\right\}_F\right] =
\left\{\left[{\cal B},{\cal A}_1 \right],\dots,{\cal A}_F\right\}_F  + \dots +
\left\{{\cal A}_1,\dots,\left[{\cal B},{\cal A}_F\right] \right\}_F \\
&&\sum\limits_{i=1}^{F+1} \left[ {\cal A}_i,\left\{{\cal A}_1,\dots,
{\cal A}_{i-1},
{\cal A}_{i+1},\dots,{\cal A}_{F+1}\right\}_F \right] =0. \nonumber
\eeqa

\noindent
Les deux premi\`eres identit\'es sont identiques \`a celles obtenues
pour les superalg\`ebres de Lie, la troisi\`eme est d\'eriv\'ee en utilisant
la r\`egle de Leibniz avec ${\cal B}$ et $\{ \cdots \}$ et la derni\`ere est
donn\'ee par un calcul direct. Vu la structure particuli\`ere
de l'alg\`ebre, il n'y a pas d'autre identit\'es.

Pour d\'eterminer le reste de la structure engendr\'ee par $\left\{ {\cal A},
{ \cal B} \right\}$ on postule la forme la plus g\'en\'erale pour la 
premi\`ere \'equation de (\ref{eq:PQ0}) 

$$ \left[\-|\cal{A},\cdots, \cal{A}|\-\right]_F = \alpha. P + \beta. J,$$

\noindent
o\`u $ \left[\-| \cdots | \-\right]_F$ repr\'esente une combinaison de produits
sym\'etriques  $\{ \cdots \}_F$ d'ordre $F$   \`a d\'etermi\-ner.

A partir de la premi\`ere identit\'e de Jacobi avec ${\cal B}=P$,
en  utilisant $\left[P,Q\right]=0$,
on montre que $\beta =0$, alors que la m\^eme identit\'e avec  ${\cal B}=J_0$
nous indique que les deux membres de l'\'equation ont la m\^eme h\'elicit\'e.
Enfin, l'action des op\'erateurs $J_\pm$ impose que l'on ait \cite{fsusy3d}

\beqa
\label{eq:PQ}
&&\left\{Q^\pm_{-{1\over F}},\dots,Q^\pm_{-{1\over F}} \right\}_F =
\left(Q^\pm_{-{1 \over F}}\right)^F = P_\mp
\nonumber \\
&&\left\{Q^\pm_{-{1\over F}},\dots,Q^\pm_{-{1\over F}},Q^\pm_{1-{1\over F}}
\right\}_F
 = \pm i \sqrt{{2 \over F}} P^0 \nonumber \\
&& -(F-1)
\left\{Q^\pm_{-{1\over F}},\dots,Q^\pm_{-{1\over F}},Q^\pm_{1-{1\over F}},
Q^\pm_{1-{1\over F}} \right\}_F
\pm i \sqrt{F-2} 
\left\{Q^\pm_{-{1\over F}},\dots,Q^\pm_{-{1\over F}},Q^\pm_{2-{1\over F}}
\right\}_F
=   P_\pm \nonumber \\
&&\left[J_\pm,\left[J_\pm,\left[J_\pm, \left(Q^\pm_{-{1 \over F}}\right)^F
\right]\right]\right]=0  \\
&&~~~~~~~~~~~~~~~~ \vdots \nonumber \\ 
&&\left[J_\pm,\left[J_\pm,\left[J_\pm, \cdots \left[ 
\left(Q^\pm_{-{1 \over F}}\right)^F\right]\right]\right] \cdots \right]=0
\nonumber \\
&&~~~~~~~~~~~~~~~~ \vdots~~~, \nonumber
\eeqa

\noindent
($P_\pm = P^1 \mp i P^2$).
Partant d'un vecteur primitif $\left(Q_{-1/F}^\pm\right)^F$
$\Big(\left[J_\mp,\left(Q_{-1/F}^\pm\right)^F\right]=0\Big)$ , que l'on 
identifie avec $P_\mp$, on construit la repr\'esen\-tation vectorielle
\`a partir de ${\cal D}_{-1/F;\pm}^\pm$ par action successive de $J_\pm$.
Si on suppose que $\Big[J_+,\Big[J_+,$
$\Big[J_+,\left(Q^+_{-{1 \over F}}\right)^F\Big]\Big]\Big] \neq 0$, 
c'est-\`a-dire que partant d'un vide d'h\'elicit\'e $h=-1$, on peut
obtenir un \'etat d'h\'elicit\'e $h=2$, 
utilisant les m\'ethode usuelles de  th\'eorie des groupes, on arrive \`a 
une contradiction. En effet, une telle hypoth\`ese conduirait \`a
$\left[J_\mp, \left(Q^\pm_{-{1 \over F}}\right)^F
\right] \ne 0$, contredisant  le fait que
$\left(Q^\pm_{-{1 \over F}}\right)^F$ est un  vecteur primitif (voir
(\ref{eq:Q})).

Les relations (\ref{eq:PQ}) nous montrent explicitement que nous avons 
construit une application d'un sous-espace de 
${\cal S}^F({\cal D}_{-{1 \over F};\pm}^\pm)$ (le produit sym\'etrique
d'ordre $F$) dans ${\cal D}_{-1}$ 

$${\cal D}_{-1}=\left\{\left(Q^{\pm}_{-{1 \over F}}\right)^F,
\left[J_\pm,\left(Q^{\pm}_{-{1 \over F}}\right)^F\right],
\left[J_\pm,\left[J_\pm,\left(Q^{\pm}_{-{1 \over F}}\right)^F\right]\right] 
\right\}\equiv \left\{P_0, P_\pm\right\}.$$
Si maintenant on suppose que l'on puisse d\'ecomposer la 
troisi\`eme \'equation de (\ref{eq:PQ}) en deux termes ind\'ependants, on
peut montrer, en utilisant les identit\'es de Jacobi, que l'on obtient
une contradiction car on d\'ebouche sur une incoh\'erence. Ainsi donc, au fur
et \`a mesure que l'h\'elicit\'e du membre de droite va augmenter, on aura
de plus en plus de termes dans la somme. Le fait que l'on
obtienne des produits sym\'etriques est tout simplement une cons\'equence
de la r\`egle de Leibniz. La raison de ce comportement s'explique:
on veut construi\-re \`a partir de 
${\cal S}^F \left\{ {\cal D}_{-1/F;\pm}^\pm \right\}$ (le produit sym\'etrique
d'ordre $F$) la repr\'esentation vectorielle de $SO(1,2)$. La premi\`ere est
de dimension infinie et la seconde de dimension $3$. En fait, (\ref{eq:PQ})
ne repr\'esente que ${\cal S}^F \left\{ {\cal D}_{-1/F;\pm}^\pm \right\} \sim
{\cal D}_{-1} \oplus \cdots .$
Cependant, on observe qu'il est impossible de trouver une d\'ecomposition
de ${\cal S}^F\left({\cal D}^\pm_{-1/F;\pm}\right)$

\beq
\label{eq:decomp}
{\cal S}^F\left({\cal D}^\pm_{-1/F;\pm}\right) = {\cal D}_{-1} \oplus V,
\eeq

\noindent
pour laquelle $V$ soit stable par action de $SO(1,2)$. En effet, si tel
\'etait le cas, nous pourrions mettre en \'evidence une projection $SO(1,2)-$
\'equivariante

\beq
\label{eq:equiva}
\pi:~{\cal S}^F\left({\cal D}^\pm_{-1/F;\pm}\right) \longrightarrow 
{\cal D}_{-1}.
\eeq

\noindent
De ce fait,  les vecteurs $X^\pm=\pi\left( {\cal S}^F\left(Q^\pm_{-1/F},
\cdots,Q^\pm_{-1/F},Q^\pm_{3-1/F} \right) \right) \in {\cal D}_{-1}$ 
satisferaient

$$\big[J_\mp,\big[J_\mp,\big[J_\mp,X^\pm \big]\big]\big]= 
\pm i\sqrt{2/F}\sqrt{2(1-2/F)} \sqrt{3(2-2/F)} P_-\ne 0,$$

\noindent
ce qui est en contradiction avec le fait que $J_\pm^3$ est repr\'esent\'e
par z\'ero pour la repr\'esentation vectorielle.

Si $F=2$, on obtient ${\cal S}^2 \left\{ {\cal D}_{-1/2} \right\}=
{\cal D}_{-1}$ et une telle extension de l'alg\`ebre de Poincar\'e 
constitue l'extension supersym\'etrique que l'on peut exprimer en termes de 
matrices de Pauli. 

\subsection{Propri\'et\'es}

\noindent
{\large {\bf 3.3.3.1. Structure de l'alg\`ebre}}\\
\addcontentsline{toc}{subsection}{\numberline{}  
3.3.3.1. Structure de l'alg\`ebre }

Il est  \'egalement possible, lorsque l'ordre de la supersym\'etrie 
fractionnaire n'est pas un nombre premier ($F^\prime = fF$), de 
construire une extension de FSUSY d'ordre $F$ \`a partir 
d'une extension d'ordre
$fF$. En reprenant le langage et les notations 
que nous avons utilis\'ees, on 
construit une 
identification  ${\cal S}^f \left\{ {\cal D}_{-1/fF;\pm}^\pm \right\} \sim
{\cal D}_{-1/F;\pm}^\pm  \oplus \cdots$. Bien entendu, les restrictions 
\'enonc\'ees
resteront vraies

\vfill \eject
\beqa
\label{eq:incl3d}
&&\left(Q^\pm_{-{1/ fF}}\right)^f = Q^\pm_{-{1/F}}  \nonumber \\
&&~~~~~~~~~~~~ \vdots \\
&& \left[J_\mp, \cdots , \left[J_\pm,\left(Q^\pm_{-{1/ fF}}\right)^f
\right] \cdots \right] \sim Q^\pm_{-{1/F} +n} \nonumber \\
&&~~~~~~~~~~~~ \vdots~. \nonumber
\eeqa 

\noindent
Toutes les propri\'et\'es mentionn\'ees lors du plongement de
${\cal S}^F \left\{ {\cal D}_{-1/F;\pm}^\pm \right\}$ dans la repr\'esentation
vectorielle restent vraies. Si en outre $F$ est pair et $f=F/2$,  
(\ref{eq:incl3d}) conduit aux charges de la supersym\'etrie et on obtient
une repr\'esentation de dimension finie; donc comme pr\'ec\'edem\-ment,
le premier membre de
l'\'equation s'annule (apr\`es l'application de l'op\'erateur $J_\pm$ plus de
deux fois).\\

L'existence de sous-sym\'etries peut \^etre mise en relation avec les
r\'esultats de \cite{sem} o\`u une supersym\'etrie entre particules
de spin $1/4$ et $3/4$ (semions) a \'et\'e mise en \'evidence.\\

\noindent\\
{\large {\bf 3.3.3.2. Extensions avec charges centrales}}\\
\addcontentsline{toc}{subsection}{\numberline{}  
 3.3.3.2. Extensions avec charges centrales}

Tout comme dans les th\'eories supersym\'etriques usuelles, il est possible
de construire des extensions de l'alg\`ebre de Poincar\'e avec 
$N$ s\'eries de supercharges dans la repr\'esentation ${\cal D}_{-1/F;\pm}^\pm$
de $SO(1,2)$. On peut montrer que l'on obtient des extensions avec des
charges centrales.\\

%\vfill \eject
\noindent
{\large {\bf 3.3.3.3. Propri\'et\'es g\'en\'erales}}\\
\addcontentsline{toc}{subsection}{\numberline{}  
3.3.3.3. Propri\'et\'es g\'en\'erales }

Dans la prochaine section, nous allons rechercher les repr\'esentations de 
FSUSY. Cependant, on peut d\'ej\`a tirer un certain nombre de conclusions.
Il est facile de se rendre compte que $P^2$ est un op\'erateur de Casimir
donc tous les \'etats apparaissant dans une repr\'esentation irr\'eductible
auront la m\^eme masse. Ensuite, si on introduit
un op\'erateur nombre anyo\-nique ${\cal N}_{{\cal A}}$ satisfaisant
$\exp({\cal N}_{{\cal A}}) Q_s=q^s Q_s \exp({\cal N}_{{\cal A}})$, par une
technique analogue \`a celle utilis\'ee en supersym\'etrie
(en montrant que ${\mathrm{tr}} \left( \exp({2i\pi{\cal N}_A})
\left\{Q^+_{-{1\over F}},\dots, Q^+_{-{1\over F}},Q^+_{1-{1\over F}}\right\}_F
\right)=0$), on obtient
que $ {\mathrm{tr}} \left(\exp({\cal N}_{{\cal A}})\right)=0$. Ceci nous permet
de conclure que chaque repr\'esentation irr\'eductible contient $F$ 
statistiques diff\'erentes $(\lambda,\lambda-1/F, \cdots,\lambda -(F-1/F))$
--$\lambda$ sera sp\'ecifi\'e
ult\'erieurement--  et que le nombre d'\'etats d'une statistique donn\'ee est
ind\'ependant de sa statistique; ceci revient \`a dire que la dimension
de $E_{\lambda -a/F}$ ne d\'epend pas de $a$ ($E_{ \lambda-a/F}$ 
\'etant l'espace des \'etats de statistique $ \lambda -a/F$).
 Un tel r\'esultat
a \'et\'e obtenu en dimension deux par Saidi {\it et al} \cite{fsusy2d}.
Il faut cependant mentionner que la trace doit \^etre construite avec 
pr\'ecaution car nous avons affaire \`a des alg\`ebres de dimension infinie.

\subsection{Repr\'esentations de l'alg\`ebre}

Pour des raisons de commodit\'e, nous allons rechercher les repr\'esentations
de l'alg\`ebre dans le cas le plus simple. Nous allons choisir une s\'erie
de charges que nous fixons dans la repr\'esentation ${\cal D}_{-1/F;\pm}^\pm$.
Nous allons voir que l'unitarit\'e de la repr\'esentation impose que l'on 
prenne des charges dans les repr\'esentations ${\cal D}_{-1/F;+}^+$ 
et  ${\cal D}_{-1/F;-}^-$.

Les repr\'esentations massives sont obtenues en \'etudiant la sous-alg\`ebre
pr\'eservant l'impul\-sion au repos $p^\alpha=(m,0,0)$. De ce fait, 
les \'etats 
\`a une particule seront d\'etermin\'es  par la valeur propre associ\'ee aux
rotations dans le plan $(x^1,x^2)$, 
c'est-\`a-dire l'h\'elicit\'e. Si on regarde
la structure du second membre de (\ref{eq:PQ}), on observe que les relations
invoquant les charges $Q_{-1/F+1 + n}^+, n > 0$ admettent toujours un 
second membre nul. Ceci nous sugg\`ere fortement de repr\'esenter ces 
charges par z\'ero.
Ce faisant, apr\`es un choix de normalisation appropri\'e, on obtient

\beqa
\label{eq:PQr}
&&\left\{Q^+_{-{1 \over F}}, \dots,Q^+_{-{1 \over F}},Q^+_{1-{1 \over F}}
 \right\}_F = 1/F  \\
&&\left\{Q^+_{s_1},, \dots,Q^+_{s_F} \right\}_F = 0,~~ ,
i_1,\cdots i_F=-1/F,1-1/F~~ \mathrm {and },~~i_1 + \dots + i_F \neq 0.
\nonumber
\eeqa

Si on se reporte au chapitre 1, on voit que $Q^+_{-1/F}$ et $Q^+_{1 -1/F}$ 
engendrent
l'alg\`ebre de Clifford du polyn\^ome $x^{F-1}y$, justifiant a posteriori 
notre choix $Q_{-1/F+1 + n}=0, n > 0$. Tout comme en dimension $1$ et $2$,
l'alg\`ebre FSUSY est associ\'ee \`a l'alg\`ebre de Clifford de $x^{F-1}y$.
De ce fait, pour obtenir les repr\'esentations de (\ref{eq:P}), ({\ref{eq:Q}) 
et (\ref{eq:PQ}), il suffit de conna\^{\i}tre celles de $\C_{x^{F-1}y}$ 
\cite{fsusy3d}.
De telles repr\'esentations ont \'et\'e mises en \'evidence dans la section
(1.1.3). On voit donc transpara\^{\i}tre une certaine unit\'e dans cette
s\'erie d'extensions alg\'ebriques et on peut comparer l'\'etude de ses 
repr\'esentations \`a celles d\'eriv\'ees en supersym\'etrie \cite{repsusy}. 
On obtient donc

\beq
\label{eq:Q+}
Q^+_{-{1\over F}}=\pmatrix{0&0&0&\ldots&0&0& \cr
                         1&0&0&\ldots&0&0& \cr
                         0&1&0&\ldots&0&0& \cr
                         &\cr
                         \vdots&\vdots&&\ddots&\ddots&\vdots   \cr
                         0&0&\ldots&0&1&0& }
~~\mathrm{et}~~ Q^+_{1-{1\over F}}=
                  \pmatrix{0&0&0&\ldots&0&1 \cr
                          0&0&0&\ldots&0&0& \cr
                          0&0&0&\ldots&0&0& \cr
                          &\cr\
                          \vdots&\vdots&&&\ddots&\vdots& \cr 
                          0&0&0&\ldots&0&0&}.
\eeq

Cependant, les matrices obtenues ci-dessus ne sont pas 
appropri\'ees pour mettre
en \'evidence que la repr\'esentation de l'alg\`ebre FSUSY est unitaire.
Autrement dit, elles ne conduisent pas \`a des relations quadratiques
simples entre $Q^+$ et $\left(Q^+\right)^\dag=Q^-$. De ce fait, il est
pr\'ef\'erable de choisir d'autres repr\'esentations matricielles (reli\'ees
\`a (\ref{eq:Q+}) par un changement d'\'echelle)

{\tiny
\beqa
\label{eq:Q+1}
\begin{array}{ll}
Q^+_{-{1\over F}}=\left\{ 
\begin{array}{l}\pmatrix{0&0&0&\ldots&0&0& \cr
                         \sqrt{[1]}&0&0&\ldots&0&0& \cr
                         0&\sqrt{[2]} &0&\ldots&0&0& \cr
                         &\cr
                         \vdots&\vdots&&\ddots&\ddots&\vdots   \cr
                         0&0&\ldots&0&\sqrt{[F-1]}&0& } \cr
                      ~~~~~~ \cr
                 \pmatrix{0&0&0&\ldots&0&0& \cr
                         \sqrt{1(F-1)}&0&0&\ldots&0&0& \cr
                         0& \sqrt{2(F-2)}&0&\ldots&0&0& \cr
                         &\cr
                         \vdots&\vdots&&\ddots&\ddots&\vdots   \cr
                         0&0&\ldots&0&\sqrt{(F-1)1}&0& } 
\end{array} 
\right.
Q^+_{1-{1\over F}}=\left\{
\begin{array}{ll} 
                  \pmatrix{0&0&0&\ldots&0&\left\{\sqrt{[F-1]!}\right\}^{-1} \cr
                          0&0&0&\ldots&0&0& \cr
                          0&0&0&\ldots&0&0& \cr
                          &\cr\
                          \vdots&\vdots&&&\ddots&\vdots& \cr 
                           0&0&0&\ldots&0&0&} \cr
                        ~~~~~~ \cr
                \pmatrix{0&0&0&\ldots&0&1/(F-1)!\cr
                          0&0&0&\ldots&0&0& \cr
                          0&0&0&\ldots&0&0& \cr
                          &\cr\
                          \vdots&\vdots&&&\ddots&\vdots& \cr 
                           0&0&0&\ldots&0&0&} 
  
\end{array} 
\right.
\end{array} \nonumber \\
\eeqa  
}                      

\noindent
Il est clair que les trois s\'eries de matrices (\ref{eq:Q+},\ref{eq:Q+1})
sont reli\'ees par un changement de base (ou un changement d'\'echelle).
A partir de $Q^+$, on construit alors $Q^-$

\beqa
\label{eq:qdag}
Q^-_{-{1\over F}}&=&\left(Q^+_{-{1\over F}}\right)^\dag \\
Q^-_{1-{1\over F}}&=&\left(Q^+_{1-{1\over F}}\right)^\dag  \nonumber           
\eeqa  

\noindent
On observe alors deux cons\'equences de la repr\'esentation obtenue:

\begin{enumerate}
\item
Un calcul direct nous prouve que les matrices $Q^+_{-1/F}$ et $Q^-_{-1/F}$
satisfont des relations quadratiques
\begin{enumerate}
\item
Avec le premier choix propos\'e en (\ref{eq:Q+1}), on montre que l'on obtient
l'alg\`ebre des $q-$oscillateurs

\beqa
\label{eq:q+os}
&&Q^-_{-1/F} Q^+_{-1/F} - q^{\pm 1/2} Q^+_{-1/F} Q^-_{-1/F} = q^{\mp N/2} \\
\nonumber
&&[N,Q^+_{-1/F}]=Q^+_{-1/F} \\
&&[N,Q^-_{-1/F}]=-Q^-_{-1/F},  \nonumber
\eeqa
\noindent
et $N={\mathrm{diag}}(0,1,\cdots,F-1)$ l'op\'erateur nombre.
\item
Alors qu'avec le second, on obtient
\beqa
\label{eq:sl}
&&[Q^-_{-1/F}, Q^+_{-1/F}]  =  N =\mathrm{diag} (F-1,F-3,\cdots,
1-F) \\
&&[ N,Q^\pm_{-1/F}]= \mp 2  Q^\pm_{-1/F}, \nonumber
\eeqa
et  $Q^\pm$ engendrent la repr\'esentation de dimension $F$ de $SL(2,\RR)$. 
\end{enumerate}

Parmi ces  s\'eries de matrices, nous n'avons pas pu trouver d'argument
permettant de choisir l'une des s\'eries  plut\^ot qu'une autre.
Autrement, dit nous ne sommes pas parvenus \`a mettre en \'evidence de 
relations
quadratiques {\it intrins\`eques} (ind\'ependamment de toute r\'ealisation
matricielle)  de l'alg\`ebre FSUSY.

Les relations quadratiques (\ref{eq:q+os}) ou (\ref{eq:sl}) nous permettent 
de montrer que la repr\'esentation de l'alg\`ebre FSUSY est unitaire.
C'est-\`a-dire que si l'on d\'efinit un vide $|0>$ annihil\'e par
$Q^-_{-1/F}$ (voir ci-apr\`es), les \'etats $\left(Q^+_{-1/F}\right)^n
|0>,~ n=0,
\cdots,F-1$ forment une base orthonorm\'ee de l'espace de Hilbert.

\item
Deuxi\`emement, nous avons
$$Q^+_{1-1/F}= {1 \over [F-1]!} \left(Q^-_{-1/F}\right)^{F-1};$$
\noindent dans le cas des $q-$oscillateurs et
$$Q^+_{1-1/F}={1 \over ((F-1)!)^2} \left(Q^-_{-1/F}\right)^{F-1}$$
pour $SL(2,\RR)$. De ce fait la repr\'esentation est construite uniquement
avec les matrices $Q_{-1/F}^\pm$.
On peut se poser la question des autres choix possibles,  
pour $Q^+_{1-1/F}$,
analogues \`a la matrice $Y$  pour $F=3$ (voir section (1.3.1)).
Etant donn\'e que $Q^+_{1-1/F}$ augmente l'h\'elicit\'e de $1-1/F$ et que
$Q^+_{-1/F}$ l'augmente de $-1/F$, l'unique choix compatible avec la 
structure du groupe
de Poincar\'e est celui propos\'e.
 Autrement dit, le fait
d'imposer que les matrices $Q$ soient dans une repr\'esentation du petit
groupe s\'electionne automatiquement la repr\'esentation la plus simple parmi
toutes les repr\'esentations de $\C_{x^{F-1}y}$.
\end{enumerate}

Pour construire la repr\'esentation de FSUSY,
on choisit un vide $\Omega_\lambda$ de spin $\lambda$,  
ce qui donne, d'apr\`es les r\'esultats de Jackiw et Nair \cite{jn},
plus le fait que les repr\'esentations de $SO(2)$ sont r\'eelles

\beq
\label{eq:vide}
\Omega_\lambda = \Omega_{\lambda}^+ \oplus  \Omega_{-\lambda}^-,
\eeq

\noindent
c'est-\`a-dire que nous devons consid\'erer un vide d'h\'elicit\'e $h=\lambda$
et d'\'energie positive ainsi qu'un vide d'h\'elicit\'e $h=-\lambda$
et d'\'energie n\'egative. Ces deux \'etats sont $CPT-$conjugu\'es. 

Le vide $\Omega_{\lambda}^+$ est 
annihil\'e par $Q_{-1/F}^-$, les divers \'etats sont engendr\'es par
l'action successive de $Q^+_{-1/F}$.
L'invariance
par conjugaison $CPT$ nous force \`a choisir $Q^+_{-1/F}\Omega_{\lambda}^- =0$.
On obtient donc la repr\'esentation de l'alg\`ebre \cite{fsusy3d}
(avec les normalisations des $q-$oscillateurs)

$$\vbox{\offinterlineskip \halign{
\tv# & \cc{#} & \tv# & \cc{#}  & \tv# &
\cc{#} & \tv# & \cc{#} & \tv# & \cc{#}& \tv# \cr
\noalign{\hrule}
&\cc{states}&&\cc{helicity} &&\cc{states}&&\cc{helicity} & \cr
\noalign{\hrule}
&$\Omega_{\lambda}^+$&&$\lambda$&
&$\Omega_{-\lambda}^-$&&$-\lambda$& \cr
\noalign{\hrule}
&$Q^+_{-1/F}\Omega_{\lambda}^+$&&$\lambda-1/F$&
&$Q^-_{-1/F}\Omega_{-\lambda}^-$&
&$-\lambda+1/F$& \cr
\noalign{\hrule}
&$\vdots$&& && &&$\vdots$& \cr
\noalign{\hrule}
&${\left(Q^+_{-1/F}\right)^a \over \sqrt{[a]!}}\Omega_{\lambda}^+$&
&$\lambda-a/F$&
&${\left(Q^-_{-1/F}\right)^a \over \sqrt{[a]!}}\Omega_{-\lambda}^-$&
&$-\lambda+a/F$ &\cr
\noalign{\hrule}
&$\vdots$&& && &&$\vdots$& \cr
\noalign{\hrule}
&${\left(Q^+_{-1/F}\right)^{F-1} \over \sqrt{[F-1]!}}\Omega_{\lambda}^+$&
&$\lambda-(F-1)/F$&
&${\left(Q^-_{-1/F}\right)^{F-1}\over \sqrt{[F-1]!}}\Omega_{-\lambda}^-$&
&$-\lambda+(F-1)/F$ &\cr
\noalign{\hrule}
}}$$

\noindent
On illustre  
explicitement la  propri\'et\'e qui affirme que l'on a autant d'\'etats
de spin $0$,  de spin $1/F$ $\cdots$ et  de spin
$(F-1)/F)$ (modulo $\lambda$, le spin du vide).
Les \'etats d'\'energie positive et d'h\'elicit\'e ($\lambda,\lambda -
{1 \over F}, \dots,\lambda -{F-1 \over F}$), sont $CPT-$conjugu\'es des
\'etats d'\'energie n\'egative et  d'h\'elicit\'e ($-\lambda,-\lambda +{1 \over
F}, \dots,-\lambda +{F-1 \over F})$. 
Outre le fait que la repr\'esentation est unitaire, le second choix
pour les matrices $Q$ (\ref{eq:Q+1}) permet de conclure que la repr\'esentation
de la FSUSY est dans une repr\'esentation de dimension $F$ de $SU(2)$.

Une telle m\'ethode, lorsque l'on consid\`ere un nombre arbitraire $N$ de 
repr\'esentations ${\cal D}_{-1/F;\pm}^\pm$, est transposable. Dans le cas 
le plus
simple 
\footnote{Il n'est pas toujours possible d'\'ecrire $\left\{Q_{-1/F}^{i_1},
\cdots,Q_{-1/F}^{i_F} \right\}= \delta^{i_1 \cdots i_F} P_-$ o\`u $Q^{i}$ 
correspond aux diff\'erentes familles de supercharges. En g\'en\'eral, \`a la
diff\'erence des th\'eories supersym\'etriques usuelles, 
le coefficient
multiplicatif de $P_-$ est un tenseur sym\'etrique d'ordre $F$ 
non-forc\'ement \'equivalent
\`a $\delta^{i_1 \cdots i_F}$, le tenseur de Kronecker g\'en\'eralis\'e.},
et en l'absence de charge centrale, on doit mettre en \'evidence les
repr\'esentations de l'alg\`ebre de Clifford du polyn\^ome 
$x_1^{F-1}y_1 + \cdots + x_{N}^{F-1}y_{N}^{~}$
(ceci renforce encore l'analogie avec la supersym\'etrie \cite{repsusy}).
Un choix probable consisterait \`a prendre $N$ s\'eries de matrices
$q-$mutantes les unes par rapport aux autres (voir chapitre 1).

Enfin, notons que lorsque $F$ n'est pas premier, comme en dimension $1$ et 
$2$, on peut d\'ecomposer le supermultiplet en sous-supermultiplets
(voir eq.(\ref{eq:fmult})).

\subsection{Perspectives}

Nous avons \'et\'e capables de mettre en \'evidence les repr\'esentations
unitaires
de l'alg\`ebre FSUSY en $3D$. Est-il possible d'exhiber des repr\'esentations
hors couche de masse et de d\'efinir des champs auxiliaires 
(comme en dimension $2$)? Dans cette voie, est-il possible de relier
les repr\'esentations de $SO(1,2)$ \`a l'approche exploit\'ee dans la section
(3.1), c'est-\`a-dire profiter de l'existence de repr\'esentations de dimension
infinie pour l'alg\`ebre de Clifford du polyn\^ome 
$x_0^2y_0-x_1^2y_1-x_2^2y_2$,
de  telle sorte que les variables de Grassmann g\'en\'eralis\'ees 
apparaissant soient susceptibles de d\'ecrire les degr\'es 
de libert\'e (infinis) de spin?
Comme \'etape ult\'erieure dans la compr\'ehension, on 
pourrait vouloir d\'efinir un mod\`ele de Wess-Zumino. Comme
point de d\'epart, on peut prendre les lagrangiens propos\'es dans 
\cite{jn,plu}
ou \'eventuellement l'\'equation (\ref{eq:3rac}) apr\`es avoir montr\'e sa
correspondance avec un anyon.
On peut \'egalement se poser  la  question  suivante: est-il possible de jauger
la FSUSY en $3D$, et de d\'efinir des repr\'esentations non-massives
de (\ref{eq:P}), (\ref{eq:Q}) et (\ref{eq:PQ})? Et a plus long terme, on peut
s'interroger sur la pertinence d'un tel formalisme et de ses relations avec 
la physique \`a $3$ dimensions.

\vskip 1truecm
Pour conclure cette \'etude g\'en\'erale de la FSUSY en dimension $1,2,3$,
il serait int\'eressant de relier ces r\'esultats \`a ceux de \cite{bl} o\`u il
a \'et\'e montr\'e que la FSUSY en dimension $1$ est un cas
particulier de la ligne quantique. Il serait \'egalement int\'eressant de
construire un formalisme g\'en\'eral, analogue aux alg\`ebres de Lie
$\ZZ_2-$gradu\'ees, adapt\'e pour $F$ arbitraire.

\chapter*{\center{ Conclusion}}
\stepcounter{chapter}{}
%pagestyle{empty}
\addcontentsline{toc}{chapter}{\numberline{} \hskip -.78cm Conclusion}

\renewcommand{\chaptermark}[1]{}{\markboth{}{}}
\renewcommand{\sectionmark}[1]{\markright{}}

Nous avons montr\'e, tout au long de ce travail, comment une classe 
particuli\`ere d'extensions d'alg\`ebres de Clifford et de Grassmann permettait
de munir naturellement une structure non-commutative
d'une diff\'erentielle ou, ce qui revient au m\^eme, de d\'efinir des 
$q-$oscillateurs. Profitant des propri\'et\'es sp\'ecifiques des espaces de
dimension un, deux ou trois, nous avons  construit une extension 
non-triviale des th\'eories supersym\'etriques. Celles-ci
ont ensuite engendr\'e une th\'eorie des champs appropri\'ee. 
Pour chacune des dimensions consid\'er\'ees $(D=1,2,3)$, nous avons mis en 
avant une s\'erie de  probl\`emes non r\'esolus, ce qui nous a
conduit \`a  envisager un certain nombre de perspectives.

Mentionnons celles qui nous paraissent les plus prometteuses. Devant
la richesse des th\'eories conformes, une connexion de la FSUSY, dans sa
pr\'esente version ou pas, avec les syst\`emes int\'egrables, nous semble
riche de perspectives. Par ailleurs, en marge des 
th\'eories sur les $q-$oscillateurs,
s'est d\'evelopp\'ee  l'\'etude des statistiques $q-$d\'eform\'ees.
Devant la profonde similitude des relations de $q-$mutation 
entre les oscillateurs et de celles propos\'ees par Greenberg pour
les quons \cite{gm}, on peut envisager  une \'etude de la FSUSY en relation 
avec ce nouveau  type de statistique. Il serait \'egalement int\'eressant
de pouvoir relier la FSUSY en dimension $1$ et $3$,
en relation avec une approche bas\'ee sur la th\'eorie des groupes pour
la description  des \'etats de spin fractionnaire ou les anyons.  
A plus long terme, on pourrait  \'etudier les relations de la FSUSY
avec la physique en $3d$. Tout au long de cette \'etude sur
les alg\`ebres de Clifford et la FSUSY, nous avons per\c cu des points
communs avec les groupes quantiques, r\'esultats qui ont \'et\'e renforc\'es
par les travaux de Azcarraga {\it et al} \cite{bl}. Il serait peut-\^etre
souhaitable d'envisager une approche de la FSUSY sous l'angle des groupes
quantiques, de fa\c con \`a d\'egager une assise et un \'eclairage diff\'erent
(et unitaire ?). Au cours du dernier chapitre, on a relev\'e deux 
probl\`emes,
r\'esolus partiellement et certainement corr\'el\'es: celui du statut des
g\'en\'erateurs et de la construction d'une th\'eorie invariante locale. 
En effet, bien que l'alg\`ebre fondamentale n'admette pas
de g\'en\'erateurs de toutes les graduations, ce n'est
pas le cas au niveau de ses repr\'esentations, o\`u les champs de toutes
les graduations sont pr\'esents. Cette particularit\'e a compliqu\'e 
quelque peu la
construction d'une th\'eorie pr\'esentant une invariance locale; il
serait vivement souhaitable d'\'etudier l'extension aux structures 
non-quadratiques des m\'ethodes conduisant \`a des sym\'etries locales. 

La structure alg\'ebrique de la FSUSY est principalement bas\'ee sur
des crochets sym\'etri\-ques d'ordre $F$ se 
refermant sur des champs bosoniques, structure
pr\'esentant des similitudes avec la d\'efinition de l'alg\`ebre de Clifford
d'un polyn\^ome (voir eq. (\ref{eq:g})). Or nous avons, par le biais de 
repr\'esentations non-fid\`eles, remplac\'e, pour l'\'etude
de l'alg\`ebre de Clifford d'un polyn\^ome, les relations de fermeture 
d'ordre $F$ par
des relations quadratiques. Si on applique le m\^eme raisonnement \`a
(\ref{eq:PQ}), on pourrait d'une part, d\'efinir des op\'erateurs de toutes
les graduations, et d'autre part, introduire des relations de commutation
analogues \`a celles \'etablies pour les alg\`ebres de Lie g\'en\'eralis\'ees
\cite{gla} et de ce fait, on pourrait alors \'etudier les relations entre
la FSUSY et les r\'esultats de Wills Toro \cite{wt}.

L'approche que nous avons suivie est tr\`es sp\'ecifique car nous avons 
consid\'er\'e une repr\'esen\-tation non-fid\`ele d'un polyn\^ome particulier
($x^{F-1}y$).
A la diff\'erence des th\'eories quadratiques o\`u les
polyn\^omes sont tous \'equivalents, d\`es que le degr\'e 
est sup\'erieur \`a deux, ce
n'est plus le cas. On peut donc se demander si, \`a chaque famille de 
polyn\^omes et pour chaque repr\'esentation de son alg\`ebre de Clifford
associ\'ee, on peut construire une extension diff\'erente des th\'eories
supersym\'etriques. A titre d'illustration, la parasupersym\'etrie 
\cite{psusy1} ou les consid\'erations des r\'ef\'erences \cite{kern} sont
bas\'ees sur d'autres repr\'esentations de l'alg\`ebre $n-$ext\'erieure que
les variables de Grassmann g\'en\'eralis\'ees. Il semble donc qu'une
classification des diverses familles de polyn\^omes, et des
repr\'esentations de son alg\`ebre de Clifford associ\'ee,  permettrait
de mettre un peu d'ordre dans les extensions de SUSY envisag\'ees.
A cet \'egard, mentionnons les travaux de Durand \cite{dur1} o\`u une connexion
entre les th\'eories FSUSY et PSUSY est mise en \'evidence.

On voit donc que d'un point de vue math\'ematique et physique, l'\'etude
des th\'eories FSUSY, ou de ses homologues,  est riche de perspectives, et que 
ceci permettra d'\'elucider un certain nombre de probl\`emes fondamentaux qui 
ne sont toujours pas r\'esolus.

\stepcounter{chapter}

\end{document}